\documentclass[a4paper,fleqn]{cas-sc}

\ExplSyntaxOn
\cs_set:Npn \__first_footerline:
{
  \group_begin:
  \small
  \sffamily
  \ifnum\theblind>0\relax
  \else 
    \__short_authors: :~
  \fi
  { \rmfamily \itshape A~preprint}
  \group_end:
}
\ExplSyntaxOff

\usepackage[numbers]{natbib}

\RequirePackage{amsthm,amsmath,amsfonts,amssymb}
\RequirePackage{graphicx}
\usepackage{bigints}
\usepackage{tikz}

\usepackage{subcaption}
\usepackage{caption} 
\usetikzlibrary{shapes.geometric}
\usetikzlibrary{backgrounds}

\usepackage{siunitx}

\theoremstyle{plain}

\newtheorem{theorem}{Theorem}[section]
\newtheorem{lemma}[theorem]{Lemma}
\newtheorem{corollary}[theorem]{Corollary}

\theoremstyle{remark}
\newtheorem{example}[theorem]{Example}

\newcommand{\Csixcomplement}{\begin{tikzpicture}[thick, scale=0.6, every node/.style={circle, minimum size = 3pt, scale=0.8}]
		\foreach \i [evaluate=\i as \label using int(\i+1)] in {0,...,5} {
			\coordinate (v\label) at ({360/6*\i}:2);
		}
		
		\draw [dashed] (v1) -- (v2)  node[midway, right] {$t_1$};
		\draw (v1) -- (v3);
		\draw (v1) -- (v4);
		\draw (v1) -- (v5);
		\draw [dashed] (v2) -- (v3)  node[midway, above] {$t_2$};
		\draw (v2) -- (v4);
		\draw (v2) -- (v5);
		\draw (v2) -- (v6);
		\draw [dashed] (v3) -- (v4)  node[midway, left] {$t_3$};
		\draw (v3) -- (v5);
		\draw (v3) -- (v6);
		\draw (v4) -- (v6);
		
		\foreach \i in {1,...,6} {
			\node at (v\i) [circle, fill=white, draw=black] {$v_\i$};
		}
\end{tikzpicture}}
\newcommand{\CfourA}{\begin{tikzpicture}[thick, scale=0.6, every node/.style={circle, minimum size = 3pt, scale=0.8}]
		\node[shape=circle,draw=black] (1) at (0,2) {$SL$};
		\node[shape=circle,draw=black] (2) at (2,2) {$SW$};
		\node[shape=circle,draw=black] (3) at (0,0) {$PL$};
		\node[shape=circle,draw=black] (4) at (2,0) {$PW$};
		
		\draw (2) -- (4);
		\draw (3) -- (4);
		\draw (1) -- (3);
		\draw (2) -- (1);
		
\end{tikzpicture}}
\newcommand{\CfourB}{\begin{tikzpicture}[thick, scale=0.6, every node/.style={circle, minimum size = 3pt, scale=0.8}]
		\node[shape=circle,draw=black] (1) at (0,2) {$SL$};
		\node[shape=circle,draw=black] (2) at (2,2) {$SW$};
		\node[shape=circle,draw=black] (3) at (0,0) {$PL$};
		\node[shape=circle,draw=black] (4) at (2,0) {$PW$};
		
		\draw (1) -- (4);
		\draw (1) -- (3);
		\draw (2) -- (3);
		\draw (2) -- (4);
\end{tikzpicture}}
\newcommand{\CfourC}{\begin{tikzpicture}[thick, scale=0.6, every node/.style={circle, minimum size = 3pt, scale=0.8}]
		\node[shape=circle,draw=black] (1) at (0,2) {$SL$};
		\node[shape=circle,draw=black] (2) at (2,2) {$SW$};
		\node[shape=circle,draw=black] (3) at (0,0) {$PL$};
		\node[shape=circle,draw=black] (4) at (2,0) {$PW$};
		
		\draw (1) -- (4);
		\draw (3) -- (4);
		\draw (2) -- (3);
		\draw (2) -- (1);
\end{tikzpicture}}
\newcommand{\Tsixthree}{\begin{tikzpicture}[thick, scale=0.6, every node/.style={circle, minimum size = 3pt, scale=0.8}]
		\foreach \i [evaluate=\i as \label using int(\i+1)] in {0,...,5} {
			\coordinate (v\label) at ({360/6*\i}:2);
		}
		
		\draw (v1) -- (v2);
		\draw (v3) -- (v4);
		\draw (v4) -- (v5);
		\draw (v1) -- (v6);
		\draw (v5) -- (v3);
		\draw (v2) -- (v6);
		\draw (v2) -- (v5);
		\draw (v3) -- (v6);
		\draw (v1) -- (v3);
		\draw (v2) -- (v4);
		\draw (v1) -- (v5);
		\draw (v4) -- (v6);
		\draw[dashed] (v2) -- (v3);
		\draw[dashed] (v4) -- (v1);
		\draw[dashed] (v5) -- (v6);
		
		\foreach \i in {1,...,6} {
			\node at (v\i) [circle, fill=white, draw=black] {$v_\i$};
		}
\end{tikzpicture}}
\newcommand{\CsixA}{\begin{tikzpicture}[thick, scale=0.6, every node/.style={circle, minimum size = 3pt, scale=0.8}]
		\foreach \i [evaluate=\i as \label using int(\i+1)] in {0,...,5} {
			\coordinate (v\label) at ({360/6*\i}:2);
		}
		
		\draw (v1) -- (v2);
		\draw (v2) -- (v3);
		\draw (v3) -- (v4);
		\draw (v4) -- (v5);
		\draw (v5) -- (v6);
		\draw (v1) -- (v6);
		\draw[dashed] (v1) -- (v3) node[midway, above, yshift = -0.5mm] {$t_1$};
		\draw[dashed] (v1) -- (v4) node[midway, above, yshift = -1mm] {$t_2$};
		\draw[dashed] (v1) -- (v5) node[midway, above, yshift = -0.5mm] {$t_3$};
		
		\foreach \i in {1,...,6} {
			\node at (v\i) [circle, fill=white, draw=black] {$v_\i$};
		}
\end{tikzpicture}}
\newcommand{\CsixB}{\begin{tikzpicture}[thick, scale=0.6, every node/.style={circle, minimum size = 3pt, scale=0.8}]
		\foreach \i [evaluate=\i as \label using int(\i+1)] in {0,...,5} {
			\coordinate (v\label) at ({360/6*\i}:2);
		}
		
		\draw (v1) -- (v2);
		\draw (v2) -- (v3);
		\draw (v3) -- (v4);
		\draw (v4) -- (v5);
		\draw (v5) -- (v6);
		\draw (v1) -- (v6);
		\draw[dashed] (v1) -- (v3) node[midway, above] {$t_1$};
		\draw[dashed] (v1) -- (v5) node[midway, above] {$t_2$};
		\draw[dashed] (v3) -- (v5) node[midway, left, xshift = 1.5mm] {$t_3$};
		
		\foreach \i in {1,...,6} {
			\node at (v\i) [circle, fill=white, draw=black] {$v_\i$};
		}
\end{tikzpicture}}
\newcommand{\CsixC}{\begin{tikzpicture}[thick, scale=0.6, every node/.style={circle, minimum size = 3pt, scale=0.8}]
		\foreach \i [evaluate=\i as \label using int(\i+1)] in {0,...,5} {
			\coordinate (v\label) at ({360/6*\i}:2);
		}
		
		\draw (v1) -- (v2);
		\draw (v2) -- (v3);
		\draw (v3) -- (v4);
		\draw (v4) -- (v5);
		\draw (v5) -- (v6);
		\draw (v1) -- (v6);
		\draw[dashed] (v1) -- (v4) node[midway, above, yshift = -1.5mm] {$t_2$};
		\draw[dashed] (v2) -- (v4) node[midway, above, yshift = -0.8mm] {$t_1$};
		\draw[dashed] (v1) -- (v5) node[midway, above, yshift = -1mm] {$t_3$};
		
		\foreach \i in {1,...,6} {
			\node at (v\i) [circle, fill=white, draw=black] {$v_\i$};
		}
\end{tikzpicture}}
\newcommand{\GstarA}{\begin{tikzpicture}[thick, scale=0.6, every node/.style={circle, minimum size = 3pt, scale=0.8}]
		\foreach \i [evaluate=\i as \label using int(\i+1)] in {0,...,7} {
			\coordinate (v\label) at ({360/8*\i}:2);
		}
		
		\draw (v1) -- (v2);
		\draw [dashed] (v1) -- (v3);
		\draw (v1) -- (v4);
		\draw (v1) -- (v5);
		\draw (v1) -- (v6);
		\draw (v1) -- (v7);
		\draw (v1) -- (v8);
		\draw [dashed] (v2) -- (v3);
		\draw (v2) -- (v4);
		\draw (v2) -- (v5);
		\draw (v2) -- (v6);
		\draw (v2) -- (v7);
		\draw (v2) -- (v8);
		\draw [dashed] (v3) -- (v4);
		\draw (v3) -- (v5);
		\draw (v3) -- (v6);
		\draw (v3) -- (v7);
		\draw (v3) -- (v8);
		\draw (v4) -- (v5);
		\draw (v4) -- (v6);
		\draw (v4) -- (v7);
		\draw (v4) -- (v8);
		\draw [dashed](v5) -- (v6);
		\draw (v5) -- (v7);
		\draw (v5) -- (v8);
		\draw [dashed] (v6) -- (v7);
		\draw (v6) -- (v8);
		\draw (v7) -- (v8);
		
		\foreach \i in {1,...,8} {
			\node at (v\i) [circle, fill=white, draw=black] {$v_\i$};
		}
\end{tikzpicture}}
\newcommand{\GstarB}{\begin{tikzpicture}[thick, scale=0.6, every node/.style={circle, minimum size = 3pt, scale=0.8}]
		\foreach \i [evaluate=\i as \label using int(\i+1)] in {0,...,7} {
			\coordinate (v\label) at ({360/8*\i}:2);
		}
		
		\draw [dashed] (v1) -- (v2);
		\draw [dashed] (v1) -- (v3);
		\draw [dashed] (v1) -- (v4);
		\draw (v1) -- (v5);
		\draw (v1) -- (v6);
		\draw (v1) -- (v7);
		\draw (v1) -- (v8);
		\draw (v2) -- (v3);
		\draw (v2) -- (v4);
		\draw (v2) -- (v5);
		\draw (v2) -- (v6);
		\draw (v2) -- (v7);
		\draw (v2) -- (v8);
		\draw (v3) -- (v4);
		\draw (v3) -- (v5);
		\draw (v3) -- (v6);
		\draw (v3) -- (v7);
		\draw (v3) -- (v8);
		\draw (v4) -- (v5);
		\draw (v4) -- (v6);
		\draw (v4) -- (v7);
		\draw (v4) -- (v8);
		\draw [dashed] (v5) -- (v6);
		\draw [dashed] (v5) -- (v7);
		\draw (v5) -- (v8);
		\draw (v6) -- (v7);
		\draw (v6) -- (v8);
		\draw (v7) -- (v8);
		
		\foreach \i in {1,...,8} {
			\node at (v\i) [circle, fill=white, draw=black] {$v_\i$};
		}
\end{tikzpicture}}
\newcommand{\GstarC}{\begin{tikzpicture}[thick, scale=0.6, every node/.style={circle, minimum size = 3pt, scale=0.8}]
		\foreach \i [evaluate=\i as \label using int(\i+1)] in {0,...,7} {
			\coordinate (v\label) at ({360/8*\i}:2);
		}
		
		\draw [dashed] (v1) -- (v2);
		\draw (v1) -- (v3);
		\draw (v1) -- (v4);
		\draw (v1) -- (v5);
		\draw (v1) -- (v6);
		\draw (v1) -- (v7);
		\draw (v1) -- (v8);
		\draw (v2) -- (v3);
		\draw (v2) -- (v4);
		\draw (v2) -- (v5);
		\draw (v2) -- (v6);
		\draw (v2) -- (v7);
		\draw (v2) -- (v8);
		\draw [dashed] (v3) -- (v4);
		\draw (v3) -- (v5);
		\draw (v3) -- (v6);
		\draw (v3) -- (v7);
		\draw (v3) -- (v8);
		\draw (v4) -- (v5);
		\draw (v4) -- (v6);
		\draw (v4) -- (v7);
		\draw (v4) -- (v8);
		\draw [dashed] (v5) -- (v6);
		\draw (v5) -- (v7);
		\draw (v5) -- (v8);
		\draw (v6) -- (v7);
		\draw (v6) -- (v8);
		\draw [dashed] (v7) -- (v8);
		
		\foreach \i in {1,...,8} {
			\node at (v\i) [circle, fill=white, draw=black] {$v_\i$};
		}
\end{tikzpicture}}
\newcommand{\KfiveKsix}{\begin{tikzpicture}[thick, scale=0.6, every node/.style={circle, minimum size = 3pt, scale=0.8}, rotate = 90]
		\foreach \i [evaluate=\i as \label using int(\i+1)] in {0,...,5} {
			\coordinate (v\label) at ({360/6*\i + 30}:2);
		}
		\coordinate (v7) at (-1,3.5); 
		\coordinate (v8) at (1,3.5); 
		
		\draw (v1) -- (v2);
		\draw (v1) -- (v3);
		\draw (v1) -- (v4);
		\draw (v1) -- (v5);
		\draw [dashed, red] (v1) -- (v6) node[pos=0.5, above] {$t_1$};
		\draw [dashed, blue] (v2) -- (v3) node[pos=0.45, right, xshift=-1.5mm] {$s_1$};
		\draw (v2) -- (v4);
		\draw (v2) -- (v5);
		\draw (v2) -- (v6);
		\draw [dashed, blue] (v3) -- (v4) node[pos=0.5, above, yshift=-1.5mm] {$s_2$};
		\draw (v3) -- (v5);
		\draw (v3) -- (v6);
		\draw (v4) -- (v5);
		\draw (v4) -- (v6);
		\draw [dashed, red] (v5) -- (v6) node[pos=0.5, right] {$t_2$};
		\draw (v1) -- (v7);
		\draw [dashed, cyan] (v1) -- (v8) node[pos=0.5, above] {$r_1$};
		\draw (v2) -- (v7);
		\draw (v2) -- (v8);
		\draw (v3) -- (v7);
		\draw (v3) -- (v8);
		\draw [dashed, cyan] (v7) -- (v8) node[pos=0.5, left] {$r_2$};
		
		\foreach \i in {1,...,8} {
			\node at (v\i) [circle, fill=white, draw=black] {$v_\i$};
		}
\end{tikzpicture}}
\newcommand{\KfiveKsixA}{\begin{tikzpicture}[thick, scale=0.6, every node/.style={circle, minimum size = 3pt, scale=0.8}, , rotate = 90]
		\foreach \i [evaluate=\i as \label using int(\i+1)] in {0,...,5} {
			\coordinate (v\label) at ({360/6*\i + 30}:2);
		}
		\coordinate (v7) at (-1,3.5); 
		\coordinate (v8) at (1,3.5); 
		
		\draw (v1) -- (v2);
		\draw (v1) -- (v3);
		\draw (v1) -- (v4);
		\draw (v1) -- (v5);
		\draw (v2) -- (v3);
		\draw (v2) -- (v4);
		\draw (v2) -- (v5);
		\draw (v2) -- (v6);
		\draw (v3) -- (v4);
		\draw (v3) -- (v5);
		\draw (v3) -- (v6);
		\draw (v4) -- (v5);
		\draw (v4) -- (v6);
		\draw (v1) -- (v7);
		\draw (v1) -- (v8);
		\draw (v2) -- (v7);
		\draw (v2) -- (v8);
		\draw (v3) -- (v7);
		\draw (v3) -- (v8);
		\draw (v7) -- (v8);
		\draw [dashed, red] (v5) -- (v6) node[pos=0.5, right] {$t_2$};
		\draw [dashed, red] (v1) -- (v6) node[pos=0.5, above] {$t_1$};
		
		\foreach \i in {1,...,8} {
			\node at (v\i) [circle, fill=white, draw=black] {$v_\i$};
		}
\end{tikzpicture}}
\newcommand{\KfiveKsixB}{\begin{tikzpicture}[thick, scale=0.6, every node/.style={circle, minimum size = 3pt, scale=0.8}, rotate = 90]
		\foreach \i [evaluate=\i as \label using int(\i+1)] in {0,...,5} {
			\coordinate (v\label) at ({360/6*\i + 30}:2);
		}
		\coordinate (v7) at (-1,3.5); 
		\coordinate (v8) at (1,3.5); 
		
		\draw (v1) -- (v2);
		\draw (v1) -- (v3);
		\draw (v1) -- (v4);
		\draw (v1) -- (v5);
		\draw (v1) -- (v6);
		\draw (v2) -- (v3);
		\draw (v2) -- (v4);
		\draw (v2) -- (v5);
		\draw (v2) -- (v6);
		\draw (v3) -- (v4);
		\draw (v3) -- (v5);
		\draw (v3) -- (v6);
		\draw (v4) -- (v5);
		\draw (v4) -- (v6);
		\draw (v5) -- (v6);
		\draw (v1) -- (v7);
		\draw (v2) -- (v7);
		\draw (v2) -- (v8);
		\draw (v3) -- (v7);
		\draw (v3) -- (v8);
		\draw [dashed, cyan] (v7) -- (v8) node[pos=0.5, right] {$r_2$};
		\draw [dashed, cyan] (v1) -- (v8) node[pos=0.5, above] {$r_1$};
		
		\foreach \i in {1,...,8} {
			\node at (v\i) [circle, fill=white, draw=black] {$v_\i$};
		}
\end{tikzpicture}}
\newcommand{\KfiveKsixC}{\begin{tikzpicture}[thick, scale=0.6, every node/.style={circle, minimum size = 3pt, scale=0.8}, rotate = 90]
		\foreach \i [evaluate=\i as \label using int(\i+1)] in {0,...,5} {
			\coordinate (v\label) at ({360/6*\i + 30}:2);
		}		
		\coordinate (v7) at (-1,3.5); 
		\coordinate (v8) at (1,3.5); 
		
		\draw (v1) -- (v2);
		\draw (v1) -- (v3);
		\draw (v1) -- (v4);
		\draw (v1) -- (v5);
		\draw (v1) -- (v6);
		\draw (v2) -- (v4);
		\draw (v2) -- (v5);
		\draw (v2) -- (v6);
		\draw (v3) -- (v5);
		\draw (v3) -- (v6);
		\draw (v4) -- (v5);
		\draw (v4) -- (v6);
		\draw (v5) -- (v6);
		\draw (v1) -- (v7);
		\draw (v1) -- (v8);
		\draw (v2) -- (v7);
		\draw (v2) -- (v8);
		\draw (v3) -- (v7);
		\draw (v3) -- (v8);
		\draw (v7) -- (v8);
		\draw [dashed, blue] (v3) -- (v4) node[pos=0.5, above, yshift=-1.5mm] {$s_2$};
		\draw [dashed, blue] (v2) -- (v3) node[pos=0.4, right, xshift=-1mm] {$s_1$};
		
		\foreach \i in {1,...,8} {
			\node at (v\i) [circle, fill=white, draw=black] {$v_\i$};
		}
\end{tikzpicture}}
\newcommand{\Gfivetwooneone}{\begin{tikzpicture}[thick, scale=0.6, every node/.style={circle, minimum size = 3pt, scale=0.8}]		
		\foreach \i [evaluate=\i as \label using int(\i+1)] in {0,...,4} {
			\coordinate (v\label) at ({360/5*\i+90}:2); 
		}
		\node[shape=circle,draw=black] (v6) at ({90-360/10}:2.5) {$u_3$};
		\node[shape=circle,draw=black] (v7) at ({90+25}:3) {$u_1$};
		\node[shape=circle,draw=black] (v8) at ({90+50}:3) {$u_2$};
		\node[shape=circle,draw=black] (v9) at (0:0) {$u_4$};
		
		\draw [dashed] (v1) -- (v2);
		\draw [dashed] (v1) -- (v3);
		\draw (v1) -- (v4);
		\draw [dashed] (v1) -- (v5);
		\draw (v2) -- (v3);
		\draw (v2) -- (v4);
		\draw (v2) -- (v5);
		\draw (v3) -- (v4);
		\draw (v3) -- (v5);
		\draw (v4) -- (v5);
		\draw (v7) -- (v8);
		\draw (v1) -- (v7);
		\draw (v1) -- (v8);
		\draw (v2) -- (v7);
		\draw (v2) -- (v8);
		\draw (v1) -- (v6);
		\draw (v5) -- (v6);
		\draw (v1) -- (v9);
		\draw (v3) -- (v9);
		
		\node at (v1) [circle, fill=white, draw=black] {$v_0$};
		\node at (v2) [circle, fill=white, draw=black] {$v_1$};
		\node at (v5) [circle, fill=white, draw=black] {$v_2$};
		\node at (v3) [circle, fill=white, draw=black] {$v_3$};
		\node at (v4) [circle, fill=white, draw=black] {$v_4$};
\end{tikzpicture}}
\newcommand{\Csixplusone}{\begin{tikzpicture}[thick, scale=0.6, every node/.style={circle, minimum size = 3pt, scale=0.8}]
		\node[shape=circle,draw=black] (v1) at (240:2) {$v_1$};
		\node[shape=circle,draw=black] (v2) at (120:2) {$v_2$};
		\node[shape=circle,draw=black] (v3) at (0,0) {$v_3$};
		\node[shape=circle,draw=black] (v4) at (180:2) {$v_4$};
		\node[shape=circle,draw=black] (v5) at (60:2) {$v_5$};
		\node[shape=circle,draw=black] (v6) at (360:2) {$v_6$};
		\node[shape=circle,draw=black] (v7) at (300:2) {$v_7$};
		
		\draw (v1) -- (v4);
		\draw (v2) -- (v4);
		\draw (v2) -- (v5);
		\draw (v6) -- (v5);
		\draw (v7) -- (v6);
		\draw (v1) -- (v7);
		\draw (v4) -- (v3);
		\draw (v3) -- (v6);
		\draw (v7) -- (v5);
		\draw[dashed] (v4) to[out=25,in=155,looseness=1] (v6);
		\draw[dashed] (v5) -- (v4);
		\draw[dashed] (v7) -- (v4);
\end{tikzpicture}}
\newcommand{\Cfive}{\begin{tikzpicture}[thick, scale=0.6, every node/.style={circle, minimum size = 3pt, scale=0.8}]
        \coordinate (v1) at (18:2);
		\foreach \i [evaluate=\i as \label using int(\i+1)] in {1,...,4} {
			\coordinate (v\label) at ({360/5*\i+18}:2);
		}
		
		\draw (v1) -- (v2);
		\draw (v2) -- (v3);
		\draw (v3) -- (v4);
		\draw (v4) -- (v5);
		\draw (v5) -- (v1);
		\draw[dashed] (v1) -- (v3) node[midway, above] {$t_1$};
		\draw[dashed] (v1) -- (v4) node[midway, above] {$t_2$};
		
		\foreach \i in {1,...,5} {
			\node at (v\i) [circle, fill=white, draw=black] {$v_\i$};
		}
\end{tikzpicture}}
\newcommand{\gearthree}{\begin{tikzpicture}[thick, scale=0.6, every node/.style={circle, minimum size = 3pt, scale=0.8}]
		\node[shape=circle,draw=black] (v1) at (0:0) {$v_0$};
		\node[shape=circle,draw=black] (v2) at (90:2) {$v_1$};
		\node[shape=circle,draw=black] (v3) at (210:2) {$v_3$};
		\node[shape=circle,draw=black] (v4) at (330:2) {$v_5$};
		\node[shape=circle,draw=black] (v5) at (150:2) {$v_2$};
		\node[shape=circle,draw=black] (v6) at (30:2) {$v_6$};
		\node[shape=circle,draw=black] (v8) at (270:2) {$v_4$};
		
		\draw (v2) -- (v5);
		\draw (v3) -- (v5);
		\draw (v2) -- (v6);
		\draw (v4) -- (v6);
		\draw (v3) -- (v8);
		\draw (v4) -- (v8);
		\draw (v1) -- (v5);
		\draw (v1) -- (v6);
		\draw (v1) -- (v8);
		\draw [dashed] (v6) -- (v5);
		\draw [dashed] (v8) -- (v6);
		\draw [dashed] (v5) -- (v8);
\end{tikzpicture} }
\newcommand{\gearfour}{\begin{tikzpicture}[thick, scale=0.6, every node/.style={circle, minimum size = 3pt, scale=0.8}]
		\node[shape=circle,draw=black] (v1) at (0,0) {$v_6$};
		\node[shape=circle,draw=black] (v2) at (0,2) {$v_7$};
		\node[shape=circle,draw=black] (v3) at (0,4) {$v_8$};
		\node[shape=circle,draw=black] (v4) at (2,0) {$v_5$};
		\node[shape=circle,draw=black] (v5) at (2,2) {$v_9$};
		\node[shape=circle,draw=black] (v6) at (2,4) {$v_1$};
		\node[shape=circle,draw=black] (v7) at (4,0) {$v_4$};
		\node[shape=circle,draw=black] (v8) at (4,2) {$v_3$};
		\node[shape=circle,draw=black] (v9) at (4,4) {$v_2$};
		
		\draw (v3) -- (v6);
		\draw (v6) -- (v9);
		\draw (v2) -- (v5);
		\draw (v5) -- (v8);
		\draw (v1) -- (v4);
		\draw (v4) -- (v7);
		\draw (v3) -- (v2);
		\draw (v1) -- (v2);
		\draw (v5) -- (v6);
		\draw (v5) -- (v4);
		\draw (v9) -- (v8);
		\draw (v7) -- (v8);
		\draw [dashed] (v6) -- (v8);
		\draw [dashed] (v2) -- (v6);
		\draw [dashed] (v2) -- (v4);
		\draw [dashed] (v4) -- (v8);
		\draw[dashed] (v6) to[out=-45,in=45,looseness=0.8] (v4) ;
\end{tikzpicture} }
\newcommand{\HA}{\begin{tikzpicture}[thick, scale=0.6, every node/.style={circle, minimum size = 3pt, scale=0.8}]
		\node[shape=circle,draw=black] (v2) at (90:2) {$v_2$};
		\node[shape=circle,draw=black] (v3) at (210:2) {$v_1$};
		\node[shape=circle,draw=black] (v4) at (330:2) {$v_3$};
		\node[shape=circle,draw=black] (v5) at (150:2.5) {$u_1$};
		\node[shape=circle,draw=black] (v6) at (15:2.5) {$w_1$};
		\node[shape=circle,draw=black] (v7) at (50:2.5) {$w_2$};
		\node[shape=circle,draw=black] (v8) at (270:2.8) {$x_1$};
		\node[shape=circle,draw=black] (v9) at (240:2.5) {$x_2$};
		\node[shape=circle,draw=black] (v10) at (300:2.5) {$x_3$};
		
		\draw[dashed] (v2) -- (v3) node[midway, left] {$t_1$};
		\draw[dashed] (v2) -- (v4) node[midway, right] {$t_2$};
		\draw[dashed] (v3) -- (v4) node[midway, above] {$t_3$};
		\draw (v2) -- (v5);
		\draw (v3) -- (v5);
		\draw (v2) -- (v6);
		\draw (v2) -- (v7);
		\draw (v4) -- (v6);
		\draw (v4) -- (v7);
		\draw (v6) -- (v7);
		\draw (v3) -- (v8);
		\draw (v3) -- (v9);
		\draw (v3) -- (v10);
		\draw (v4) -- (v8);
		\draw (v4) -- (v9);
		\draw (v4) -- (v10);
		\draw (v9) -- (v8);
		\draw (v10) -- (v9);
		\draw (v8) -- (v10);
\end{tikzpicture}}
\newcommand{\HB}{\begin{tikzpicture}[thick, scale=0.6, every node/.style={circle, minimum size = 3pt, scale=0.8}]
		\node[shape=circle,draw=black] (v2) at (90:2) {$v_2$};
		\node[shape=circle,draw=black] (v3) at (210:2) {$v_1$};
		\node[shape=circle,draw=black] (v4) at (330:2) {$v_3$};
		\node[shape=circle,draw=black] (v5) at (150:2.5) {$u_1$};
		\node[shape=circle,draw=black] (v6) at (15:2.5) {$w_1$};
		\node[shape=circle,draw=black] (v7) at (50:2.5) {$w_2$};
		\node[shape=circle,draw=black] (v8) at (270:2.8) {$x_1$};
		\node[shape=circle,draw=black] (v9) at (240:2.5) {$x_2$};
		\node[shape=circle,draw=black] (v10) at (300:2.5) {$x_3$};
		
		\draw[dashed] (v3) -- (v6) node[midway, above, yshift = -1mm] {$t_3$};
		\draw[dashed] (v3) -- (v7) node[midway, above] {$t_2$};
		\draw[dashed] (v3) -- (v2) node[midway, left] {$t_1$};
		\draw[dashed] (v3) -- (v4) node[midway, above, yshift = -1.5mm, xshift = 1mm] {$t_4$};
		\draw (v2) -- (v5);
		\draw (v3) -- (v5);
		\draw (v2) -- (v6);
		\draw (v2) -- (v7);
		\draw (v4) -- (v6);
		\draw (v4) -- (v7);
		\draw (v6) -- (v7);
		\draw (v3) -- (v8);
		\draw (v3) -- (v9);
		\draw (v3) -- (v10);
		\draw (v4) -- (v8);
		\draw (v4) -- (v9);
		\draw (v4) -- (v10);
		\draw (v9) -- (v8);
		\draw (v10) -- (v9);
		\draw (v8) -- (v10);
\end{tikzpicture} }

\newcommand{\GtwoA}{\begin{tikzpicture}[thick, scale=0.8, every node/.style={circle, minimum size = 3pt, scale=0.8}]
		\node[shape=circle,draw=black] (v1) at (1,0) {};
		\node[shape=circle,draw=black] (v2) at (-1,0) {};
		
		\draw (v1) -- (v2);
\end{tikzpicture}}
\newcommand{\GthreeA}{\begin{tikzpicture}[thick, scale=0.8, every node/.style={circle, minimum size = 3pt, scale=0.8}]
		\foreach \i [evaluate=\i as \label using int(\i+1)] in {0,...,2} {
			\node[shape=circle,draw=black] (v\label) at ({360/3*\i+90}:1) {};
		}
		
		\draw (v1) -- (v2);
		\draw (v1) -- (v3);
		\draw (v2) -- (v3);
\end{tikzpicture}}
\newcommand{\GfourA}{\begin{tikzpicture}[thick, scale=0.8, every node/.style={circle, minimum size = 3pt, scale=0.8}]
		\foreach \i [evaluate=\i as \label using int(\i+1)] in {0,...,3} {
			\node[shape=circle,draw=black] (v\label) at ({360/4*\i+135}:1) {};
		}
		
		\draw (v1) -- (v2);
		\draw (v1) -- (v3);
		\draw (v1) -- (v4);
		\draw (v2) -- (v3);
		\draw (v2) -- (v4);
		\draw (v3) -- (v4);
\end{tikzpicture}}
\newcommand{\GfourB}{	\begin{tikzpicture}[thick, scale=0.8, every node/.style={circle, minimum size = 3pt, scale=0.8}]
		\foreach \i [evaluate=\i as \label using int(\i+1)] in {0,...,3} {
			\node[shape=circle,draw=black] (v\label) at ({360/4*\i+135}:1) {};
		}		
		
		\draw (v1) -- (v2);
		\draw (v1) -- (v4);
		\draw (v2) -- (v3);
		\draw (v3) -- (v4);
		\draw[dashed] (v1) -- (v3) ;
\end{tikzpicture}}
\newcommand{\GfiveA}{\begin{tikzpicture}[thick, scale=0.8, every node/.style={circle, minimum size = 3pt, scale=0.8}]
		\foreach \i [evaluate=\i as \label using int(\i+1)] in {0,...,4} {
			\node[shape=circle,draw=black] (v\label) at ({360/5*\i+90}:1) {};
		}		
		
		\draw (v1) -- (v2);
		\draw (v1) -- (v3);
		\draw (v1) -- (v4);
		\draw (v1) -- (v5);
		\draw (v2) -- (v3);
		\draw (v2) -- (v4);
		\draw (v2) -- (v5);
		\draw (v3) -- (v4);
		\draw (v3) -- (v5);
		\draw (v4) -- (v5);
\end{tikzpicture}}
\newcommand{\GfiveB}{\begin{tikzpicture}[thick, scale=0.8, every node/.style={circle, minimum size = 3pt, scale=0.8}]
		\foreach \i [evaluate=\i as \label using int(\i+1)] in {0,...,4} {
			\node[shape=circle,draw=black] (v\label) at ({360/5*\i+90+72}:1) {};
		}		
		
		\draw (v1) -- (v2);
		\draw (v1) -- (v3);
		\draw (v1) -- (v4);
		\draw (v1) -- (v5);
		\draw [dashed] (v2) -- (v3);
		\draw (v2) -- (v4);
		\draw (v2) -- (v5);
		\draw (v3) -- (v4);
		\draw (v3) -- (v5);
\end{tikzpicture}}
\newcommand{\GfiveC}{\begin{tikzpicture}[thick, scale=0.8, every node/.style={circle, minimum size = 3pt, scale=0.8}]
		\foreach \i [evaluate=\i as \label using int(\i+1)] in {0,...,4} {
			\node[shape=circle,draw=black] (v\label) at ({360/5*\i+90}:1) {};
		}		
		
		\draw (v1) -- (v3);
		\draw (v1) -- (v4);
		\draw (v2) -- (v3);
		\draw (v2) -- (v4);
		\draw (v2) -- (v5);
		\draw [dashed] (v3) -- (v4);
		\draw (v3) -- (v5);
		\draw (v4) -- (v5);
\end{tikzpicture}}
\newcommand{\GfiveD}{\begin{tikzpicture}[thick, scale=0.8, every node/.style={circle, minimum size = 3pt, scale=0.8}]
		\foreach \i [evaluate=\i as \label using int(\i+1)] in {0,...,4} {
			\node[shape=circle,draw=black] (v\label) at ({360/5*\i+90}:1) {};
		}		
		
		\draw (v1) -- (v3);
		\draw (v1) -- (v4);
		\draw (v2) -- (v3);
		\draw (v2) -- (v4);
		\draw [dashed] (v3) -- (v4);
		\draw (v3) -- (v5);
		\draw (v4) -- (v5);
\end{tikzpicture}}
\newcommand{\GfiveE}{\begin{tikzpicture}[thick, scale=0.8, every node/.style={circle, minimum size = 3pt, scale=0.8}]
		\foreach \i [evaluate=\i as \label using int(\i+1)] in {0,...,4} {
			\node[shape=circle,draw=black] (v\label) at ({360/5*\i+90}:1) {};
		}		
		
		\draw (v1) -- (v2);
		\draw [dashed] (v1) -- (v3);
		\draw [dashed] (v1) -- (v4);
		\draw (v1) -- (v5);
		\draw (v2) -- (v3);
		\draw (v3) -- (v4);
		\draw (v4) -- (v5);
\end{tikzpicture}}
\newcommand{\GsixA}{\begin{tikzpicture}[thick, scale=0.8, every node/.style={circle, minimum size = 3pt, scale=0.8}]
		\foreach \i [evaluate=\i as \label using int(\i+1)] in {0,...,5} {
			\node[shape=circle,draw=black] (v\label) at ({360/6*\i+60}:1) {};
		}		
		
		\draw (v1) -- (v2);
		\draw (v1) -- (v3);
		\draw (v1) -- (v4);
		\draw (v1) -- (v5);
		\draw (v1) -- (v6);
		\draw (v2) -- (v3);
		\draw (v2) -- (v4);
		\draw (v2) -- (v5);
		\draw (v2) -- (v6);
		\draw (v3) -- (v4);
		\draw (v3) -- (v5);
		\draw (v3) -- (v6);
		\draw (v4) -- (v5);
		\draw (v4) -- (v6);
		\draw (v5) -- (v6);
\end{tikzpicture}}
\newcommand{\GsixB}{\begin{tikzpicture}[thick, scale=0.8, every node/.style={circle, minimum size = 3pt, scale=0.8}]
		\foreach \i [evaluate=\i as \label using int(\i+1)] in {0,...,5} {
			\node[shape=circle,draw=black] (v\label) at ({360/6*\i+60+60}:1) {};
		}		
		
		\draw (v1) -- (v2);
		\draw (v1) -- (v3);
		\draw (v1) -- (v4);
		\draw (v1) -- (v5);
		\draw (v1) -- (v6);
		\draw (v2) -- (v3);
		\draw (v2) -- (v4);
		\draw (v2) -- (v5);
		\draw (v2) -- (v6);
		\draw (v3) -- (v5);
		\draw (v3) -- (v6);
		\draw (v4) -- (v5);
		\draw (v4) -- (v6);
		\draw [dashed] (v3) -- (v4);
\end{tikzpicture}}
\newcommand{\GsixC}{\begin{tikzpicture}[thick, scale=0.8, every node/.style={circle, minimum size = 3pt, scale=0.8}]
		\foreach \i [evaluate=\i as \label using int(\i+1)] in {0,...,5} {
			\node[shape=circle,draw=black] (v\label) at ({360/6*\i+60+60}:1) {};
		}		
		
		\draw (v1) -- (v2);
		\draw (v1) -- (v3);
		\draw (v1) -- (v4);
		\draw (v1) -- (v5);
		\draw (v2) -- (v3);
		\draw (v2) -- (v4);
		\draw (v2) -- (v5);
		\draw (v2) -- (v6);
		\draw [dashed] (v3) -- (v4);
		\draw (v3) -- (v5);
		\draw (v3) -- (v6);
		\draw (v4) -- (v5);
		\draw (v4) -- (v6);
\end{tikzpicture}}
\newcommand{\GsixD}{\begin{tikzpicture}[thick, scale=0.8, every node/.style={circle, minimum size = 3pt, scale=0.8}]
		\foreach \i [evaluate=\i as \label using int(\i+1)] in {0,...,5} {
			\node[shape=circle,draw=black] (v\label) at ({360/6*\i+60+60}:1) {};
		}		
		
		\draw (v1) -- (v2);
		\draw (v1) -- (v3);
		\draw (v1) -- (v4);
		\draw (v1) -- (v5);
		\draw (v2) -- (v3);
		\draw (v2) -- (v4);
		\draw (v2) -- (v5);
		\draw [dashed] (v3) -- (v4);
		\draw (v3) -- (v5);
		\draw (v3) -- (v6);
		\draw (v4) -- (v5);
		\draw (v4) -- (v6);
\end{tikzpicture}}
\newcommand{\GsixE}{\begin{tikzpicture}[thick, scale=0.8, every node/.style={circle, minimum size = 3pt, scale=0.8}]
		\foreach \i [evaluate=\i as \label using int(\i+1)] in {0,...,5} {
			\node[shape=circle,draw=black] (v\label) at ({360/6*\i+60+60}:1) {};
		}		
		
		\draw (v1) -- (v3);
		\draw (v1) -- (v4);
		\draw (v1) -- (v5);
		\draw (v2) -- (v3);
		\draw (v2) -- (v4);
		\draw (v2) -- (v5);
		\draw (v2) -- (v6);
		\draw [dashed] (v3) -- (v4);
		\draw (v3) -- (v5);
		\draw (v3) -- (v6);
		\draw (v4) -- (v5);
		\draw (v4) -- (v6);
\end{tikzpicture}}
\newcommand{\GsixF}{\begin{tikzpicture}[thick, scale=0.8, every node/.style={circle, minimum size = 3pt, scale=0.8}]
		\foreach \i [evaluate=\i as \label using int(\i+1)] in {0,...,5} {
			\node[shape=circle,draw=black] (v\label) at ({360/6*\i+60+60}:1) {};
		}		
		
		\draw (v1) -- (v3);
		\draw (v1) -- (v4);
		\draw (v1) -- (v5);
		\draw (v2) -- (v3);
		\draw (v2) -- (v4);
		\draw (v2) -- (v6);
		\draw [dashed] (v3) -- (v4);
		\draw (v3) -- (v5);
		\draw (v3) -- (v6);
		\draw (v4) -- (v5);
		\draw (v4) -- (v6);
\end{tikzpicture}}
\newcommand{\GsixG}{\begin{tikzpicture}[thick, scale=0.8, every node/.style={circle, minimum size = 3pt, scale=0.8}]
		\foreach \i [evaluate=\i as \label using int(\i+1)] in {0,...,5} {
			\node[shape=circle,draw=black] (v\label) at ({360/6*\i+60+60}:1) {};
		}		
		
		\draw (v1) -- (v3);
		\draw (v1) -- (v4);
		\draw (v2) -- (v3);
		\draw (v2) -- (v4);
		\draw (v2) -- (v5);
		\draw [dashed] (v3) -- (v4);
		\draw (v3) -- (v5);
		\draw (v3) -- (v6);
		\draw (v4) -- (v5);
		\draw (v4) -- (v6);
\end{tikzpicture}}
\newcommand{\GsixH}{\begin{tikzpicture}[thick, scale=0.8, every node/.style={circle, minimum size = 3pt, scale=0.8}]
		\foreach \i [evaluate=\i as \label using int(\i+1)] in {0,...,5} {
			\node[shape=circle,draw=black] (v\label) at ({360/6*\i+60+60}:1) {};
		}		
		
		\draw (v1) -- (v3);
		\draw (v1) -- (v4);
		\draw (v2) -- (v3);
		\draw (v2) -- (v4);
		\draw [dashed] (v3) -- (v4);
		\draw (v3) -- (v5);
		\draw (v3) -- (v6);
		\draw (v4) -- (v5);
		\draw (v4) -- (v6);
\end{tikzpicture}}
\newcommand{\GsixI}{\begin{tikzpicture}[thick, scale=0.8, every node/.style={circle, minimum size = 3pt, scale=0.8}]
		\foreach \i [evaluate=\i as \label using int(\i+1)] in {0,...,5} {
			\node[shape=circle,draw=black] (v\label) at ({360/6*\i+60+60}:1) {};
		}		
		
		\draw (v1) -- (v3);
		\draw (v1) -- (v4);
		\draw (v1) -- (v5);
		\draw (v2) -- (v3);
		\draw (v2) -- (v4);
		\draw (v2) -- (v5);
		\draw [dashed] (v3) -- (v4);
		\draw (v3) -- (v5);
		\draw (v3) -- (v6);
		\draw (v4) -- (v5);
		\draw (v4) -- (v6);
		\draw (v5) -- (v6);
\end{tikzpicture}}
\newcommand{\GsixJ}{\begin{tikzpicture}[thick, scale=0.8, every node/.style={circle, minimum size = 3pt, scale=0.8}]
		\foreach \i [evaluate=\i as \label using int(\i+1)] in {0,...,5} {
			\node[shape=circle,draw=black] (v\label) at ({360/6*\i+60+60}:1) {};
		}		
		
		\draw (v1) -- (v3);
		\draw (v1) -- (v4);
		\draw (v1) -- (v5);
		\draw (v2) -- (v3);
		\draw (v2) -- (v4);
		\draw [dashed] (v3) -- (v4);
		\draw (v3) -- (v5);
		\draw (v3) -- (v6);
		\draw (v4) -- (v5);
		\draw (v4) -- (v6);
		\draw (v5) -- (v6);
\end{tikzpicture}}
\newcommand{\GsixK}{\begin{tikzpicture}[thick, scale=0.8, every node/.style={circle, minimum size = 3pt, scale=0.8}]
		\foreach \i [evaluate=\i as \label using int(\i+1)] in {0,...,5} {
			\node[shape=circle,draw=black] (v\label) at ({360/6*\i+60}:1) {};
		}		
		
		\draw (v1) -- (v3);
		\draw (v1) -- (v4);
		\draw (v2) -- (v4);
		\draw (v2) -- (v5);
		\draw (v2) -- (v6);
		\draw [dashed] (v3) -- (v4);
		\draw (v3) -- (v5);
		\draw (v3) -- (v6);
		\draw (v4) -- (v5);
		\draw (v4) -- (v6);
		\draw [dashed] (v5) -- (v6);
		
\end{tikzpicture}}
\newcommand{\GsixL}{\begin{tikzpicture}[thick, scale=0.8, every node/.style={circle, minimum size = 3pt, scale=0.8}]
		\foreach \i [evaluate=\i as \label using int(\i+1)] in {0,...,5} {
			\node[shape=circle,draw=black] (v\label) at ({360/6*\i+60}:1) {};
		}		
		
		\draw (v1) -- (v3);
		\draw (v1) -- (v4);
		\draw (v1) -- (v5);
		\draw (v2) -- (v4);
		\draw (v2) -- (v5);
		\draw (v2) -- (v6);
		\draw [dashed] (v3) -- (v4);
		\draw (v3) -- (v5);
		\draw (v3) -- (v6);
		\draw (v4) -- (v5);
		\draw (v4) -- (v6);
		\draw [dashed] (v5) -- (v6);
\end{tikzpicture}}
\newcommand{\GsixM}{\begin{tikzpicture}[thick, scale=0.8, every node/.style={circle, minimum size = 3pt, scale=0.8}]
		\foreach \i [evaluate=\i as \label using int(\i+1)] in {0,...,5} {
			\node[shape=circle,draw=black] (v\label) at ({360/6*\i+60}:1) {};
		}		
		
		\draw (v1) -- (v3);
		\draw (v1) -- (v4);
		\draw (v2) -- (v5);
		\draw (v2) -- (v6);
		\draw [dashed] (v3) -- (v4);
		\draw (v3) -- (v5);
		\draw (v3) -- (v6);
		\draw (v4) -- (v5);
		\draw (v4) -- (v6);
		\draw [dashed] (v5) -- (v6);
\end{tikzpicture}}
\newcommand{\GsixN}{\begin{tikzpicture}[thick, scale=0.8, every node/.style={circle, minimum size = 3pt, scale=0.8}]
		\foreach \i [evaluate=\i as \label using int(\i+1)] in {0,...,5} {
			\node[shape=circle,draw=black] (v\label) at ({360/6*\i+60}:1) {};
		}		
		
		\draw (v1) -- (v3);
		\draw (v1) -- (v4);
		\draw (v1) -- (v5);
		\draw (v1) -- (v6);
		\draw (v2) -- (v3);
		\draw (v2) -- (v4);
		\draw (v2) -- (v5);
		\draw (v2) -- (v6);
		\draw [dashed] (v3) -- (v4);
		\draw (v3) -- (v5);
		\draw (v3) -- (v6);
		\draw (v4) -- (v5);
		\draw (v4) -- (v6);
		\draw [dashed] (v5) -- (v6);
\end{tikzpicture}}
\newcommand{\GsixO}{\begin{tikzpicture}[thick, scale=0.8, every node/.style={circle, minimum size = 3pt, scale=0.8}]
		\foreach \i [evaluate=\i as \label using int(\i+1)] in {0,...,5} {
			\node[shape=circle,draw=black] (v\label) at ({360/6*\i+60}:1) {};
		}		
		
		\draw (v1) -- (v3);
		\draw (v1) -- (v4);
		\draw (v1) -- (v5);
		\draw (v2) -- (v3);
		\draw (v2) -- (v4);
		\draw (v2) -- (v5);
		\draw (v3) -- (v6);
		\draw (v4) -- (v6);
		\draw (v5) -- (v6);
\end{tikzpicture}}
\newcommand{\GsixP}{\begin{tikzpicture}[thick, scale=0.8, every node/.style={circle, minimum size = 3pt, scale=0.8}]
		\foreach \i [evaluate=\i as \label using int(\i+1)] in {0,...,5} {
			\node[shape=circle,draw=black] (v\label) at ({360/6*\i+60}:1) {};
		}		
		
		\draw (v1) -- (v4);
		\draw (v1) -- (v5);
		\draw (v2) -- (v5);
		\draw (v2) -- (v6);
		\draw (v3) -- (v4);
		\draw (v3) -- (v5);
		\draw (v3) -- (v6);
		\draw [dashed] (v4) -- (v5);
		\draw (v4) -- (v6);
		\draw [dashed] (v5) -- (v6);
\end{tikzpicture}}
\newcommand{\GsixQ}{\begin{tikzpicture}[thick, scale=0.8, every node/.style={circle, minimum size = 3pt, scale=0.8}]
		\foreach \i [evaluate=\i as \label using int(\i+1)] in {0,...,5} {
			\node[shape=circle,draw=black] (v\label) at ({360/6*\i+60}:1) {};
		}		
		
		\draw (v1) -- (v3);
		\draw (v1) -- (v4);
		\draw (v1) -- (v5);
		\draw (v2) -- (v5);
		\draw (v2) -- (v6);
		\draw (v3) -- (v4);
		\draw (v3) -- (v5);
		\draw (v3) -- (v6);
		\draw [dashed] (v4) -- (v5);
		\draw (v4) -- (v6);
		\draw [dashed] (v5) -- (v6);
\end{tikzpicture}}
\newcommand{\GsixR}{\begin{tikzpicture}[thick, scale=0.8, every node/.style={circle, minimum size = 3pt, scale=0.8}]
		\foreach \i [evaluate=\i as \label using int(\i+1)] in {0,...,5} {
			\node[shape=circle,draw=black] (v\label) at ({360/6*\i+60}:1) {};
		}		
		
		\draw (v1) -- (v3);
		\draw (v1) -- (v4);
		\draw (v1) -- (v5);
		\draw (v2) -- (v5);
		\draw (v2) -- (v6);
		\draw (v3) -- (v4);
		\draw (v3) -- (v5);
		\draw [dashed] (v4) -- (v5);
		\draw (v4) -- (v6);
		\draw [dashed] (v5) -- (v6);
\end{tikzpicture}}
\newcommand{\GsixS}{\begin{tikzpicture}[thick, scale=0.8, every node/.style={circle, minimum size = 3pt, scale=0.8}]
		\foreach \i [evaluate=\i as \label using int(\i+1)] in {0,...,5} {
			\node[shape=circle,draw=black] (v\label) at ({360/6*\i+60}:1) {};
		}		
		
		\draw (v1) -- (v4);
		\draw (v1) -- (v5);
		\draw (v1) -- (v6);
		\draw (v2) -- (v4);
		\draw (v2) -- (v5);
		\draw (v2) -- (v6);
		\draw (v3) -- (v4);
		\draw (v3) -- (v5);
		\draw (v3) -- (v6);
		\draw [dashed] (v4) -- (v5);
		\draw (v4) -- (v6);
		\draw [dashed] (v5) -- (v6);
\end{tikzpicture}}
\newcommand{\GsixT}{\begin{tikzpicture}[thick, scale=0.8, every node/.style={circle, minimum size = 3pt, scale=0.8}]
		\foreach \i [evaluate=\i as \label using int(\i+1)] in {0,...,5} {
			\node[shape=circle,draw=black] (v\label) at ({360/6*\i+60}:1) {};
		}		
		
		\draw (v1) -- (v4);
		\draw (v1) -- (v5);
		\draw (v2) -- (v4);
		\draw (v2) -- (v5);
		\draw (v3) -- (v5);
		\draw (v3) -- (v6);
		\draw [dashed] (v4) -- (v5);
		\draw (v4) -- (v6);
		\draw [dashed] (v5) -- (v6);
\end{tikzpicture}}
\newcommand{\GsixU}{\begin{tikzpicture}[thick, scale=0.8, every node/.style={circle, minimum size = 3pt, scale=0.8}]
		\foreach \i [evaluate=\i as \label using int(\i+1)] in {0,...,5} {
			\node[shape=circle,draw=black] (v\label) at ({360/6*\i+60}:1) {};
		}		
		
		\draw (v1) -- (v2);
		\draw (v1) -- (v4);
		\draw (v1) -- (v5);
		\draw (v1) -- (v6);
		\draw (v2) -- (v4);
		\draw (v2) -- (v5);
		\draw (v2) -- (v6);
		\draw (v3) -- (v4);
		\draw (v3) -- (v5);
		\draw (v3) -- (v6);
		\draw [dashed] (v4) -- (v5);
		\draw (v4) -- (v6);
		\draw [dashed] (v5) -- (v6);
\end{tikzpicture}}
\newcommand{\GsixV}{\begin{tikzpicture}[thick, scale=0.8, every node/.style={circle, minimum size = 3pt, scale=0.8}]
		\foreach \i [evaluate=\i as \label using int(\i+1)] in {0,...,5} {
			\node[shape=circle,draw=black] (v\label) at ({360/6*\i+60}:1) {};
		}		
		
		\draw (v1) -- (v3);
		\draw (v1) -- (v4);
		\draw (v1) -- (v5);
		\draw (v2) -- (v3);
		\draw (v2) -- (v5);
		\draw (v2) -- (v6);
		\draw (v3) -- (v4);
		\draw (v3) -- (v5);
		\draw (v3) -- (v6);
		\draw [dashed] (v4) -- (v5);
		\draw (v4) -- (v6);
		\draw [dashed] (v5) -- (v6);
\end{tikzpicture}}
\newcommand{\GsixW}{\begin{tikzpicture}[thick, scale=0.8, every node/.style={circle, minimum size = 3pt, scale=0.8}]
		\foreach \i [evaluate=\i as \label using int(\i+1)] in {0,...,5} {
			\node[shape=circle,draw=black] (v\label) at ({360/6*\i+60}:1) {};
		}		
		
		\draw (v1) -- (v2);
		\draw (v1) -- (v6);
		\draw (v2) -- (v3);
		\draw [dashed] (v2) -- (v4);
		\draw [dashed] (v2) -- (v5);
		\draw [dashed] (v2) -- (v6);
		\draw (v3) -- (v4);
		\draw (v4) -- (v5);
		\draw (v5) -- (v6);
\end{tikzpicture}}
\newcommand{\GsixX}{\begin{tikzpicture}[thick, scale=0.8, every node/.style={circle, minimum size = 3pt, scale=0.8}]
		\foreach \i [evaluate=\i as \label using int(\i+1)] in {0,...,5} {
			\node[shape=circle,draw=black] (v\label) at ({360/6*\i+60}:1) {};
		}		
		
		\draw [dashed] (v1) -- (v2);
		\draw (v1) -- (v3);
		\draw (v1) -- (v4);
		\draw (v1) -- (v5);
		\draw [dashed] (v1) -- (v6);
		\draw [dashed] (v2) -- (v3);
		\draw (v2) -- (v4);
		\draw (v2) -- (v5);
		\draw (v2) -- (v6);
		\draw (v3) -- (v5);
		\draw (v3) -- (v6);
		\draw (v4) -- (v6);
\end{tikzpicture}}

\newcommand{\given}{\,|\,}
\newcommand{\const}{\mathcal C}

\newcommand{\R}[2]{\mathbb R^{{#1} \times {#2}}}
\newcommand{\dd}{\,\mathrm{d}}
\newcommand{\s}{\quad\,\,}
\newcommand{\T}{\mathsf{T}}
\newcommand{\spt}{\mathsf S}
\newcommand{\com}{\mathcal K}
\newcommand{\G}{\mathcal G}
\newcommand{\E}{\mathcal E}
\newcommand{\HH}{\mathcal H}
\newcommand{\st}{\mathfrak S}
\newcommand{\p}{\mathsf P}
\newcommand{\cc}{\mathsf C}
\newcommand{\V}{\mathcal V}

\DeclareMathOperator{\tr}{tr}
\DeclareMathOperator{\NB}{\mathcal {N}}
\DeclareMathOperator{\Iss}{Iss}
\DeclareMathOperator{\diag}{diag}
\DeclareMathOperator{\student}{t}

\begin{document}
\let\WriteBookmarks\relax
\def\floatpagepagefraction{1}
\def\textpagefraction{.001}
\shorttitle{$\G$-Wishart normalising constants}
\shortauthors{C. Wong et~al.}

\newcommand{\rev}[1]{#1}

\title[mode = title]{A new way to evaluate $\G$-Wishart normalising constants via Fourier analysis}                      

\author[1]{Ching Wong}[orcid = 0009-0009-9342-2731]
\ead{ching.wong@unibas.ch}

\author[1]{Giusi Moffa}[orcid = 0000-0002-2739-0454]
\ead{giusi.moffa@unibas.ch}

\author[2]{Jack Kuipers}[orcid = 0000-0001-5357-2705]
\ead{jack.kuipers@bsse.ethz.ch}

\affiliation[1]{organization={Department of Mathematics and Computer Science, University of Basel},
                country={Switzerland}}

\affiliation[2]{organization={Department of Biosystems Science and Engineering, ETH Zurich},
                country={Switzerland}}

\begin{abstract}
The $\G$-Wishart distribution is \rev{a core} component for the Bayesian analysis of Gaussian graphical models as the conjugate prior for the precision matrix. Evaluating the marginal likelihood of such models usually requires computing high-dimensional integrals to determine the $\G$-Wishart normalising constant. Closed-form results are known for decomposable or chordal graphs, while an explicit representation as a formal series expansion has been derived recently for general graphs. The nested infinite sums, however, do not lend themselves to computation, remaining of limited practical value. Borrowing techniques from random matrix theory and Fourier analysis, we provide novel exact results well suited to the numerical evaluation of the normalising constant for classes of graphs beyond chordal graphs. \rev{We additionally develop a Monte Carlo scheme for general graphs, which can be orders of magnitude more efficient than current approaches.} 
\end{abstract}


\maketitle

\section{Introduction}

The Wishart distribution \cite{wishart1928generalised} plays a key role in Bayesian statistics \cite{gelman1995bayesian} as the conjugate prior for the precision (inverse covariance) matrix of multivariate Gaussians. Given independently drawn observations from a centred multivariate normal, the product of the data matrix and its transpose constitute a sample matrix from a Wishart distribution, which represents the distribution of scatter or sample covariance matrices. Since the sample data is itself a random matrix, the Wishart distribution is a classical distribution in random matrix theory \cite{mehta2004random, livan2018introduction}, where it is known as the Wishart-Laguerre, or Laguerre ensemble (which also extends beyond the real case to complex and quaternionic data matrices).

Random matrices were first used by Wigner \cite{wigner1955characteristic} as simple models of complex quantum systems, like nuclear reactions, where physical observables are related to the eigenvalue spectrum. The random matrix approach is predicted to be applicable when the underlying classical dynamics are chaotic, and the inverse eigenvalues from the Wishart-Laguerre ensemble correspond to the Wigner time delays in quantum chaotic scattering \cite{brouwer1997quantum}. Agreement between statistics of a random matrix and quantum spectra can be derived through diagrammatic perturbation theory \cite{kuipers2014efficient} and understood via intermediate matrix integrals \cite{novaes2023semiclassical}. For the related problem of quantum transport and the Jacobi ensemble (which can be obtained from a combination of two Wisharts), full equivalence has been proven \cite{berkolaiko2012universality}. The Wishart-Laguerre ensemble is also a key model for quantum chromodynamics \cite{verbaarschot2000random} and entanglement  \cite{majumdar2015extreme}, while the eigenvalue distribution is also important for principal component analysis \cite{johnstone2001distribution}.  

For \rev{multivariate} statistics, the Wishart distribution is instrumental in aiding the modelling of continuous data with probabilistic graphical models. These are popular and powerful tools \cite{Lauritzen1996, KollerEtAl2009} for compactly representing data and their dependencies with a graph, where each node encodes a variable and the edges encode conditional independence relationships. The most common types are Markov random fields, represented as \emph{undirected graphs}, and Bayesian networks, represented as \emph{directed acyclic graphs (DAGs)}. Evaluating the marginal likelihood of each structure is a key ingredient to enable Bayesian analyses.

\subsection{Directed graphs}

The factorisation of Bayesian networks into components involving each child and its parents allows us to leverage the properties of the conjugate Wishart prior to easily evaluate the Gaussian integrals and express them as ratios of the graph normalising constants of the prior and posterior distributions \cite{GeigerEtAl2002,kmh14}. Increasingly efficient MCMC schemes have been developed to create Bayesian samplers \cite{my95,GiudiciEtAl2003,GrzegorczykEtAl2008,km17,kuipers2022efficient} and exact samplers have been built for smaller networks \cite{talvitie2020exact}. With the current advances in Bayesian sampling of DAGs, we can also propagate the uncertainty in both structure and parameters to obtain the posterior distribution of causal intervention effects in fully Bayesian analyses \cite{MoffaEtAl2017,viinikka2020towards}.
In contrast, the persistent intractability of the integrals needed to compute the marginal likelihood has hampered Bayesian inference for undirected graphical models despite the space being simpler. Defining the integrals we need to compute explicitly will help us formulate the problem more precisely.

\subsection{The $\G$-Wishart normalising constant}

We denote by $\mathbb{S}^p$ the set of all $p$ by $p$ real symmetric matrices, and by $\mathbb S^p_{++}$ the set of all symmetric positive definite matrices in $\R p p$. For a graph $\G = (\V(\G), \E(\G))$ with vertex set $\V(\G) = \{v_1, \ldots, v_p\}$, let
\begin{equation}
	\mathbb S^p_{++}(\G) := \{M \in \mathbb S^p_{++} : \{v_\mu, v_\nu\} \not \in \E(\G) \implies m_{\mu,\nu} = 0\} \nonumber
\end{equation}
be the set of $p$ by $p$ real symmetric positive definite matrices whose entries corresponding to a pair of non-adjacent vertices are zero. The \textit{Gaussian graphical model} with respect to $\G$ is
\begin{equation}
	\mathcal M_\G = \{\mathcal N_p(0, \Sigma) : \Sigma^{-1} \in \mathbb S^p_{++}(\G)\}, \nonumber
\end{equation}
the set of $p$-variate Gaussian distributions with mean zero and variance $\Sigma$ such that the \textit{precision matrix} $K = \Sigma^{-1}$ is in $\mathbb S^p_{++}(\G)$. A common choice for the prior distribution for $K \in \mathbb S^p_{++}(\G)$ is
\begin{equation}
	f(K\given \G) \sim \det(K)^{\frac{\delta-2}{2}} e^{-\frac{\tr (KD)}2}, \quad \text{where $\delta > 0$ and $D \in \mathbb S^p_{++}$,}\nonumber
\end{equation}
as it is conjugate \cite{Roverato}. Let $Z \in \mathbb R^{p \times n}$ be a dataset with $n$ samples and $p$ variables, the marginal likelihood for the Gaussian graphical model above is then
\begin{equation}
	p(Z \given \G) = \dfrac{2^{-\frac{p(p-1)}{2}}}{(2\pi)^{\frac{pn}2}} \dfrac{\const_{\G}(\delta+n, U+D)}{\const_{\G}(\delta, D)}, \qquad U = \sum_{j=1}^n (\boldsymbol{z}_j - \overline{\boldsymbol{z}}) (\boldsymbol{z}_j - \overline{\boldsymbol{z}})^\T , \nonumber 
\end{equation}
where $U$ is the scatter matrix (an unnormalised sample covariance), and
\begin{equation}
	\const_\G(\delta, D) := \int_{\mathbb S^p_{++}(\G)} \det(K)^{\frac{ \delta-2}{2}} e^{-\frac{\tr(KD)}2} \dd K, \s \s \textrm{for $\delta > 0$ and $D \in \mathbb S^p_{++}$}, \nonumber
\end{equation}
is the \textit{$\G$-Wishart normalising constant}. As in \cite{Uhler}, a change of variables $K \to 2K$ allows us to simplify 
\begin{equation}
	\const_\G(\delta, D) = 2^{\frac{p \delta}{2}+|\E(\G)|} \mathcal I_\G\left(\dfrac{\delta-2}{2}, D\right), \nonumber
\end{equation}
where
\begin{equation}
	\mathcal I_\G(\beta, D) := \int_{\mathbb S^p_{++}(\G)} \det(K)^\beta e^{-\tr(KD)} \dd K, \s \s \textrm{for $\beta > -1$ and $D \in \mathbb S^p_{++}$}.\label{eq:Iintegraldef}
\end{equation}

Evaluating the integral $\mathcal I_\G(\beta, D)$ for a general graph $\G$ is challenging. Roverato \cite{Roverato} proved that the normalising constant for $\G$ can be factorised according to the \textit{prime components} of $\G$, see equation (\ref{eq:WishartPrimeFactors}) in Section~\ref{sec:chordal}. Consequently, we only need to evaluate $\mathcal I_\G(\beta, D)$ for prime graphs $\G$.  Apart from this, the only progress has come more recently from Uhler et al.~\cite{Uhler} as an iterative method for evaluating the integral. This, however, only offers layers of series expansions rather than a closed-form result, and the highly intricate nested infinite sums do not appear to offer a viable path for evaluation.
Currently, the only approaches for evaluating the $\G$-Wishart normalising constant for general graphs are Monte Carlo algorithms \cite{Roverato, Atay-Kayis, mw19}, which become increasingly computationally intensive for larger networks. 

\subsection{Known results}

There are a few classes of prime graphs for which the normalising constant is known explicitly. To describe them, we need some graph theory terminology. A $p$-vertex graph is \textit{complete}, denoted by $\com_p$, if every pair of vertices is adjacent. For $k \geq 2$, a graph is \textit{complete $k$-partite}, denoted by $\com_{p_1, \ldots, p_k}$, if its vertex set can be partitioned into $k$ sets, $\V_1, \ldots, \V_k$, with $|\V_\mu| = p_\mu$ for $1 \leq \mu \leq k$, such that $v$ is adjacent to $u$ if and only if $v$ and $u$ belong to different parts $\V_\mu$ and $\V_\nu$. When $k = 2$, it is \textit{complete bipartite}. A graph is \textit{chordal} if it does not possess any induced cycle of length longer than three. Note that every graph has a \textit{chordal completion} (i.e., a chordal supergraph on the same vertex set), as the complete graphs are chordal. The \textit{minimum fill-in} of a graph is the smallest number of edges that need to be added to turn it into a chordal graph.

If $\G$ is a $p$-vertex \textit{prime graph} (see Section~\ref{sec:prime} for the definition) and $\beta > -1$ is a real number, an explicit formula (involving gamma functions) for $\mathcal I_\G(\beta, D)$ is known when
\begin{itemize}
	\item[(A1)] $\G$ is complete and $D \in \mathbb S^p_{++}$ \cite{Dawid, Eaton}, or
	\item[(A2)] $\G$ has minimum fill-in 1 and $D = I_p$ \cite{Uhler}, or
	\item [(A3)] $\G$ is complete bipartite and $D = I_p$ \cite{Uhler}.
\end{itemize}

As an interesting aside, Roverato \cite{Roverato} observed, for a graph $\G$ with $p$ vertices, $D \in \mathbb S^p_{++}$ and real number $\delta > 0$, that $\mathcal C_\G(\delta, D)$ seemed to be expressible in terms of $\tilde D \in \mathbb S^p_{++}$, the \textit{PD-completion} of $D$ with respect to $\G$,
\begin{equation}
	2^{\frac{p}{2}} \det(\Iss_\G(\tilde D))^{-\frac12} \det(\tilde D)^{-\frac{\delta-2}2} \mathcal C_\G(\delta, I_p), \label{eq:conj}
\end{equation}
where $\Iss_\G(\tilde D)$ is the \textit{Isserlis matrix} of $\tilde D$ with respect to $\G$, as holds for complete graphs. Such a transformation would reduce the computation of the $\G$-Wishart normalising constant for general matrices $D$ down to the simpler case where $D$ is the identity matrix.  Though appealing, and conjectured \cite{Roverato} to hold for general graphs, this conjecture was recently disproved \cite{wong2025conjecture}, though the transformation can still offer very accurate approximations.

\subsection{Bayesian inference for undirected graphs}

Efficient sampling methods exist for the restricted class of decomposable (complete) graphs \cite{GreenThomas, GiudiciGreen, Olsson}, thanks to their known results (A1).
For general undirected graphs, without a handle on evaluating the marginal likelihood and hence the posterior of each network, Bayesian approaches \cite{vogels2024bayesian} have progressed by avoiding evaluating $\mathcal I_\G(\beta, D)$ and working instead in the joint (un-marginalised) space of networks and elements of the precision matrices \cite{wl12, lenkoski2013direct, mw15}. Due to the different-sized parameter spaces for different networks, these approaches build on trans-dimensional MCMC \cite{green2003trans}. Notably, Bayesian sampling \cite{mw15, vogels2024bayesian} outperforms the simple point estimate, which can be obtained from regularised optimisation \cite{friedman2008sparse}. Although current Bayesian methods avoid evaluating equation (\ref{eq:Iintegraldef}) directly, they can still simplify the sampling by using the evaluation for the case where $D = I_p$. This highlights how results for the identity matrix can be useful. Indeed, approximations for $D = I_p$ were developed and shown to offer computational advantages for Bayesian inference for undirected networks \cite{mml23}.

Alternatively, \rev{MCMC schemes have been developed that replace the use of the $\G$-Wishart prior with a simpler spike-and-slab prior \cite{wang2015scaling}, removing conjugacy for computational efficiency, especially for sparse networks \cite{sulem2025bayesian}. While Bayesian approaches for undirected graphs have focussed on the network structure \cite{vogels2024bayesian}, effectively neglecting downstream analysis tasks, studies with directed networks \cite{MoffaEtAl2017, kuipers2018mutational} demonstrate their importance. For such tasks, parameter uncertainty is paramount and directly accessible with the conjugate $\G$-Wishart prior. Keeping conjugacy, while avoiding} the normalising constant even for the identity matrix, exchange or auxiliary variable approaches eliminate normalising constants from acceptance ratios \cite{mw15, Willem}. These schemes still require sampling from the $\G$-Wishart distribution \cite{lenkoski2013direct}, \rev{to implicitly estimate normalising constants}, so exact computable results for $\G$-Wishart normalising constants could still reduce the computational burden of sampling.

The $\G$-Wishart distribution can also be extended, for example, to having multivariate shape parameters rather than a single parameter $\delta$, which offers more flexibility in prior design and has been studied in detail for the decomposable case \cite{letac2007wishart}. More recently, a Gibbs sampler has been developed allowing Bayesian inference for a certain type of generalised $\G$-Wishart distribution and for certain classes of undirected graphs \cite{khare2018bayesian}. 

Though allowing Bayesian inference, these sampling-based approaches do not directly tackle evaluating the $\G$-Wishart normalising constant beyond Monte Carlo integration \cite{Atay-Kayis, Roverato, mw19}. Therefore, we focus on developing an alternative approach to the integrals themselves.

\subsection{Our contribution} \label{sec:contribution}

Borrowing inspiration from random matrix theory \cite{janik2003wishart} and using tools from Fourier analysis, we show how to transform the integral $\mathcal I_\G(\beta, D)$ in a way that enables us to derive an explicit formula (involving gamma functions and generalised hypergeometric functions ${}_3F_2$) for the $\G$-Wishart normalising constant when	
\begin{itemize}
	\item [(B1)] $\G$ has a chordal completion $\G^*$ in which every triangle contains at most one edge from $\E(\G^*) \setminus \E(\G)$ and $D = I_{|\V(\G)|}$ (Corollary~\ref{thm:disjoint}), or
	\item [(B2)] $\G$ has minimum fill-in 2 and $D = I_{|\V(\G)|}$ (Section~\ref{sec:mfi2}), or
	\item [(B3)] $\G$ is complete $k$-partite and $D = I_{|\V(\G)|}$ (Example~\ref{ex:completekpartite}).
\end{itemize}

Note that (B1) contains every \textit{Tur\'an graph} $\mathcal T(2p,p)$, which is a prime graph with minimum fill-in $p-1$ when $p \geq 2$ (Example~\ref{ex:turan}).
Moreover, we show that the integral of interest can be reduced to a one-dimensional real integral when
\begin{itemize}
	\item [(C1)] $\G$ is isomorphic to some $\G(m; k_1, \ldots, k_\ell)$ (defined in Section~\ref{sec:Gmk}), where $m \geq 4$, $3 \leq \ell \leq m-1$ and $k_1, \ldots, k_\ell \geq 1$, and $D = I_{m+k_1+\cdots+k_\ell}$ (Corollary~\ref{coro:ell>2}), or
	\item [(C2)] $\G$ has a chordal completion with 3 added edges that form a triangle and $D = I_{|\V(\G)|}$ (Section~\ref{sec:missingTriangle}), or
	\item [(C3)] $\G$ is the cycle of length 6 or its complement and $D = I_6$ (Example~\ref{ex:C6complement}, Example~\ref{ex:C6}).
\end{itemize}
We remark that in (C1), there are infinitely many prime graphs of any given minimum fill-in greater than 2.

All graphs with up to 6 vertices are included in (B1)--(C3) so their $\mathcal I_\G(\beta, I)$ can be easily computed, and in Supplementary Section~\ref{sec:smallPrime} we list all small graphs including the 24 connected prime graphs on 6 vertices. For larger graphs, many sparse networks fall into the cases above we can handle, and we display in Table~\ref{tab:graph_fraction} the fraction of random networks where $\mathcal I_\G(\beta, I)$ can accordingly be evaluated without needed Monte Carlo approximations.

\begin{table}[t!] 
	\caption{Fraction of random graphs that fall into cases (B1)--(C3) of different sizes and densities ($\zeta$: expected neighbourhood size)}
	\begin{tabular}{l|S[table-format=1.3]|S[table-format=1.3]|S[table-format=1.3]|S[table-format=1.3]|S[table-format=1.3]}
		& {$p=10$} & {$p=20$} & {$p=30$} & {$p=40$} & {$p=50$}\\
		\hline
		{$\zeta = 0.50$} & 0.994 & 0.976 & 0.956 & 0.943 & 0.930 \\
		{$\zeta = 0.75$} & 0.948 & 0.791 & 0.656 & 0.540 & 0.451 \\
        {$\zeta = 1.00$} & 0.817 & 0.432 & 0.224 & 0.116 & 0.062\\
		{$\zeta = 1.25$} & 0.619 & 0.151 & 0.036 & 0.008 & 0.002
	\end{tabular} \label{tab:graph_fraction}
\end{table}

Furthermore, we show that the $\G$-Wishart normalising constant can be written as an integral of the form
\begin{equation}
	\int_{\mathbb R^\tau} \prod_{\{\alpha_1, \ldots, \alpha_k\} \in \mathfrak I} (1+t_{\alpha_1}^2 + \cdots + t_{\alpha_k}^2)^{-\gamma_{\alpha_1, \ldots, \alpha_k}} \dd t_1 \cdots \dd t_\tau, \s \textrm{ where } \mathfrak I \subseteq \mathcal P(\{1, \ldots, \tau\}), \label{eq:specialForm}
\end{equation}
\rev{involving subsets of missing edges where the $\gamma$ are powers related to the parameter $\beta$ and the graph structure, when} 
\begin{itemize}
	\item [(D1)] $\G$ has \textit{starry fill-ins} and $D = I_{|\V(\G)|}$ (Corollary~\ref{thm:star}), or
	\item [(D2)] $\G$ is a \textit{gear graph} and $D = I_{|\V(\G)|}$ (Corollary~\ref{coro:gear}).
\end{itemize}
The definitions of \textit{starry fill-in} and \textit{gear graphs} can be found in Section~\ref{sec:starryfillins} and Supplementary Section~\ref{sec:gear}, respectively. We remark that (D1) contains all the cycle graphs. An example of (\ref{eq:specialForm}) is
\begin{equation}
	\int_{\mathbb R^3} (1+t_1^2)^{-\frac12} (1+t_2^2)^{\frac92} (1+t_3^2)^{-\frac12} (1+t_1^2+t_2^2)^{-5}(1+t_2^2+t_3^2)^{-5} \dd t_1 \dd t_2 \dd t_3, \nonumber
\end{equation}
with $\tau = 3$ and $\mathfrak I = \{\{1\}, \{2\}, \{3\}, \{1, 2\}, \{2, 3\}\}$, which we encounter later in Example~\ref{ex:C6} for the cycle of length 6, before reducing it down to one dimension as in (C3). \rev{These results are connected to relationships between the normalising constants of different graphs, which we explore in Supplementary Section~\ref{sec:map}.}

Finally, for general $D$, we obtain a one-dimensional integral when
\begin{itemize}
	\item [(E1)] $\G$ has minimum fill-in 1 and $D \in \mathbb S^{|\V(\G)|}_{++}$ (Section~\ref{sec:generalA}).
\end{itemize}

Together with the prime factorisation (Equation (\ref{eq:WishartPrimeFactors}) later) due to Roverato \cite{Roverato}, the above results simplify the computation for the $\G$-Wishart normalising constants for many graphs. For instance, for a $2$ by $m$ grid (whose prime components are $m-1$ cycles of length 4, each with minimum fill-in of 1) and a matrix $D \in \mathbb S^{2m}_{++}$, the $\G$-Wishart normalising constant can be written as the product of $m-1$ one-dimensional integrals. Section~\ref{sec:generalA} delves deeper into the details of such integrals.

\rev{Finally, larger fill-ins lead to higher dimensional integrals, and we develop a new Monte Carlo importance sampling scheme to evaluate these integrals in Section~\ref{sec:MC}. We implement our methods in the \textit{GWnorm}\footnote{\url{https://CRAN.R-project.org/package=GWnorm}} R package, while scripts running the examples and simulations are also available\footnote{\url{https://github.com/jackkuipers/GWnorm}}.}

All the results mentioned above have been built upon the following theorem, in which we transform the integral $\mathcal I_\G(\beta, D)$ over the restricted (relative to $\mathbb R^{|\E(\G)| + |\V(\G)|}$) space $\mathbb S^{|\V(\G)|}_{++}(\G)$ into an integral over the Euclidean space $\mathbb R^\tau$, where $\tau$ is the number of edges needed for a known chordal completion of $\G$. The matrix Dirac delta function (Fourier transform of 1), as used in \cite{janik2003wishart}, is employed. The proof of this result can be found in Section~\ref{sec:Diracdelta}.	
\begin{theorem} \label{thm:main}
	Let $\G$ be a proper subgraph of $\G^*$, both on the vertex set $\{v_1, \ldots, v_p\}$. 
	Let $\beta > -1$ be a real number and $D \in \mathbb S^p_{++}$. Then,
	\begin{equation}
		\mathcal I_{\G} (\beta, D) = \dfrac{1}{\pi^{|\E(\G^*)| - |\E(\G)|}} \int_{\mathbb S(\G, \G^*)} \mathcal I_{\G^*}(\beta, D+iT) \dd T, \nonumber
	\end{equation}
	where
	\begin{equation}
		\mathbb S(\G, \G^*) := \{ T \in \mathbb S^p : \mu = \nu \textrm{ or } \{v_\mu, v_\nu\} \in \E(\G) \textrm{ or } \{v_\mu, v_\nu\} \not\in \E(\G^*) \implies t_{\mu, \nu} = 0\}. \nonumber
	\end{equation}
	
	In particular, if $\G^*$ is a chordal completion of $\G$, then
	\begin{equation}
		\mathcal I_{\G} (\beta, D) = \dfrac{\mathcal I_{\G^*}(\beta, I_p)}{\pi^{|\E(\G^*)| - |\E(\G)|}} \bigintss_{\mathbb S(\G, \G^*)} \dfrac{\prod\limits_{\mu=1}^m \det((D+iT)[\cc_\mu])^{-\beta-\frac{|\cc_\mu|+1}{2}}}{\prod\limits_{\nu=1}^{m-1} \det((D+iT)[\spt_\nu])^{-\beta-\frac{|\spt_\nu|+1}{2}}} \dd T, \nonumber
	\end{equation}
	where $\cc_1, \ldots, \cc_m \subseteq \V(\G)$ are the \textit{maximal cliques} of $\G^*$ and $\spt_1, \ldots, \spt_{m-1} \subseteq \V(\G)$ are the \textit{minimal separators} (see Section~\ref{sec:chordal} for definitions).
\end{theorem}

Note that $\mathbb S(\G, \G^*)$ is the set of $p$ by $p$ symmetric real matrices, in which every entry is equal to zero unless it corresponds to an edge in $\E(\G^*) \setminus \E(\G)$. Thus, the number of variables in ${\mathbb S(\G, \G^*)}$ is $|\E(\G^*)| - |\E(\G)|$. It follows that an integral with domain ${\mathbb S(\G, \G^*)}$ (Lebesgue measure) is the same as an integral over $\mathbb R^{|\E(\G^*)| - |\E(\G)|}$.

While the normalising constant can be factorised according to the prime components of a graph, under certain assumptions, the integral $\mathcal I_\G(\beta, I_{|\V(\G)|})$ can be further factorised for some prime graphs $\G$, since $\mathcal I_{\G^*}(\beta, I_{|\V(\G)|}+iT)$ can be written as the product of some separable functions. The following theorem characterises these graphs. In Section~\ref{sec:partition} we illustrate this result using a toy example (Example~\ref{ex:K5K6}), while a formal proof is given in Appendix~\ref{sec:appPartition}. 
\begin{theorem} \label{thm:partition}
	Let $\G$ be a graph on $p$ vertices and let $\G^*$ be a chordal completion.
	Suppose that there exists a partition of the missing edges $\E(\G^*) \setminus \E(\G) = \E_1 \sqcup \cdots \sqcup \E_k$ such that for $1 \leq \mu < \nu \leq k$, either 
	\begin{itemize}
		\item $\V_\mu \cap \V_\nu = \emptyset$, or 
		\item $|\V_\mu \cap \V_\nu| = 1$ and there is no edge in $\G^*$ between the sets $\V_\mu \setminus \V_\nu$ and $\V_\nu \setminus \V_\mu$,
	\end{itemize}
	where $\V_\xi \subseteq \V(\G)$ is the vertex set associated with the edges in $\E_\xi$, for $\xi = 1, \ldots, k$.
	
	Let $\beta > -1$ be a real number. Then,
	\begin{equation}
		\mathcal I_{\G}(\beta, I_p) = \mathcal I_{\G^*}(\beta, I_p)^{1-k}\prod_{\xi=1}^k \mathcal I_{\G_\xi}(\beta, I_p), \nonumber
	\end{equation}
	where $\G_\xi$ is the graph obtained from $\G^*$ by removing the edges in $\E_\xi$, for $1 \leq \xi \leq k$.
\end{theorem}

\section{Preliminaries} \label{sec:chordal}

Our focus in this paper is in evaluating the $\G$-Wishart normalising constant for simple and undirected graphs $\G = (\V(\G), \E(\G))$. We denote by $\NB_\G(v) \subseteq \V(\G)$ the set of all the neighbours of the vertex $v$ in graph $\G$. A set $\cc \subseteq \V(\G)$ is a \textit{clique} of $\G$ if two vertices $u$ and $v$ are adjacent whenever they are both in $\cc$. A clique $\cc$ of $\G$ is considered \textit{maximal} if there is no clique in $\G$ that strictly contains $\cc$. A graph $\G$ is \textit{connected} if for any pair of vertices $u, v \in \V(\G)$, one can travel from $u$ to $v$ via edges of $\G$. For a subset $\V' \subseteq \V(\G)$, we use $\G[\V']$ to denote the \textit{induced subgraph} of $\G$ formed from the vertices in $\V'$. A set $\spt \subseteq \V(\G)$ is a \textit{separator} of a (connected) graph $\G$ if the induced subgraph $\G[\V(\G)\setminus \spt]$ is not connected. A separator $\spt$ is considered \textit{minimal} if there is no separator in $\G$ that is strictly contained in $\spt$. For a graph $\G$ with vertex set $\V(\G) = \{v_1, \ldots, v_p\}$ and a matrix $D$ in $\mathbb S^p_{++}$, the matrix $D[\cc] \in \mathbb S^{|\cc|}_{++}$, where $\cc \subseteq \V(\G)$, is the submatrix of $D$ obtained by the corresponding rows and columns.

In this section, we highlight some definitions and known results about chordal graphs. For a comprehensive overview, we refer to \cite{Lauritzen1996}.

\subsection{Graph decomposition} \label{sec:prime}

A graph $\G$ is \textit{prime} if it contains no separator that is a clique. Examples of prime graphs include complete graphs, cycle graphs and grid graphs.
Given a non-prime graph $\G$, one can select a minimal clique $\cc$ that separates the graph into two non-empty sets $\mathsf A$ and $\mathsf B$. Then, we say that $\G$ is \textit{decomposed} into two components $\G[\mathsf A \cup \cc]$ and $\G[\mathsf B \cup \cc]$. Continuing this process,  the graph $\G$ can be uniquely decomposed into its prime components. For example, the 2 by $m$ grid can be decomposed into $m-1$ cycles of length 4.

\subsection{Perfect sequence of prime components} \label{sec:perfectSeq}

Let $\G$ be a graph with $m$ prime components. An ordering of the prime components of $\G$, say $\G[\p_1], \ldots, \G[\p_m]$, is a \textit{perfect sequence of prime components} if, for every $1 \leq \mu \leq m-1$, there exists $\nu \leq \mu$ such that
\begin{equation}
	(\p_1 \cup \cdots \cup \p_\mu) \cap \p_{\mu+1} \subseteq \p_\nu, \nonumber
\end{equation}
which is known as the \textit{running intersection property}. For simplicity in notation, we slightly abuse the representation by referring to $\p_1, \ldots, \p_m$ as a perfect sequence of prime components, as opposed to using $\G[\p_1], \ldots, \G[\p_m]$. It is known that each graph possesses at least one perfect sequence of prime components.

Given a perfect sequence of prime components $\p_1, \ldots, \p_m$ of a graph $\G$, the sets
\begin{equation}
	\spt_\nu := (\p_1 \cup \cdots \cup \p_\nu) \cap \p_{\nu+1}, \qquad \textrm{where $1 \leq \nu \leq m-1$}, \nonumber
\end{equation}
are the corresponding set of separators. It is well known that the induced subgraphs $\G[\spt_\nu]$ are complete and that the \textit{multiset} $\{\spt_1, \ldots, \spt_{m-1}\}$ is invariant to the choice of the perfect sequence of prime components. In this paper, keeping in mind that some of these sets might be identical, we refer to the sets $\spt_1, \ldots, \spt_{m-1}$ as \textit{the minimal separators} of $\G$.

\subsection{Chordal graphs}
A graph is considered \textit{decomposable} if its prime components are all complete. These graphs can be characterised by their graphical properties; specifically, they are graphs in which all induced cycles have length 3. Equivalently, every cycle in these graphs with length greater than three contains at least one \textit{chord}, which refers to an edge connecting two vertices within the cycle that is not part of the cycle itself. Thus, decomposable graphs are also known as \textit{chordal graphs}, and here we will use the term chordal graphs.

For chordal graphs, we write \textit{perfect sequence of cliques} $\cc_1, \ldots, \cc_m$ in place of \textit{perfect sequence of prime components} $\p_1, \ldots, \p_m$, to emphasise that the vertex sets of the prime components are cliques. We remark that the prime components of a chordal graph are precisely its maximal cliques. 

\subsection{Chordal completion} \label{sec:chordalcompletion}

Given any graph $\G$, a graph $\G^*$ on the same vertex set is a \textit{chordal completion} of $\G$ if $\G^*$ is chordal, and $\G$ is a subgraph of $\G^*$. When there is no ambiguity, we refer to the edges $\E(\G^*) \setminus \E(\G)$ the \textit{missing edges}. Our results later rely heavily on finding a chordal completion (of a given graph) with some nice properties. In particular, the fewer the edges in the chordal completion, the lower the dimension of the integral in Theorem~\ref{thm:main}. Finding a chordal completion of a given graph with the minimum number of edges is known as the \textit{minimum chordal completion problem} or the \textit{minimum fill-in problem}, which is NP-complete \cite{Yannakakis}. Several linear-time algorithms, including Lexicographic Breadth-First Search \cite{Corneil, Rose} and Maximum Cardinality Search \cite{Tarjan}, have been proposed for finding chordal completions of graphs, though they do not always find one with minimum fill-in.

\subsection{Factorisation of $\G$-Wishart normalising constant}

For a complete graph $\com_p$, the $\G$-Wishart normalising constant is the normalising constant for a Wishart distribution:
\begin{eqnarray}
	\mathcal I_{\com_p}(\beta, D) & = & \int_{\mathbb S^p_{++}} \det(K)^\beta e^{-\tr(KD)} \, \dd K =  \det(D)^{-\beta-\frac{p+1}{2}} \Gamma_p\left(\beta+\dfrac{p+1}{2}\right), \nonumber
\end{eqnarray}
where $\Gamma_k(a)$ is the \textit{multivariate gamma function}
\begin{equation}
	\Gamma_k(a) := \pi^{\frac{k(k-1)}{4}} \prod_{\mu=1}^k \Gamma\left(a+\dfrac{1-\mu}{2}\right), \qquad \textrm{for $a > \dfrac{k-1}{2}$}. \nonumber
\end{equation}
Let $\G^*$ be a chordal graph. Let $\cc_1, \ldots, \cc_m$ be the maximal cliques and $\spt_1, \ldots, \spt_{m-1}$ be the minimal separators. Then, it is known \cite{Dawid} that the integral $\mathcal I_{\G^*}(\beta, D)$ can be factorised as
\begin{equation}
	\mathcal I_{\G^*}(\beta, D) = \dfrac{\prod\limits_{\mu=1}^m \mathcal I_{\G^*[\cc_\mu]}(\beta, D[\cc_\mu])}{\prod\limits_{\nu=1}^{m-1} \mathcal I_{\G^*[\spt_\nu]}(\beta, D[\spt_\nu])} = \dfrac{\prod\limits_{\mu=1}^m \det(D[\cc_\mu])^{-\beta - \frac{|\cc_\mu|+1}{2}} \Gamma_{|\cc_\mu|}\left(\beta + \frac{|\cc_\mu|+1}{2}\right)}{\prod\limits_{\nu=1}^{m-1} \det(D[\spt_\nu])^{-\beta - \frac{|\spt_\nu|+1}{2}}\Gamma_{|\spt_\nu|}\left(\beta + \frac{|\spt_\nu|+1}{2}\right)} \label{eq:chordalA}.
\end{equation}
Roverato \cite{Roverato} generalised this factorisation to general graphs $\G$:
\begin{equation}
	\mathcal I_{\G}(\beta, D) = \dfrac{\prod\limits_{\mu=1}^m \mathcal I_{\G[\p_\mu]}(\beta, D[\p_\mu])}{\prod\limits_{\nu=1}^{m-1} \mathcal I_{\G[\spt_\nu]}(\beta, D[\spt_\nu])} = \dfrac{\prod\limits_{\mu=1}^m \mathcal I_{\G[\p_\mu]}(\beta, D[\p_\mu])}{\prod\limits_{\nu=1}^{m-1} \det(D[\spt_\nu])^{-\beta - \frac{|\spt_\nu|+1}{2}}\Gamma_{|\spt_\nu|}\left(\beta + \frac{|\spt_\nu|+1}{2}\right)} \label{eq:WishartPrimeFactors},
\end{equation}
where $\p_1, \ldots, \p_m$ are the prime components of $\G$ and $\spt_1, \ldots, \spt_{m-1}$ are the minimal separators. This is why we only need to find the normalising constants for prime graphs. We remark that (\ref{eq:WishartPrimeFactors}) can be easily derived from Theorem~\ref{thm:main}, by separating the variables.

\section{Transforming the integral using Fourier analysis} \label{sec:Diracdelta}

We provide an example after proving Theorem~\ref{thm:main}.

\begin{proof}[Proof of Theorem~\ref{thm:main}]
	We begin with a parametrised form of the domain $\mathbb S^p_{++}(\G)$ in the integral $\mathcal I_\G(\beta, D)$ (defined in (\ref{eq:Iintegraldef})) and write
	\begin{eqnarray}
		\mathcal I_\G(\beta, D) & = & \int_{\mathbb S^p_{++}(\G)} \det(K)^\beta e^{-\tr(KD)} \dd K \nonumber \\
		& = & \bigintsss_{\mathbb R^{|\E(\G)|}} \bigintsss_{\rho_1(K)}^\infty \cdots \bigintsss_{\rho_p(K)}^\infty \det(K)^\beta e^{-\tr(KD)} \dd k_{p,p} \cdots \dd k_{1,1} \prod_{\substack{1 \leq \mu < \nu \leq p \\ \{v_\mu, v_\nu\} \in \E(\G)}} \dd k_{\mu, \nu}, \nonumber
	\end{eqnarray}
	where $\rho_1(K) = 0$, and for $2 \leq \mu \leq p$, 
	\begin{equation}
		\rho_\mu(K) := \begin{pmatrix}
			k_{1,\mu} & \cdots & k_{\mu-1,\mu}
		\end{pmatrix} (K[1,\mu-1])^{-1} \begin{pmatrix}
			k_{1,\mu} \\
			\vdots \\
			k_{\mu-1,\mu}
		\end{pmatrix} \geq 0. \nonumber
	\end{equation}
	
	We use the following version of the Fourier inversion theorem:
	\begin{equation}
		f(\boldsymbol{k}',\boldsymbol{0}) = \int_{\mathbb{R}^\tau}
		\hat{f}(\boldsymbol{k}',\boldsymbol{t}) \dd\boldsymbol{t} = \int_{\mathbb{R}^{\tau}}\int_{\mathbb{R}^{\tau}} f(\boldsymbol{k}',\boldsymbol{k}'') e^{2\pi i\boldsymbol{k}'' \cdot \boldsymbol{t}} \dd\boldsymbol{k}''\dd\boldsymbol{t}, \quad \forall \, \boldsymbol{k}' \in \mathbb{R}^{p + |\E(\G)|}. \label{eq:F-inv}
	\end{equation}
	Alternatively, this can be considered as an application of the Plancherel theorem applied to the integrand together with the $\tau$-dimensional Dirac delta function:
	\begin{equation}
		\delta(\boldsymbol{k}'') =\int_{\mathbb R^\tau} e^{2\pi i \boldsymbol{k}'' \cdot \boldsymbol{t} } \dd \boldsymbol{t}, 	\quad \forall \, \boldsymbol{k}'' \in \mathbb{R}^{\tau}. \nonumber
	\end{equation}
	
	Let $\tau = |\E(\G^*)| - |\E(\G)|$. On the entries $k_{\mu, \nu}$, where $\{v_\mu, v_\nu\} \in \E(\G^*)\setminus \E(\G)$, we use Equation (\ref{eq:F-inv}) and Fubini's Theorem to obtain 
	\begin{eqnarray}
		\mathcal I_\G(\beta, D) & = &\bigintsss_{\mathbb R^{\tau}} \bigintsss_{\mathbb R^{|\E(\G^*)|}} \bigintsss_{\rho_1(K)}^\infty \cdots \bigintsss_{\rho_p(K)}^\infty \det(K)^\beta \exp\left(-\tr(KD)+2\pi i\sum\limits_{\substack{1 \leq \mu < \nu \leq p\\ \{v_\mu, v_\nu\} \in \E(\G^*) \setminus \E(\G)}}k_{\mu,\nu} t_{\mu,\nu} \right) \nonumber\\
		& & \quad \dd k_{p,p} \cdots \dd k_{1,1} \prod_{\substack{1 \leq \mu < \nu \leq p\\ \{v_\mu, v_\nu\} \in \E(\G^*)}} \dd k_{\mu, \nu} \prod_{\substack{1 \leq \mu < \nu \leq p\\ \{v_\mu, v_\nu\} \in \E(\G^*) \setminus \E(\G)}} \dd t_{\mu, \nu} \nonumber \\
		& = &\dfrac{1}{\pi^{\tau}} \int_{\mathbb S(\G, \G^*)} \int_{\mathbb S^p_{++}(\G^*)} \det(K)^\beta e^{-\tr(K(D+iT))} \dd K \dd T = \dfrac{1}{\pi^{\tau}} \int_{\mathbb S(\G, \G^*)} \mathcal I_{\G^*}(\beta, D+iT) \dd T. \nonumber
	\end{eqnarray}
	When the graph $\G^*$ is chordal, we use analytic continuation of (\ref{eq:chordalA}) to write the integrand as
	\begin{eqnarray}
		\mathcal I_{\G^*}(\beta, D+iT) & = &\dfrac{\prod\limits_{\mu=1}^m \det((D+iT)[\cc_\mu])^{-\beta - \frac{|\cc_\mu|+1}{2}} \Gamma_{|\cc_\mu|}\left(\beta + \frac{|\cc_\mu|+1}{2}\right)}{\prod\limits_{\nu=1}^{m-1} \det((D+iT)[\spt_\nu])^{-\beta - \frac{|\spt_\nu|+1}{2}}\Gamma_{|\spt_\nu|}\left(\beta + \frac{|\spt_\nu|+1}{2}\right)} = \mathcal I_{\G^*}(\beta, I_p) \dfrac{\prod\limits_{\mu=1}^m \det((D+iT)[\cc_\mu])^{-\beta - \frac{|\cc_\mu|+1}{2}} }{\prod\limits_{\nu=1}^{m-1} \det((D+iT)[\spt_\nu])^{-\beta - \frac{|\spt_\nu|+1}{2}}}, \nonumber
	\end{eqnarray}
	where $\det((D+iT)[\cc])^{-\beta-\frac{|\cc|+1}{2}} \in \mathbb C$ is taken to be continuous in $T$, and it is a real number when $T$ is the zero matrix, for any subset $\cc$ of $\V(\G^*)$.
\end{proof}

Let $\G$ be a graph on $p$ vertices, with a chordal completion $\G^*$ with $\tau$ added edges. Theorem~\ref{thm:main} provides an alternative integral for the $\G$-Wishart normalising constant whose domain is $\mathbb R^\tau$. It may be possible to simplify the integral further (especially when $D = I_p$) into an even lower dimensional one, as shown in the following example.

\begin{example} \label{ex:C6complement}
	Let $\G$ be the complement of the cycle of length 6, as shown in Figure~\ref{fig:C6_complementb}. $\G$ has minimum fill-in 3, and an example chordal completion with 3 extra edges is also shown in the same figure.
	\begin{figure}
        \begin{subfigure}{0.03\textwidth}
			\centering
			\caption{} \label{fig:C6_complementb}
		\end{subfigure}
		\begin{subfigure}{0.27\textwidth}
			\centering
			\Csixcomplement
		\end{subfigure}
       \begin{subfigure}{0.03\textwidth}
            \centering
		      \caption{} \label{fig:K5K6b}
		\end{subfigure}
		\begin{subfigure}{0.27\textwidth}
			\centering
			\KfiveKsix
		\end{subfigure}
        \begin{subfigure}{0.03\textwidth} \centering
			\caption{} \label{fig:T63b}
		\end{subfigure}
		\begin{subfigure}{0.27\textwidth}
			\centering
			\Tsixthree
		\end{subfigure}

		\begin{subfigure}{0.3\textwidth}
			\centering
			\includegraphics[height=2.5cm]{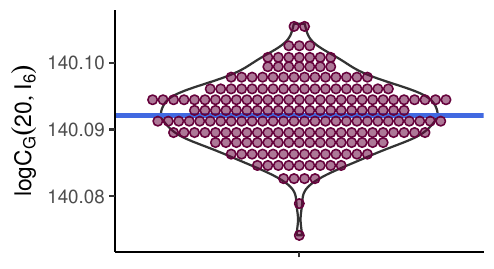}
		\end{subfigure}
		\begin{subfigure}{0.3\textwidth}\centering
			\includegraphics[height=2.5cm]{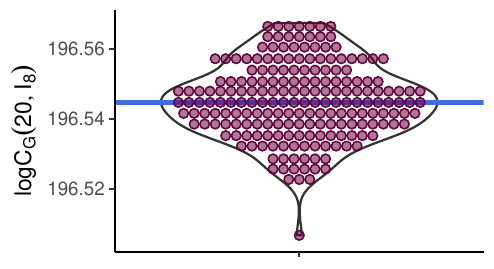}
		\end{subfigure}
		\begin{subfigure}{0.3\textwidth} \centering
			\includegraphics[height=2.5cm]{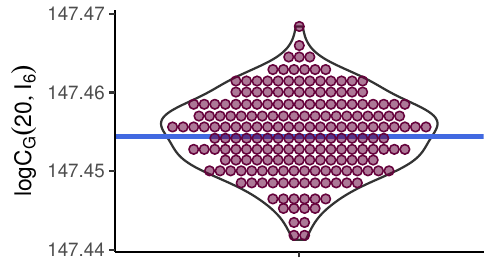}
		\end{subfigure}
		
		\begin{subfigure}{0.03\textwidth} \centering
			\caption{} \label{fig:C6b}
		\end{subfigure}
        \begin{subfigure}{0.27\textwidth}
			\centering
			\CsixA
		\end{subfigure}
        \begin{subfigure}{0.03\textwidth} \centering
			\caption{} \label{fig:C6+1b}
		\end{subfigure}
		\begin{subfigure}{0.27\textwidth}
			\centering
			\Csixplusone
		\end{subfigure}
        \begin{subfigure}{0.03\textwidth} \centering
			\caption{} \label{fig:C5b}
		\end{subfigure}
		\begin{subfigure}{0.27\textwidth}
			\centering
			\Cfive
		\end{subfigure}
		
		\begin{subfigure}{0.3\textwidth} \centering
			\includegraphics[height=2.4cm]{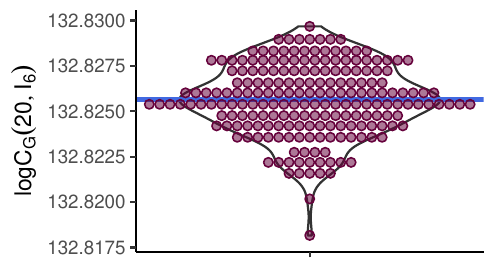}
		\end{subfigure}
		\begin{subfigure}{0.3\textwidth} \centering
			\includegraphics[height=2.4cm]{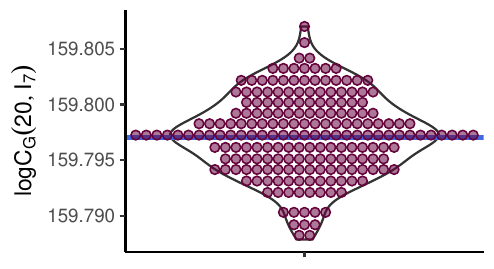}
		\end{subfigure}
		\begin{subfigure}{0.3\textwidth} \centering
			\includegraphics[height=2.4cm]{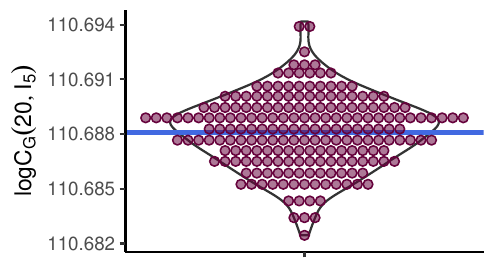}
		\end{subfigure}
		
		\caption{In the graphs, solid edges represent the edges, while dashed edges represent missing edges. In the violin plots, the dots represent the estimates of the value of $\log\const_{\G}(20, I_p)$, where $\G$ is the corresponding graph with $p$ vertices, using Monte Carlo integration \cite{Atay-Kayis,mw19} with $1000$ samples for 200 different seeds. They agree well with our results, represented by the horizontal lines, which have the benefit of avoiding stochastic noise and higher computational efficiency.}
	\end{figure}
	Let $\beta > -1$ be a real number. By Theorem~\ref{thm:main}, $\mathcal I_\G(\beta, I_6)$ is equal to
	\begin{eqnarray}
		&  & \dfrac{\mathcal I_{\G^*}(\beta, I_6)}{\pi^3} \int_{\mathbb R^3} (1+t_1^2+t_2^2+t_3^2 + t_1^2t_3^2)^{-\beta-\frac52} (1+t_1^2+t_2^2)^{-\frac12} (1+t_2^2+t_3^2)^{-\frac12} \dd t_1 \dd t_2 \dd t_3, \nonumber
	\end{eqnarray}
	which is a 3-dimensional integral. Substituting $t_2 \to (1+t_1^2)^{\frac12}(1+t_3^2)^{\frac12}t_2$, we have
	\begin{eqnarray}
		\mathcal I_\G(\beta, I_6) 
		& = & \dfrac{\mathcal I_{\G^*}(\beta, I_6)}{\pi^3} \int_{\mathbb R} (1+t_2^2)^{-\beta-\frac52} \left(\int_{\mathbb R} (1+t_1^2)^{-\beta-\frac72} (1+t_2^2(1+t_1^2))^{-\frac12} \dd t_1\right)  \nonumber\\
		& & \s \left(\int_{\mathbb R} (1+t_3^2)^{-\beta-\frac72} (1+t_2^2(1+t_3^2))^{-\frac12} \dd t_3 \right) \dd t_2. \nonumber
	\end{eqnarray}
	
	An explicit formula for $\mathcal I_{\G^*}(\beta, I_6)$ can be found using (\ref{eq:chordalA}). The integrals with respect to $t_1$ and $t_3$ can be solved using the integral representation of the hypergeometric function ${}_2F_1$. Thus, we are left with a one-dimensional integral:
	\begin{eqnarray} \label{I6comp}
		\quad \mathcal I_\G(\beta, I_6) 
		& = & \dfrac{\pi \Gamma_4\left(\beta+\frac52\right) \Gamma\left(\beta+\frac52\right)^4}{\Gamma(\beta+3)^2} \int_0^1 t_2^{\beta+2} (1-t_2)^{-\frac12} {}_2F_1\left(\frac12, \frac12; \beta+3; t_2 \right)^2 \dd t_2.
	\end{eqnarray}
	
	Although we could reduce the 3-dimensional integral to one-dimensional analytically, the resulting integrand with squared hypergeometric functions does not immediately appear to be tractable. It is straightforward to perform a series expansion of the hypergeometric functions and integrate term-by-term, though for computation it is easiest to perform numerical integration. As a demonstration, we therefore compare the numerical integration of (\ref{I6comp}) with the values obtained by Monte Carlo integration \cite{Atay-Kayis,mw19} of Equation (\ref{eq:Iintegraldef}) in Figure~\ref{fig:C6_complementb}. While there is good agreement between these two approaches, ours is numerically exact (without stochastic noise) and much more computationally efficient.
\end{example}

\section{Partitioning the missing edges, $D = I$} \label{sec:partition}

In Theorem~\ref{thm:partition}, the integral $\mathcal I_\G(\beta, I_{|\V(\G)|})$ is factorised into the product of some lower dimensional integrals for some graphs $\G$. We show the proof for a toy example here. A formal proof of this theorem is in Appendix~\ref{sec:appPartition}.

\begin{example} \label{ex:K5K6}
	Let $\G$ be the graph shown in Figure~\ref{fig:K5K6b}. A chordal completion $\G^*$ is also shown in the same figure. Notice that the 6 missing edges can be partitioned into 3 parts satisfying the assumptions in Theorem~\ref{thm:partition}:
	\begin{equation}
		\E_1 = \{\{v_1, v_6\}, \{v_5, v_6\}\}, \s \E_2 = \{\{v_1, v_8\}, \{v_7, v_8\}\}, \s \E_3 = \{\{v_2, v_3\}, \{v_3, v_4\}\}. \nonumber
	\end{equation}
	The three graphs $\G_1, \G_2, \G_3$ are shown in Figure~\ref{fig:K5K6G123}.
	The maximal cliques of $\G^*$ are $\cc_1 = \{v_1, \ldots, v_6\}$ and $\cc_2 = \{v_1, v_2, v_3, v_7, v_8\}$, and the only minimal separator is $\spt_1 = \{v_1, v_2, v_3\}$. Let $\beta > -1$ be a real number. By Theorem~\ref{thm:main}, we have
	\begin{eqnarray}
		\mathcal I_{\G} (\beta, I_8) 
		& = & \dfrac{\mathcal I_{\G^*}(\beta, I_8)}{\pi^{6}} \bigintss_{\mathbb S(\G, \G^*)} \dfrac{\det((I_8+iT)[\cc_1])^{-\beta-\frac{|\cc_1|+1}{2}} \det((I_8+iT)[\cc_2])^{-\beta-\frac{|\cc_2|+1}{2}}}{\det((I_8+iT)[\spt_1])^{-\beta-\frac{|\spt_1|+1}{2}}} \dd T. \nonumber
	\end{eqnarray}
	We use the variables of the missing edges as shown in Figure~\ref{fig:K5K6b}. The above integrand is
	\begin{eqnarray}
		& & \dfrac{ ((1+s_1^2+s_2^2)(1+t_1^2+t_2^2))^{-\beta - \frac72}((1+r_1^2+r_2^2)(1+s_1^2))^{-\beta - 3}}{(1+s_1^2)^{-\beta - 2}}\nonumber \\
		& = & (1+t_1^2+t_2^2)^{-\beta - \frac72} \times (1+r_1^2+r_2^2)^{-\beta - 3} \times \dfrac{(1+s_1^2+s_2^2)^{-\beta - \frac72}(1+s_1^2)^{-\beta - 3}}{(1+s_1^2)^{-\beta-2}}, \label{eq:K5K6ex}
	\end{eqnarray}
	which can be factorised into 3 separable functions. Notice that these three functions correspond exactly to the integrands of $\mathcal I_{\G_\mu}(\beta, I_8)$, for $\mu = 1, 2, 3$. Indeed, by Theorem~\ref{thm:main}
	\begin{eqnarray}
		\mathcal I_{\G_1}(\beta, I_8) & = & \dfrac{\mathcal I_{\G^*}(\beta, I_8)}{\pi^2} \int_{\mathbb R^2} (1+t_1^2+t_2^2)^{-\beta - \frac72} \dd t_1 \dd t_2 = \dfrac{\mathcal I_{\G^*}(\beta, I_8)}{\pi^2} \dfrac{\pi}{\beta+\frac52}, \nonumber \\
		\mathcal I_{\G_2}(\beta, I_8) & = & \dfrac{\mathcal I_{\G^*}(\beta, I_8)}{\pi^2} \int_{\mathbb R^2} (1+r_1^2+r_2^2)^{-\beta - 3} \dd r_1 \dd r_2 = \dfrac{\mathcal I_{\G^*}(\beta, I_8)}{\pi^2} \dfrac{\pi}{\beta+2}, \nonumber \\
		\mathcal I_{\G_3}(\beta, I_8) & = & \dfrac{\mathcal I_{\G^*}(\beta, I_8)}{\pi^2} \int_{\mathbb R^2} \dfrac{(1+s_1^2+s_2^2)^{-\beta - \frac72}(1+s_1^2)^{-\beta - 3}}{(1+s_1^2)^{-\beta-2}} \dd s_1 \dd s_2 = \dfrac{\mathcal I_{\G^*}(\beta, I_8)}{\pi^2}\dfrac{\pi}{\beta+3}. \nonumber
	\end{eqnarray}
	
	Hence, in line with Theorem~\ref{thm:partition}, we have	
	\begin{eqnarray}
		\mathcal I_{\G}(\beta, I_8) & = & \dfrac{\mathcal I_{\G^*}(\beta, I_8)}{\pi^6} \dfrac{\pi^2 \mathcal I_{\G_1}(\beta, I_8)}{\mathcal I_{\G^*}(\beta, I_8)} \dfrac{\pi^2 \mathcal I_{\G_2}(\beta, I_8)}{\mathcal I_{\G^*}(\beta, I_8)} \dfrac{\pi^2 \mathcal I_{\G_3}(\beta, I_8)}{\mathcal I_{\G^*}(\beta, I_8)} = \mathcal I_{\G^*}(\beta, I_8)^{-2} \mathcal I_{\G_1}(\beta, I_8) \mathcal I_{\G_2}(\beta, I_8) \mathcal I_{\G_3}(\beta, I_8). \nonumber
	\end{eqnarray}
	While for computing the result, we simplify as follows
	\begin{equation}
		\mathcal I_{\G}(\beta, I_8) = \dfrac{\mathcal I_{\G^*}(\beta, I_8)}{\pi^6} \dfrac{\pi}{\beta+\frac52} \dfrac{\pi}{\beta+2} \dfrac{\pi}{\beta+3} = \dfrac{\pi^{\frac12}  \Gamma_6\left(\beta+\frac72\right) \Gamma\left(\beta+\frac52\right) \Gamma(\beta+3)}{\left(\beta+\frac52\right) (\beta+2) (\beta+3)}. \nonumber
	\end{equation}
	Figure~\ref{fig:K5K6b} illustrates that the above result aligns well with the values obtained using Monte Carlo integration \cite{Atay-Kayis, mw19}.

	\begin{figure}
    \centering
		\KfiveKsixA $\qquad$ \KfiveKsixB $\qquad$ \KfiveKsixC		
		\caption{Solid edges represent the three graphs $\G_1, \G_2, \G_3$ (from left to right) in Example~\ref{ex:K5K6}. Dashed edges represent missing edges.}
		\label{fig:K5K6G123}
	\end{figure}
\end{example}

We apply Theorem~\ref{thm:partition} to complete $k$-partite graphs and obtain an explicit formula for $\mathcal I_{\G}(\beta, I_p)$. This recovers a result in \cite{Uhler} (Proposition 2.1) in which the graph is complete bipartite ($k=2$), and generalises it.

\begin{example} \label{ex:completekpartite}
	Let $\G \cong \com_{p_1, \ldots, p_k}$ be a complete $k$-partite graph on $p = p_1 + \cdots + p_k$ vertices. Let $\G^* \cong \com_p$ be the complete graph on the same vertex set, which is a chordal completion of $\G$.
	Let $\beta > -1$ be a real number. For $1 \leq \xi \leq k$, let $\tau_\xi = \binom{n_\xi}{2}$. Let $\tau = \tau_1 + \cdots + \tau_k$ be the total number of missing edges. By Theorem~\ref{thm:partition}, 
	\begin{equation}
		\mathcal I_\G(\beta, I_p) = \mathcal I_{\G^*}(\beta, I_p)^{1-k}\prod_{\xi=1}^k \mathcal I_{\G_\xi}(\beta, I_p), \nonumber
	\end{equation}
	where $\G_\xi$ is the graph obtained from the complete graph $\com_p$ by removing all edges whose end points both belong to some $p_\xi$ vertices.
	Since $\G^*$ is complete, we have
	\begin{equation}
		\mathcal I_{\G^*}(\beta, I_p) = \Gamma_p\left(\beta+\frac{p+1}{2}\right). \nonumber
	\end{equation}
	It remains to find $\mathcal I_{\G_\xi}(\beta, I_p)$. Let $\tilde \G_\xi$ be the empty graph on $p_\xi$ vertices. By Theorem~\ref{thm:main},
	\begin{eqnarray}
		\mathcal I_{\G_\xi}(\beta, I_p) & = & \dfrac{\mathcal I_{\com_{p}}(\beta, I_p)}{\pi^{\tau_\xi}} \int_{\mathbb S(\G_\xi, \com_p)} \det(I_p + iT)^{-\beta-\frac{p+1}{2}} \dd T \nonumber\\
		& = & \dfrac{\mathcal I_{\com_{p}}(\beta, I_p)}{\pi^{\tau_\xi}} \int_{\mathbb S(\tilde \G_\xi, \com_{p_\xi})} \det(I_{n_\xi} + iT)^{-(\beta+\frac{p-p_\xi}{2})-\frac{p_\xi+1}{2}} \dd T \nonumber\\
		& = & \dfrac{\mathcal I_{\com_{p}}(\beta, I_p)}{\pi^{\tau_\xi}} \dfrac{\mathcal I_{\tilde \G_\xi}(\beta + \frac{p-p_\xi}{2}, I_{p_\xi}) \pi^{\tau_\xi}}{\mathcal I_{\com_{p_\xi}}(\beta + \frac{p-p_\xi}{2}, I_{p_\xi})} = \Gamma_p\left(\beta+\frac{p+1}{2}\right) \dfrac{\Gamma\left(\beta+\frac{p-p_\xi}{2}+1\right)^{p_\xi}}{\Gamma_{p_\xi}\left(\beta+\frac{p-p_\xi}{2}+\frac{p_\xi+1}{2}\right)}.\nonumber
	\end{eqnarray}
	Altogether, we have
	\begin{equation}
		\mathcal I_\G(\beta, I_p) = \Gamma_p\left(\beta+\frac{p+1}{2}\right)\prod_{\xi=1}^k \dfrac{\Gamma\left(\beta+\frac{p-p_\xi}{2}+1\right)^{p_\xi}}{\Gamma_{p_\xi}\left(\beta+\frac{p+1}{2}\right)}. \nonumber
	\end{equation}
\end{example}

\subsection{No triangle contains two missing edges}

For graphs $\G$ on $p$ vertices with minimum fill-in 1, a result in \cite{Uhler} (Theorem 2.5) states that, for real number $\beta > -1$,
\begin{equation}
	\mathcal I_\G(\beta, I_p) = \dfrac{\mathcal I_{\G^*}(\beta, I_p)}{\pi^{\frac12}} \dfrac{\Gamma\left(\beta+\frac{w+2}{2}\right)}{\Gamma\left(\beta+\frac{w+3}{2}\right)}, \label{eq:mfi1}
\end{equation}
where $\G^*$ is a chordal completion of $\G$ with one additional edge, and $w$ is the number of common neighbours of the end vertices of the added edge. We give an alternative proof of this result using Theorem~\ref{thm:main} in Appendix~\ref{sec:appmfi1}, and generalise to arbitrary $D$ in Section~\ref{sec:generalA}.

In this subsection, we combine Theorem~\ref{thm:partition} and (\ref{eq:mfi1}) to write an explicit formula for $\mathcal I_\G(\beta, I_{|\V(\G)|})$ for graphs $\G$ with a chordal completion $\G^*$ such that no triangle in $\G^*$ has more than one missing edge; \emph{i.e.}\ in every clique of $\G^*$, the missing edges are vertex-disjoint. 

\begin{corollary} \label{thm:disjoint}
	Let $\G$ be a graph with vertex set $\V(\G) = \{v_1, \ldots, v_p\}$. Let $\G^*$ be a chordal completion of $\G$.
	Suppose that every triangle in $\G^*$ contains at most one edge from $\E(\G^*) \setminus \E(\G)$.
	Let $\beta > -1$ be a real number.	Then,
	\begin{equation}
		\mathcal I_{\G} (\beta, I_p) = \dfrac{\mathcal I_{\G^*}(\beta, I_p)}{\pi^{\frac {|\E(\G^*)| - |\E(\G)|} 2}} \prod\limits_{e \in \E(\G^*) \setminus \E(\G)} \dfrac{ \Gamma\left(\beta + \frac{w_e+2}{2}\right)}{\Gamma\left(\beta + \frac{w_e+3}{2}\right)}, \nonumber
	\end{equation}
	where $w_e$ is the number of common neighbours of the end vertices of the edge $e$ in $\G^*$.
\end{corollary}

\begin{proof}	
	Notice that the assumptions stated in Theorem~\ref{thm:partition} are satisfied if we partition the missing edges into singletons. Let $\E(\G^*) \setminus \E(\G) = \{e_1\} \sqcup \cdots \sqcup \{e_\tau\}$. By Theorem~\ref{thm:partition}
	\begin{equation}
		\mathcal I_{\G}(\beta, I_p) = \mathcal I_{\G^*}(\beta, I_p) \prod_{\xi=1}^\tau \dfrac{\mathcal I_{\G_\xi}(\beta, I_p)}{\mathcal I_{\G^*}(\beta, I_p)}, \nonumber
	\end{equation}
	where $\G_\xi$ is the graph obtained from $\G^*$ by removing the edge $e_\xi$, for $1 \leq \xi \leq \tau$.
	Since the graphs $\G_\xi$ have minimum fill-in 1, one can apply (\ref{eq:mfi1}) and the proof is completed.
\end{proof}

As an example, we find $\mathcal I_\G(\beta, I_{2p})$, where $\G$ is the \textit{complete $p$-partite graph} with 2 vertices in each part. In other words, it is a graph obtained by removing $p$ pairwise vertex-disjoint edges (\emph{i.e.}\ a \textit{perfect matching}) from the complete graph $\com_{2p}$. This graph is also known as the \textit{Tur\'an graph} $\mathcal T(2p,p)$. See Figure~\ref{fig:T63b} for $\mathcal T(6, 3)$. We remark that, for $p \geq 2$, the graph $\mathcal T(2p,p)$ is prime and has minimum fill-in $(p-1)$.

\begin{example} \label{ex:turan}
	Let $\G$ be the Tur\'an graph $\mathcal T(2p,p)$, and let $\beta > -1$ be a real number. We apply Corollary~\ref{thm:disjoint} with $\G^*$ isomorphic to the complete graph $\com_{2p}$. Then, for every missing edge $e$, $w_e = 2p-2$. Thus,
	\begin{eqnarray} \label{Ituran}
		\qquad \mathcal I_\G(\beta, I_{2p}) & = & \dfrac{\mathcal I_{\com_{2p}}(\beta, I_{2p})}{\pi^{\frac p 2}} \dfrac{ \Gamma\left(\beta + \frac{2p}{2}\right)^p}{\Gamma\left(\beta + \frac{2p+1}{2}\right)^p} = \pi^{-\frac p 2} \Gamma_{2p} \left(\beta + \dfrac{2p+1}{2}\right)\dfrac{ \Gamma\left(\beta + p\right)^p}{\Gamma\left(\beta + \frac{2p+1}{2}\right)^p}. 
	\end{eqnarray}
	
	Figure~\ref{fig:T63b} illustrates for $p = 3$ that the above formula aligns well with the values obtained using Monte Carlo integration \cite{Atay-Kayis, mw19}.
\end{example}

\subsection{Starry fill-ins} \label{sec:starryfillins}

In graph theory, a \textit{star} is a connected graph with at least 2 vertices, in which all but at most one vertex have degree 1. Equivalently, a star is isomorphic to the complete bipartite graph $\com_{1,m}$, for some $m \geq 1$. If the missing edges form a star, then the corresponding normalising constant is of the form of Equation (\ref{eq:specialForm}).

\begin{lemma}
	Let $\G$ be a graph with vertex set $\V(\G) = \{v_1, \ldots, v_p\}$. Let $\G^*$ be a chordal completion of $\G$.
	Suppose that the missing edges $\E(\G^*) \setminus \E(\G)$ form a star.
	Let $\beta > -1$ be a real number. Then, the integral $\mathcal I_\G(\beta, I_p)$ has the form of Equation (\ref{eq:specialForm}).
\end{lemma}

\begin{proof}
	Let $e_1, \ldots, e_\tau$ be the missing edges. By Theorem~\ref{thm:main}, we have
	\begin{equation}
		\mathcal I_{\G}(\beta, I_p) = \dfrac{\mathcal I_{\G^*}(\beta, I_p)}{\pi^\tau} \bigintss_{\mathbb S(\G, \G^*)} \dfrac{\prod\limits_{\mu=1}^{m} \det((I_p+iT)[\cc_\mu])^{-\beta - \frac{|\cc_\mu|+1}{2}} }{\prod\limits_{\nu=1}^{m-1} \det((I_p+iT)[\spt_\nu])^{-\beta - \frac{|\spt_\nu|+1}{2}}} \dd T, \nonumber
	\end{equation}
	where $\cc_1, \ldots, \cc_m$ are the maximal cliques of $\G^*$ and $\spt_1, \ldots, \spt_{m-1}$ are the minimal separators. 
	For a matrix $T \in \mathbb S(\G, \G^*)$, let $t_\mu$ be the two entries corresponding to the missing edge $e_\mu$, for $1 \leq \mu \leq \tau$. For a set $\cc \subseteq \{v_1, \ldots, v_p\}$ containing missing edges $e_{\alpha_1}, \ldots, e_{\alpha_k}$, it is easy to see that
	\begin{equation}
		\det((I_p + iT)[\cc]) = 1 + t_{\alpha_1}^2 + \cdots + t_{\alpha_k}^2. \nonumber
	\end{equation}
\end{proof}

As an example, we find $\mathcal I_\G(\beta, I_6)$ for the cycle of length 6.

\begin{example} \label{ex:C6}
	Let $\G$ be the cycle of length 6 and $\G^*$ be a chordal completion as shown in Figure~\ref{fig:C6b}. The maximal cliques of $\G^*$ are
	\begin{equation}
		\cc_1 = \{v_1, v_2, v_3\}, \s \cc_2 = \{v_1, v_3, v_4\}, \s \cc_3 = \{v_1, v_4, v_5\}, \s \cc_4 = \{v_1, v_5, v_6\}, \nonumber
	\end{equation}
	and the corresponding separators are
	\begin{equation}
		\spt_1 = \{v_1, v_3\}, \s \spt_2 = \{v_1, v_4\}, \s \spt_3 = \{v_1, v_5\}. \nonumber
	\end{equation}
	We write $t_1, t_2, t_3$ in place of $t_{1,3}, t_{1,4}, t_{1,5}$, respectively. Let $\beta > -1$ be a real number. Then,
	\begin{equation}
		\mathcal I_\G(\beta, I_6) 
		 =   \Gamma_3(\beta+2)\Gamma(\beta+2)^3   \int_{\mathbb R^3} \dfrac{(1+t_1^2)^{-\frac12}(1+t_1^2+t_2^2)^{-\beta-2}(1+t_2^2+t_3^2)^{-\beta-2}(1+t_3^2)^{-\frac12}}{(1+t_2^2)^{-\beta-\frac32}} \dd t_1 \dd t_2 \dd t_3. \s \label{eq:C6}
	\end{equation}
	As in Example~\ref{ex:C6complement}, it can be written as a one-dimensional integral:
	\begin{equation}
		\mathcal I_\G(\beta, I_6)  =  \dfrac{ \pi \Gamma_3(\beta+2) \Gamma(\beta+2)^5}{\Gamma\left(\beta+\frac52\right)^2} \int_0^1 t^{-\frac12}(1-t)^{\beta+1} {}_2F_1\left(\beta+2, \frac12; \beta+\frac52; t \right)^2 \dd t. \nonumber
	\end{equation}
	
	We compare our result evaluated through numerical integration with values obtained by Monte Carlo integration \cite{Atay-Kayis,mw19} in Figure~\ref{fig:C6b}. Again there is good agreement though ours is much more accurate and efficient.
	
\end{example}

In light of Theorem~\ref{thm:partition}, one can have more stars among the missing edges. Let $\G$ be a graph and let $\st_1, \ldots, \st_k$ be disjoint subsets of $\V(\G)$, each of size at least 2. We say that $\G$ is \textit{star-complementary} (with respect to $\st_1, \ldots, \st_k$) if the complement of $\G$ is a disjoint union of star graphs with vertex sets $\st_1, \ldots, \st_k$, and we define $\mathcal F(\G) := \{\st_1, \ldots, \st_k\}$. Figure~\ref{fig:starComplement} shows some examples of these graphs. We say that a non-chordal graph $\G$ has \textit{starry fill-ins} if $\G$ has a chordal completion $\G^*$ such that for every maximal clique $\cc$ of $\G^*$, the induced subgraph $\G[\cc]$ is star-complementary.

Note that the graph $\G$ in Example~\ref{ex:K5K6} has starry fill-ins, and the corresponding integral has the form of (\ref{eq:specialForm}), see (\ref{eq:K5K6ex}). The following result shows that this is true for all graphs with starry fill-ins. A formal proof is in Appendix~\ref{sec:appStar}. 

\begin{figure}
\centering
	\begin{subfigure}{0.3\textwidth}
		\centering
		\GstarA
		\caption*{$\G_1$}\end{subfigure}
	\begin{subfigure}{0.3\textwidth}
		\centering
		\GstarB
		\caption*{$\G_2$} \end{subfigure}
	\begin{subfigure}{0.3\textwidth}
		\centering
		\GstarC
		\caption*{$\G_3$} \end{subfigure}
	\caption{Solid edges: All three graphs are star-complementary, with $\mathcal F(\G_1) = \mathcal F(\G_2) = \{\{v_1, v_2, v_3, v_4\}, \{v_5, v_6, v_7\}\}$ and $\mathcal F(\G_3) = \{\{v_1, v_2\}, \{v_3, v_4\}, \{v_5, v_6\}, \{v_7, v_8\}\}$. Dashed edges represent those in the complement.} \label{fig:starComplement}
\end{figure}

\begin{corollary} \label{thm:star}
	Let $\G$ be a graph with vertex set $\V(\G) = \{v_1, \ldots, v_p\}$. 
	Suppose that $\G$ has starry fill-ins with a chordal completion $\G^*$. 
	Let $\beta > -1$ be a real number. Then,
	\begin{equation}
		\mathcal I_{\G} (\beta, I_p) = \dfrac{\mathcal I_{\G^*}(\beta, I_p)}{\pi^{|\E(\G^*)| - |\E(\G)|}} \bigints_{\mathbb R^{\tau}} \dfrac{\prod\limits_{\mu=1}^m \prod\limits_{\st \in \mathcal F(\cc_\mu)} (1 + \sum\limits_{\substack{v_r, v_s \in \st \\ \{v_r, v_s\} \in \E(\G^*) \setminus \E(\G)}} t_{r,s}^2)^{-\beta - \frac{|\cc_\mu|+1}{2}}  }{\prod\limits_{\nu=1}^{m-1} \prod\limits_{\st \in \mathcal F(\spt_\nu)} (1 + \sum\limits_{\substack{v_r, v_s \in \st \\ \{v_r, v_s\} \in \E(\G^*) \setminus \E(\G)}} t_{r,s}^2)^{-\beta - \frac{|\spt_\nu|+1}{2}}} \prod_{\substack{1 \leq r < s \leq p\\ \{v_r, v_s\} \in \E(\G^*) \setminus \E(\G)}} \dd t_{r, s}, \nonumber
	\end{equation}
	where $\cc_1, \ldots, \cc_m$ are the maximal cliques and $\spt_1, \ldots, \spt_{m-1}$ the minimal separators of $\G^*$.
\end{corollary}

\subsection{The graphs $\G(m; k_1, \ldots, k_\ell)$} \label{sec:Gmk}
For integers $m \geq 4$, $3 \leq \ell \leq m-1$, and $k_1, \dots,k_\ell \geq 1$, we define the graph $\G^*(m; k_1, \ldots, k_\ell)$ as follows. The vertex set of $\G^*(m; k_1, \ldots, k_\ell)$ is $\{v_0, v_1, \ldots, v_{m-1}\} \sqcup \{u_1, u_2, \ldots, u_{k_1 +\cdots + k_\ell}\}$. In this graph, the edges form complete subgraphs on each of the $\ell+1$ sets $\{v_0, v_1, \ldots, v_{m-1}\}$, $\{v_0, v_1, u_1, \ldots, u_{k_1}\}$, $\{v_0, v_2, u_{k_1+1}, \ldots, u_{k_1+k_2}\},$
$\ldots,\{v_0, v_\ell, u_{k_1 +\cdots + k_{\ell-1}+1}, \ldots, u_{k_1 +\cdots + k_\ell}\}$, and there is no other edge. With the same parameters, let $\G(m; k_1, \ldots, k_\ell)$ to be the graph obtained from $\G^*(m; k_1, \ldots, k_\ell)$ by removing the edges $\{v_0, v_1\}, \ldots, \{v_0, v_\ell\}$. Figure~\ref{fig:Gmk} gives an example of these graphs.

Observe that the graph $\G(m; k_1, \ldots, k_\ell)$ is prime and has minimum fill-in $\ell$ and $\G^*(m; k_1, \ldots, k_\ell)$ is a chordal completion. Moreover, it has starry fill-ins and Corollary~\ref{thm:star} implies that the integral of interest has the form of
\begin{equation}
	\int_{\mathbb R^\tau} (1+t_1^2+\cdots+t_\tau^2)^{-\gamma} (1+t_1^2)^{-r_1} \cdots (1+t_\tau^2)^{-r_\tau} \dd t_1 \cdots \dd t_\tau, \nonumber
\end{equation}
which can be estimated efficiently using the fact that a Student-$t$ is a compound distribution of a $\chi^2$ and a Gaussian. We show in Appendix~\ref{sec:appell>2} that this reduces to a one-dimensional real integral. For numerical accuracy, we map the one-dimensional integral to the cumulative density function space of the chi-squared distributions and integrate over the $(0,1)$ interval.  

\begin{figure}
\centering
	\begin{subfigure}{0.3\textwidth}
		\centering
		\Gfivetwooneone
		\caption{} \label{fig:Gmk}
	\end{subfigure}
	\begin{subfigure}{0.3\textwidth}
		\centering
		\gearthree
		\caption{} \label{fig:gear}
	\end{subfigure}
	\begin{subfigure}{0.3\textwidth}
		\centering
		\gearfour
		\caption{} \label{fig:gear4}
	\end{subfigure}
	\caption{(a) Solid edges represent the graph $\G(5; 2,1,1)$. Together with the dashed ones, it is the graph $\G^*(5; 2,1,1)$. (b) Solid edges represent the graph $\G$ in Section~\ref{sec:missingTriangle}. Together with the dashed ones it is the chordal completion $\G^*$. The graph $\G$ is also the gear graph $\G_3$ defined in Section \ref{sec:gear}. (c) Solid edges represent the gear graph $\G_4$. Together with the dashed edges, it is the graph $\G_4^*$.}
\end{figure}

\begin{corollary} \label{coro:ell>2}
	Let $m \geq 4$, $3 \leq \ell \leq m-1$ and $k_1, \ldots, k_\ell \geq 1$ be integers. Let $\G = \G(m; k_1, \ldots, k_\ell)$ and $\beta > -1$ be a real number. Then,
	\begin{eqnarray}
		\frac{\mathcal I_\G(\beta, I_{m+k_1+\cdots +k_\ell})}{\pi^{k_1+\cdots + k_\ell - \frac{\ell}{2}}} & = & \Gamma_{m}\left(\beta+\frac{m+1}{2}\right) \prod\limits_{\mu=1}^\ell \Gamma_{k_\mu}\left(\beta+\frac{k_\mu+3}{2}\right) \nonumber\\
		& & \times \dfrac{1}{\Gamma\left(\beta + \frac{m+1}{2} \right) 2^{\beta+\frac{m+1}{2}} } \int_0^\infty x^{\beta+\frac{m+1}{2} - 1} e^{-\frac{x}{2}}\prod_{\mu=1}^\ell U\left(\frac12, \frac32 - \frac{k_\mu}{2}, \frac{x}{2}\right) \dd x \nonumber\\
		& = & \Gamma_{m}\left(\beta+\frac{m+1}{2}\right) \prod\limits_{\mu=1}^\ell \Gamma_{k_\mu}\left(\beta+\frac{k_\mu+3}{2}\right) \int_0^1 \prod_{\mu=1}^\ell U\left(\frac12, \frac32 - \frac{k_\mu}{2}, \frac{F^{-1}_{\chi^2(2\beta+m+1)}(x)}{2}\right) \dd x, \nonumber
	\end{eqnarray}
	where $U(a,b,x)$ is the Tricomi's confluent hypergeometric function and $F_{\chi^2(\nu)}(x)$ is the cumulative density function of a chi-squared distribution with $\nu$ degrees of freedom.
\end{corollary}

\begin{example}
	Let $\G = \G(4;1,1,1)$ and $\G^* = \G^*(4;1,1,1)$, as shown in Figure~\ref{fig:C6+1b}. Let $\beta > -1$ be a real number. Then,
	\begin{eqnarray} \label{IGweird}
		\mathcal I_\G(\beta, I_7) 
		& = & \pi^{\frac32} \Gamma_4\left(\beta+\frac52\right) \Gamma(\beta+2)^3 \int_0^1 U\left(\frac12, 1, \frac{F^{-1} _{\chi^2(2\beta+5)}(x)}{2}\right)^3 \dd x. 
	\end{eqnarray}
	We again compare our result evaluated through numerical integration with the values obtained by Monte Carlo integration \cite{Atay-Kayis,mw19} in Figure~\ref{fig:C6+1b}.
\end{example}	

\subsection{Two missing edges} \label{sec:mfi2}

In this subsection, we assume that $\G$ is a (prime) graph which has a chordal completion $\G^*$ with two extra edges.
There are 3 possibilities regarding the adjacency of the two missing edges $\{u,v\}, \{x,y\}$:
\begin{enumerate}
	\item they are vertex-disjoint, or
	\item they share a vertex, say $u = x$, and $\{v, y\}$ is not an edge in $\G$, or
	\item they share a vertex, say $u = x$, and $\{v, y\}$ is an edge in $\G$.
\end{enumerate}

For the first two cases, Corollary~\ref{thm:disjoint} provides an explicit formula of the $\G$-Wishart normalising constant when $D$ is the identity matrix. For the last case, Corollary~\ref{thm:star} implies that
\begin{equation}
	\mathcal I_{\G} (\beta, I_p) = \dfrac{\mathcal I_{\G^*}(\beta, I_p)}{\pi^2} \int_{\mathbb R^{2}} (1+t_1^2+t_2^2)^{-\gamma} (1+t_1^2)^{-r} (1+t_2^2)^{-s} \dd t_1 \dd t_2, \nonumber
\end{equation}
for some $r, s, \gamma$. By integrating, we obtain the following result. The details can be found in Appendix~\ref{sec:appmfi2}.

\begin{corollary} \label{coro:mfi2}
	Let $\G$ be a prime graph on $p$ vertices with minimum fill-in at most 2. Let $\G^*$ be a chordal completion of $\G$ with 2 added edges.
	Suppose that these two added edges form a triangle with an edge of $\G$ in $\G^*$. Let $e_1 = \{v_1, v_3\}, e_2 = \{v_2, v_3\}$ be the two missing edges.
	Let $\beta > -1$ be a real number. Then,
	\begin{equation}
		\mathcal I_{\G}(\beta, I_p) =  \dfrac{\mathcal I_{\G^*}(\beta, I_p)}{\pi} \dfrac{\Gamma\left(\beta + \frac{w_1+3}2\right) \Gamma\left(\beta + \frac{w_2+3}2\right)}{\Gamma(\beta + \frac{w_1+4}2)\Gamma(\beta + \frac{w_2+4}2)} {}_3F_2 \left( \beta + \frac{w+4}2, \frac12, \frac12; \beta + \frac{w_1+4}2, \beta + \frac{w_2+4}2; 1\right), \nonumber
	\end{equation}
	where $w$ is the number of common neighbours of $v_1, v_2$ in $\G$, and for $\mu = 1, 2$, $w_\mu$ is the number of common neighbours of $v_\mu, v_3$ or $v_1, v_2$ in $\G$.
\end{corollary}

If $w_1 = w_2 = w+1$, Dixon's identity implies
\begin{equation}
	{}_3F_2 \left( \gamma, \frac12, \frac12; \gamma+\frac12, \gamma+\frac12; 1\right) = \dfrac{\Gamma\left(\frac{\gamma+2}{2}\right) \Gamma\left(\frac{\gamma}{2}\right) \Gamma\left(\gamma+\frac12\right)^2}{\Gamma(\gamma+1) \Gamma(\gamma)\Gamma\left(\frac{\gamma+1}{2}\right)^2}. \nonumber
\end{equation}
We find $\mathcal I_\G(\beta, I_5)$ for the cycle of length 5 as an example.

\begin{example} \label{ex:C5}
	Let $\G$ be the cycle of length 5, which is isomorphic to $\G(3;1,1)$. Then,
	\begin{eqnarray}
		\mathcal I_\G(\beta, I_5) & = & \pi \Gamma(\beta+2)^2 \Gamma_{3}(\beta+2) \dfrac{\Gamma(\beta+2)^2}{\Gamma\left(\beta+\frac{5}{2}\right)^2} {}_3F_2 \left( \beta+2, \frac12, \frac12; \beta+\frac{5}{2}, \beta+\frac{5}{2}; 1\right) \nonumber\\
		& = & \dfrac{\pi\Gamma_3(\beta+2) \Gamma(\beta+2)^3 \Gamma\left(\frac{\beta+4}{2}\right) \Gamma\left(\frac{\beta+2}{2}\right) }{\Gamma(\beta+3) \Gamma\left(\frac{\beta+3}{2}\right)^2}. \label{I5cycle} 
	\end{eqnarray}
	
	Figure~\ref{fig:C5b} illustrates that the above result aligns well with Monte Carlo integration \cite{Atay-Kayis, mw19}.
\end{example}

\section{A new approach for graphs with minimum fill-in 1, arbitrary $D$} \label{sec:generalA}

Let $\G$ be a graph with $p \geq 4$ vertices which has minimum fill-in 1. For real number $\beta > -1$ and arbitrary matrix $D \in \mathbb S^p_{++}$, a complicated formula for $\mathcal I_\G(\beta, D)$ is given in \cite{Uhler} (Proposition 3.1 and Corollary 3.2). In this Section, we derive an efficient approach to evaluate the $\G$-Wishart normalising constant by reducing it to a one-dimensional integral. 
We order the vertices $\V(\G) = \{v_1, \ldots, v_p\}$ such that the graph $\G^*$ obtained by adding an edge between $v_{p-1}$ and $v_p$ in $\G$ is chordal.	

\subsection{Three assumptions}

For the sake of simplicity, we make a few assumptions.

First, we assume that $D \in \mathbb S^p_{++}$ has diagonal entries all equal to 1 since the transformation
\begin{equation}
	K \to \diag(D)^{-\frac12} K \diag(D)^{-\frac12}  \s \s \mathrm{gives} \s \s 
	\mathcal I_\G(\beta, D)
	= \mathcal I_\G(\beta, \diag(D)^{-\frac12}D\diag(D)^{-\frac12}) \prod_{\mu = 1}^p (d_{\mu, \mu})^{-\frac{\deg_\G(v_\mu)}2 - \beta - 1} , \nonumber
\end{equation}
in which the diagonal elements of $\diag(D)^{-\frac12}D\diag(D)^{-\frac12} \in \mathbb S^p_{++}$ are 1. Here, $\deg_\G(v_\mu) := |\NB_\G(v_\mu)|$ is the \textit{degree} of $v_\mu$ in $\G$.

Second, we set $d_{\mu, \nu} = 0$ whenever $\{v_\mu, v_\nu\} \not \in \E(\G)$. This can be justified since for $K \in \mathbb S^p_{++}(\G)$, the value of $\tr(KD)$, and hence $\mathcal I_\G(\beta, D)$, does not depend on those entries of $D$.

Third, we assume that $\G$ is a prime graph, see (\ref{eq:WishartPrimeFactors}). It is then easy to see that both $v_{p-1}$ and $v_p$ are adjacent to $v_\mu$, for $1 \leq \mu \leq p-2$.

\subsection{Derivation}

Let $\cc_1, \ldots, \cc_m$ be the maximal cliques of $\G^*$, and let $\spt_1, \ldots, \spt_{m-1}$ be the separators. Notice that every $\cc_\mu$, and hence every $\spt_\nu$, contains the vertices $v_{p-1}$ and $v_p$.
By Theorem~\ref{thm:main}, we have
\begin{equation}
	\mathcal I_{\G}(\beta, D) = \dfrac{\mathcal I_{\G^*} (\beta, I_{p})}{\pi} \bigintss_{{\mathbb S}(\G, \G^*)}  \dfrac{\prod\limits_{\mu=1}^m \det((D+iT)[\cc_\mu])^{-\beta-\frac{|\cc_\mu|+1}{2}}}{\prod\limits_{\nu=1}^{m-1} \det((D+iT)[\spt_\nu])^{-\beta-\frac{|\spt_\nu|+1}{2}}} \dd T. \nonumber
\end{equation}

Recall that $T \in {\mathbb S}(\G, \G^*)$ is real symmetric and has entry $t_{\mu, \nu} = 0$ unless $\mu$ and $\nu$ are exactly $p-1$ and $p$. We write $t_{p-1,p } = t_{p, p-1} = t$.
Let $\cc = \{v_{\alpha_1}, \ldots, v_{\alpha_k}, v_{p-1}, v_p\}\subseteq \V(\G)$, and $T \in {\mathbb S}(\G, \G^*)$. Then,
\begin{equation}
	\det((D+iT)[\cc]) = \det \left(\arraycolsep=3pt\def\arraystretch{1.3}
	\begin{array}{c|c}
		D[\cc \setminus \{v_{p-1}, v_p\}] & \begin{matrix}
			d_{\alpha_1, p-1} & d_{\alpha_1, p} \\
			\vdots & \vdots \\
			d_{\alpha_k, p-1} & d_{\alpha_k, p}
		\end{matrix} \\
		\hline
		\begin{matrix}
			d_{\alpha_1, p-1} & \cdots & d_{\alpha_k, p-1} \\
			d_{\alpha_1, p} & \cdots & d_{\alpha_k, p}
		\end{matrix} & \begin{matrix}
			1 & & & & it \\
			it & & & &1
		\end{matrix}
	\end{array}
	\right). \nonumber
\end{equation}
The Schur complement formula for the determinant of a 2 by 2 block matrix implies that
\begin{equation}
\mathcal I_{\G}(\beta, D) = \dfrac{\mathcal I_{\G^*}(\beta, I_p)}{\pi} \dfrac{\prod\limits_{\mu=1}^m \det(D[\cc_\mu \setminus \{v_{p-1}, v_p\}])^{-\beta-\frac{|\cc_\mu|+1}{2}}}{\prod\limits_{\nu=1}^{m-1} \det(D[\spt_\nu \setminus \{v_{p-1}, v_p\}])^{-\beta-\frac{|\spt_\nu|+1}{2}}}  \bigintss_{\mathbb R}  \dfrac{\prod\limits_{\mu=1}^m (t^2 + 2i y_{\cc_\mu} t + x_{\cc_\mu} )^{-\beta-\frac{|{\cc_\mu}|+1}{2}}}{\prod\limits_{\nu=1}^{m-1} (t^2 + 2i y_{\spt_\nu} t + x_{\spt_\nu} )^{-\beta-\frac{|\spt_\nu|+1}{2}}} \dd t, \label{eq:minfi1A}
\end{equation}
where
\begin{equation}
	x_\cc = \dfrac{\det(D[\cc])}{\det(D[\cc \setminus \{v_{p-1}, v_p\}])} , \qquad y_\cc = (d_{\alpha_1, p-1} \cdots d_{\alpha_k, p-1})D[\cc \setminus \{v_{p-1}, v_p\}]^{-1}\left(\begin{matrix}
		d_{\alpha_1, p} \\
		\vdots \\
		d_{\alpha_k, p}
	\end{matrix}\right) \nonumber
\end{equation}
and both can be evaluated more efficiently from the Cholesky decomposition of $D[\cc \setminus \{v_{p-1}, v_p\}]$ and back-substitution, without evaluating the determinants and inverse directly. 

\subsection{Evaluation}

The one dimensional integral in (\ref{eq:minfi1A}) has the form
\begin{eqnarray}
	\mathcal J & = & \mathcal J(x_1, \ldots, x_l; y_1, \dots, y_l; r_1, \ldots, r_l) \nonumber\\
	& = & \int_{\mathbb R} \prod\limits_{\mu=1}^l (t^2 + 2i y_{\mu} t + x_{\mu} )^{r_\mu} \dd t = \int_{\mathbb R} (1+t^2)^{r_1 + \cdots + r_l} \prod\limits_{\mu=1}^l \left(1 + \dfrac{2i y_{\mu} t + x_{\mu} -1}{1+t^2}\right)^{r_\mu} \dd t,\nonumber
\end{eqnarray}
where we extract the overall behaviour in $t$ as a prefactor and have correction terms which vanish as $D$ approaches the identity matrix. Next, the prefactor is related to a Student-$t$ distribution $F_{\student(\nu)}$ with $\nu = -1 - 2\sum_{\mu} r_\mu$ degrees of freedom, and under the change of variable $t \to \phi(t) = {F^{-1}_{\student(\nu)}(t)}/{\sqrt{\nu}}$ we transform to a bounded integral suitable for numerical evaluation
\begin{eqnarray}
	\mathcal J & = & \frac{\Gamma\left(\frac\nu2\right)\Gamma\left(\frac12\right)}{\Gamma\left(\frac{\nu+1}2\right)}\int_{0}^{1} \prod\limits_{\mu=1}^l \left(1 + \dfrac{2i y_{\mu} \phi(t) + x_{\mu} -1}{1+\phi(t)^2}\right)^{r_\mu} \dd t .\nonumber
\end{eqnarray}

\subsection{Fisher's Iris Virginica data} \label{sec:iris}

\begin{figure}
	\begin{subfigure}{1\textwidth}
		\centering
		\hspace{0.1\textwidth} \CfourA \hfill \CfourB \hfill \CfourC \hspace{0.1\textwidth}
	\end{subfigure}
	\\
	\begin{subfigure}{1\textwidth}
		\centering
		\includegraphics[width=0.32\textwidth]{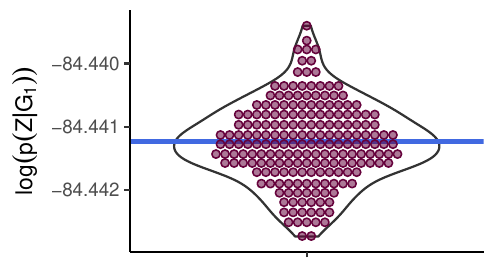} \includegraphics[width=0.32\textwidth]{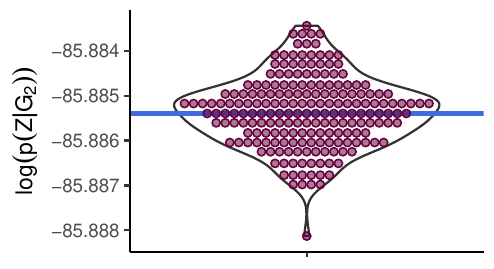} \includegraphics[width=0.32\textwidth]{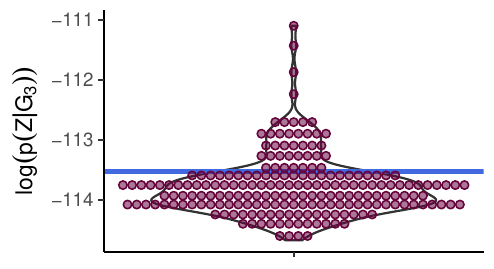}
	\end{subfigure}
	
	\caption{Above: The graphs $\G_1$, $\G_2$, $\G_3$ (from left to right), are the non-chordal graphs for Fisher's Iris Virginica dataset. Below: Violin plots of the estimates of the values of $\log(p(Z \given \G_j)))$, where $j = 1,2,3$, using Monte Carlo integration \cite{Atay-Kayis,mw19} with $10^6$ samples and for 200 different seeds. The horizontal lines represent the values obtained using our approach.} \label{fig:C4}
\end{figure}

As in \cite{Atay-Kayis, Roverato}, for a simple illustration we use Fisher's Iris Virginica dataset of measurements for four features of 50 Iris Virginica flowers: sepal length (SL), sepal width (SW), petal length (PL) and petal width (PW). We highlight here the results for the only non-chordal graphs on 4 vertices which are the three 4-cycles, $\G_1$, $\G_2$, $\G_3$ (see Figure~\ref{fig:C4}), and detail the full analysis in Supplementary Section~\ref{suppsec:iris}.

\section{A Monte Carlo approach for arbitrary $D$}
\label{sec:MC}

\rev{From Theorem~\ref{thm:main}, by comparing our graph $\G$ to the complete graph $\com_p$, and assuming our data matrix is in PD-completion form, our normalising constant is related to the following integral
\begin{equation}
\mathcal I_{\G} (\beta, \tilde{D}) \propto \int_{\mathbb S(\G, \com_p)} \det(\tilde{D}+iT)^{-\gamma} \dd T \propto \int_{\mathbb S(\G, \com_p)} \det(I_p+i\tilde{D}^{-1}T)^{-\gamma} \dd T, \quad \gamma = \beta + \frac{p+1}{2}. \nonumber
\end{equation}
To build a Monte Carlo approach for this integral, we reformulate 
\begin{equation}
\det(I_p+i\tilde{D}^{-1}T)^{-\gamma} = e^{-\gamma\log\det(I_p+i\tilde{D}^{-1}T)} = e^{-\gamma\tr\log(I_p+i\tilde{D}^{-1}T)} \approx e^{-\frac{\gamma}{2}\tr(\tilde{D}^{-1}T)^{2}}  \nonumber
\end{equation}
and approximate by expanding the logarithm up to second order. We note that the linear terms vanish since the PD-completion is 0 by construction where $T$ is non-zero. Defining $\boldsymbol{t}$ to be a vector of the non-zero element pairs of $T$
\begin{equation}
\tr(\tilde{D}^{-1}T)^{2} = 2\boldsymbol{t}^\T\Sigma^{-1}\boldsymbol{t}, \quad \Sigma^{-1}_{ij} = \tilde{D}^{-1}_{\E_{i1}, \E_{j1}}\tilde{D}^{-1}_{\E_{i2}, \E_{j2}} + \tilde{D}^{-1}_{\E_{i1}, \E_{j2}}\tilde{D}^{-1}_{\E_{i2}, \E_{j1}} \nonumber
\end{equation}
where $\E_{i1}$ and $\E_{i2}$ are the row and column position in $T$ corresponding to the $i$-th entry in $\boldsymbol{t}$. The precision matrix $\Sigma^{-1}$ is actually the Isserlis matrix of $\tilde{D}^{-1}$ for the missing edges represented in $T$, and connected \cite{wong2025conjecture} to the approximation of Equation~(\ref{eq:conj}). This approximation, $\det(I_p+i\tilde{D}^{-1}T)^{-\gamma} \approx e^{-\gamma\boldsymbol{t}^\T\Sigma^{-1}\boldsymbol{t}}$, motivates an importance sampler. While a Gaussian proposal with precision $2\gamma\Sigma^{-1}$ is natural, to account for higher-order terms we instead use a multivariate Student-$t$ distribution, yielding heavier tails. Matching the quadratic form gives precision $\frac{2\gamma\nu}{\nu+\tau}\Sigma^{-1}$, where $\tau$ is the length of $\boldsymbol{t}$ and $\nu$ is the degree of freedom. We then sample $\boldsymbol{t}$, embed it in $T$, compute importance weights using $\det(\tilde{D}+iT)^{-\gamma}$ and the proposal density, and include all the relevant constants.

To reduce dimensionality, we generalise this construction by replacing $\com_p$ with a chordal completion $\G^*$. From Theorem~\ref{thm:main}, the resulting integral factorises over cliques and separators. A second-order expansion inside the exponent again yields a quadratic form, but now linear terms arise. We eliminate these via a Newton--Raphson adjustment of $D$ on the missing entries, which we term \textit{clique-completion}, and for which we exploit the gradient and Hessian from the quadratic expansion.

The resulting quadratic form defines the importance distribution as above, leading to an efficient estimator of the $\G$-Wishart normalising constant. We implemented this approach in the R package \textit{GWnorm}, and demonstrate its performance advantages in detail in Supplementary Section~\ref{suppsec:GW_BDcomp}.}

\section{Conclusions}

Efficiently estimating the $\G$-Wishart normalising constant $\const_\G(\delta, D)$ is \rev{highly relevant} for implementing Bayesian inference of Gaussian graphical models and performing Gaussian graphical model selection. Little is known about the form of this constant for general graphs, with available estimation methods resting on computational methods based on Monte Carlo integration. While an explicit representation of the $\G$-Wishart normalising constant for general graphs has been recently derived \cite{Uhler}, its form is not well suited for practical computation.

In this paper, we provide practical results for evaluating the $\G$-Wishart normalising constant for collections of graphs beyond the well-understood class of decomposable graphs. To do so, we introduce, in Section~\ref{sec:Diracdelta}, a transformation of $\const_\G(\delta, D)$ to an unconstrained integral, which is much easier to handle than its original form. After transformation, the tractability of the normalising constant improves substantially under the assumption that $D$ is the identity matrix. For this case, we obtained closed formulae or provided numerically efficient results for $\const_\G(\delta, I_{|\V(\G)|})$ for a variety of graphs.

For many other graphs, we showed that the $\G$-Wishart normalising constant involves solving real integrals of the form of Equation (\ref{eq:specialForm}), which can sometimes be simplified and expressed in terms of special functions, or otherwise be computed numerically. Studying integrals of this particular form would constitute a valuable direction for future work.

Another promising direction for further developments follows from Section~\ref{sec:map}, where we establish a link between the constants $\const_\G(\delta, I_{|\V(\G)|})$ for different graphs (with different values of $\delta$). One might be able to connect a class of graphs with another class, whose normalising constants can be more efficiently estimated. 

Even without assuming that $D$ is the identity matrix, our transformation can facilitate the evaluation of the normalising constant. In Section~\ref{sec:generalA}, we propose an accurate and inexpensive approach for arbitrary $D$ for (prime) graphs with minimum fill-in 1, and illustrate its performance on the Fisher's Iris Virginica data. \rev{This example additionally shows that approximations \cite{mml23, mohammadi2025scalable} employed in Bayesian sampling schemes for undirected graphs can lead to significant bias, while an exchange algorithm without approximations \cite{Willem} has very slow convergence.} 

Finally, for general graphs, Theorem~\ref{thm:main} \rev{allowed us to} devise a new Monte Carlo integration approach by constructing a chordal completion with relatively few edges. \rev{When the number of missing edges is less than the sum of the number of edges and vertices in the graph, the dimension of the integral is lower than current algorithms, such as \cite{Atay-Kayis}, that sample directly in the $\G$-Wishart space. In simulations, we observed that our new approach, building on the Fourier analysis, could be orders of magnitude more accurate.}

Taken together with the exact results presented here for a larger class of graphs than previously known, this could be an interesting avenue for developing different kinds of Bayesian sampling schemes for general graphs, \rev{or allowing schemes designed for chordal graphs to relax that restriction. Indeed, schemes forced to sample from chordal graphs are expected to concentrate on minimal triangulations of the underlying graphs \cite{niu2020bayesian}, and our approach would allow the exploration of nearby non-chordal graphs by removing edges. Likewise, the bias introduced by current Bayesian sampling approaches that introduce approximations for computational efficiency could be mitigated through importance sampling using the normalising constants of the sampled graphs, or anchoring the approximations for near-chordal graphs whose posterior can be estimated well.} 

\bibliographystyle{model1-num-names}

\bibliography{bibliography}


\appendix
	
	\section{Technical proofs}
	
	\subsection{Proof of Theorem~\ref{thm:partition}} \label{sec:appPartition}
	
	For $1 \leq \xi \leq k$, let $\tau_\xi = |\E_\xi|$. Let $\tau = \tau_1 + \cdots + \tau_k$.
	By Theorem~\ref{thm:main}, we have
	\[
	\mathcal I_{\G}(\beta, I_p) = \dfrac{\mathcal I_{\G^*}(\beta, I_p)}{\pi^\tau} \bigintss_{\mathbb S(\G, \G^*)} \dfrac{\prod\limits_{\mu=1}^{m} \det((I_p+iT)[\cc_\mu])^{-\beta - \frac{|\cc_\mu|+1}{2}} }{\prod\limits_{\nu=1}^{m-1} \det((I_p+iT)[\spt_\nu])^{-\beta - \frac{|\spt_\nu|+1}{2}}} \dd T,
	\]
	where $\cc_1, \ldots, \cc_m$ are the maximal cliques and $\spt_1, \ldots, \spt_{m-1}$ the minimal separators of $\G^*$.
	
	Within each clique $\cc \in \{\cc_1, \ldots, \cc_m, \spt_1, \ldots, \spt_{m-1}\}$, the edge parts $\E_\mu$ and $\E_\nu$ are vertex-disjoint whenever $\mu \neq \nu$. For $T \in \mathbb S(\G, \G^*)$, we write $T = T_1 + \cdots + T_k$, where each $T_\xi$ represents the entries corresponding to the missing edges in $\E_\xi$. Then,
	\[
	\det((I_p+iT)[\cc]) = \prod_{\xi=1}^k \det((I_p+iT_\xi)[\cc]),
	\]
	and hence
	\begin{eqnarray}
		\mathcal I_{\G}(\beta, I_p) & = & \dfrac{\mathcal I_{\G^*}(\beta, I_p)}{\pi^\tau} \bigintss_{\mathbb S(\G, \G^*)} \prod_{\xi=1}^k \dfrac{\prod\limits_{\mu=1}^{m} \det((I_p+iT_\xi)[\cc_\mu])^{-\beta - \frac{|\cc_\mu|+1}{2}} }{\prod\limits_{\nu=1}^{m-1} \det((I_p+iT_\xi)[\spt_\nu])^{-\beta - \frac{|\spt_\nu|+1}{2}}} \dd T \nonumber \\
		& = & \dfrac{\mathcal I_{\G^*}(\beta, I_p)}{\pi^\tau} \prod_{\xi=1}^k \bigintss_{\mathbb S(\G_\xi, \G^*)}  \dfrac{\prod\limits_{\mu=1}^{m} \det((I_p+iT_\xi)[\cc_\mu])^{-\beta - \frac{|\cc_\mu|+1}{2}} }{\prod\limits_{\nu=1}^{m-1} \det((I_p+iT_\xi)[\spt_\nu])^{-\beta - \frac{|\spt_\nu|+1}{2}}} \dd T_\xi \nonumber \\
		& = & \dfrac{\mathcal I_{\G^*}(\beta, I_p)}{\pi^\tau} \prod_{\xi=1}^k \dfrac{\mathcal I_{\G_\xi}(\beta, I_p) \pi^{\tau_\xi}}{\mathcal I_{\G^*}(\beta, I_p)} = \mathcal I_{\G^*}(\beta, I_p) \prod_{\xi=1}^k \dfrac{\mathcal I_{\G_\xi}(\beta, I_p)}{\mathcal I_{\G^*}(\beta, I_p)}. \nonumber
	\end{eqnarray}
	
	\subsection{An alternative proof of (\ref{eq:mfi1})} \label{sec:appmfi1}
	
	We first use the inclusion-exclusion principle to prove the following result, which establishes a relationship between the sizes of the maximal cliques and the minimal separators of a chordal graph.
	
	\begin{lemma} \label{lem:inclusionExclusion}
		Let $\cc_1, \ldots, \cc_m$ be subsets of $\{1, \ldots, p\}$ such that $\cc_1 \cup \cdots \cup \cc_m = \{1, \ldots, p\}$.
		Suppose that the sequence $\cc_1, \ldots, \cc_m$ satisfies the running intersection property. For $1 \leq \nu \leq m-1$, let $\spt_\nu = (\cc_1 \cup \cdots \cup \cc_\nu) \cap \cc_{\nu+1}$.
		Then,
		\begin{equation}
			|\cc_1| + \cdots + |\cc_m| - (|\spt_1| + \cdots + |\spt_{m-1}|) = p. \nonumber
		\end{equation}
	\end{lemma}
	
	\begin{proof}
		By the inclusion-exclusion principle,
		\begin{eqnarray}
			p & = & \sum_{\mu=1}^m |\cc_\mu| - \sum_{1 \leq \nu < \mu \leq m} |\cc_\nu \cap \cc_\mu| + \sum_{1 \leq \nu < \xi < \mu \leq m} |\cc_\nu \cap \cc_\xi \cap \cc_\mu| - \ldots + (-1)^{m+1} |\cc_1 \cap \cdots \cap \cc_m| \nonumber\\
			& = & \sum_{\mu = 1}^m (P_{\mu, 1} + \cdots + P_{\mu, \mu}), \nonumber
		\end{eqnarray}
		where
		\begin{equation}
			P_{\mu, \ell} = (-1)^{\ell+1} \sum_{1 \leq \nu_1 < \cdots < \nu_{\ell-1} < \mu} |\cc_{\nu_1} \cap \cdots \cap \cc_{\nu_{\ell-1}} \cap \cc_\mu| \nonumber
		\end{equation}
		denotes the sum of the terms associated with the intersection of $\ell$ sets and $\mu$ is the largest index.
		
		Fix $2 \leq \mu \leq m$, we first show that $P_{\mu, 1} + \cdots + P_{\mu, \mu} = |\cc_\mu| - |\spt_{\mu-1}|$. Let $\gamma \leq \mu-1$ be an index such that $\spt_{\mu-1} = \cc_\gamma \cap \cc_\mu$. We further split $P_{\mu, \ell}$ into two parts, depending on the existence of $\cc_\gamma$, i.e., $P_{\mu, \ell} = P_{\mu, \ell}^+ + P_{\mu, \ell}^-$, where
		\[
		P_{\mu, \ell}^+ = (-1)^{\ell+1} \sum_{\substack{1 \leq \nu_1 < \cdots < \nu_{\ell-2} < \mu \\ \gamma \not \in \{\nu_1, \ldots, \nu_{\ell-2}\}}} |\cc_{\nu_1} \cap \cdots \cap \cc_{\nu_{\ell-2}} \cap \cc_\gamma \cap \cc_\mu|
		\]
		and
		\[
		P_{\mu, \ell}^- = (-1)^{\ell+1} \sum_{\substack{1 \leq \nu_1 < \cdots < \nu_{\ell-1} < \mu \\ \gamma \not \in \{\nu_1, \ldots, \nu_{\ell-1}\}}} |\cc_{\nu_1} \cap \cdots \cap \cc_{\nu_{\ell-1}} \cap \cc_\mu|.
		\]
		For $1 \leq \nu_1 < \cdots < \nu_{\ell-2} < \mu$ with $\gamma \not\in \{\nu_1, \ldots, \nu_{\ell-2}\}$, the running intersection property implies that 
		\[
		\cc_{\nu_1} \cap \cdots \cap \cc_{\nu_{\ell-2}} \cap \cc_\gamma \cap \cc_\mu = \cc_{\nu_1} \cap \cdots \cap \cc_{\nu_{\ell-2}} \cap \cc_\mu.
		\]
		Thus, $P_{\mu, \ell}^+ = - P_{\mu, \ell-1}^-$, for $3 \leq \ell \leq \mu$. Altogether with the observation that both $P_{\mu, \mu}^-$ and $P_{\mu, 1}^+$ are zero, we have
		\[
		P_{\mu, 1} + \cdots + P_{\mu, \mu} = P_{\mu, 1}^- + P_{\mu, 2}^+ = |\cc_\mu| - |\cc_\gamma \cap \cc_\mu| = |\cc_\mu| - |\spt_{\mu-1}|,
		\]
		as claimed.
		
		Finally, as desired, we have
		\[
		p = |\cc_1| + \sum_{\mu=2}^m (|\cc_\mu| - |\spt_{\mu-1}|).
		\]
	\end{proof}
	
	Let $\G$ be a graph on $p$ vertices with minimum fill-in 1, and let $\G^*$ be a minimum chordal completion of $\G$. By Theorem~\ref{thm:main}, we have
	\[
	\mathcal I_{\G}(\beta, I_p) = \dfrac{\mathcal I_{\G^*}(\beta, I_p)}{\pi} \bigintss_{\mathbb S(\G, \G^*)} \dfrac{\prod\limits_{\mu=1}^{m} \det((I_p+iT)[\cc_\mu])^{-\beta - \frac{|\cc_\mu|+1}{2}} }{\prod\limits_{\nu=1}^{m-1} \det((I_p+iT)[\spt_\nu])^{-\beta - \frac{|\spt_\nu|+1}{2}}} \dd T,
	\]
	where $\cc_1, \ldots, \cc_m$ are the maximal cliques of $\G^*$ and $\spt_1, \ldots, \spt_{m-1}$ are the minimal separators. Without loss of generality, assume that $\cc_1, \ldots, \cc_m$ forms a perfect sequence of cliques.
	
	Let $e = \{v_r, v_s\}$ be the missing edge. Note that all entries of a matrix $T \in {\mathbb S}(\G, \G^*)$ are zero, except $t_{r,s} = t_{s,r} =: t \in \mathbb R$. For a clique $\cc \subseteq \V(\G^*)$, we have
	\[
	\det(I_p + iT)[\cc] = \begin{cases}
		1 + t^2, & \textrm{ if $e$ belongs to the clique $\cc$,} \\
		1, & \textrm{ otherwise.}
	\end{cases}
	\]
	
	Now, we find the exponent of $1+t^2$ in the integrand. Suppose that the vertices $v_r$ and $v_s$ are contained in $k$ maximal cliques of $\G^*$, say $\cc_{\alpha_1}, \ldots, \cc_{\alpha_k}$, where $\alpha_1 < \cdots < \alpha_k$. Then, the vertices $v_r$ and $v_s$ are contained in $\spt_\nu$ if and only if $\nu + 1 \in \{\alpha_2, \ldots, \alpha_k\}$. By Lemma~\ref{lem:inclusionExclusion}, the exponent of interest is
	\[
	-\beta - \dfrac12  - \dfrac{|\cc_{\alpha_1}| + \cdots + |\cc_{\alpha_k}|}{2} + \dfrac{|\spt_{\alpha_2-1}| + \cdots + |\spt_{\alpha_k-1}|}{2} = -\beta - \frac{w+3}{2},
	\]
	where $w$ is the number of common neighbours of the end vertices of $e$.
	
	As a result,
	\begin{eqnarray}
		\mathcal I_{\G} (\beta, I_p) & = & \dfrac{\mathcal I_{\G^*}(\beta, I_p)}{\pi} \int_{\mathbb R} (1+t^2)^{-\beta - \frac{w + 3}{2}} \dd t \dfrac{\mathcal I_{\G^*}(\beta, I_p)}{\pi^{\frac 1 2}} \dfrac{ \Gamma\left(\beta + \frac{w+2}{2}\right)}{\Gamma\left(\beta + \frac{w+3}{2}\right)}. \nonumber
	\end{eqnarray}
	
	\subsection{Proof of Corollary~\ref{thm:star}} \label{sec:appStar}
	
	By Theorem~\ref{thm:main}, we have
	\begin{equation}
		\mathcal I_{\G} (\beta, D) = \dfrac{\mathcal I_{\G^*}(\beta, I_p)}{\pi^{|\E(\G^*)| - |\E(\G)|}} \bigintss_{\mathbb S(\G, \G^*)} \dfrac{\prod\limits_{\mu=1}^m \det((D+iT)[\cc_\mu])^{-\beta-\frac{|\cc_\mu|+1}{2}}}{\prod\limits_{\nu=1}^{m-1} \det((D+iT)[\spt_\nu])^{-\beta-\frac{|\spt_\nu|+1}{2}}} \dd T, \nonumber
	\end{equation}
	where $\cc_1, \ldots, \cc_m \subseteq \V(\G)$ are the maximal cliques of $\G^*$ and $\spt_1, \ldots, \spt_{m-1} \subseteq \V(\G)$ are the minimal separators.
	Let $T \in {\mathbb S}(\G, \G^*)$. By assumption, the induced subgraph $\G[\cc_\mu]$, where $1 \leq \mu \leq m$, is star-complementary with respect to the sets $\mathcal F(\G[\cc_\mu])$, and hence
	\begin{equation}
		\label{eq:detStar}
		\det(I_p + iT)[\cc_\mu] = \prod_{\st \in \mathcal F(\G[\cc_\mu])} \det(I_p + iT)[\st].
	\end{equation}
	Since each separator $\spt_\nu$, where $1 \leq \nu \leq m-1$, is the intersection of two maximal cliques, the induced subgraph $\G[\spt_\nu]$ is also star-complementary, which implies
	\begin{equation}
		\label{eq:detStarSep}
		\det(I_p + iT)[\spt_\nu] = \prod_{\st \in \mathcal F(\G[\spt_\nu])} \det(I_p + iT)[\st].
	\end{equation}
	
	Now, let $\st \in \mathcal F(\G[\cc_1]) \cup \cdots \cup \mathcal F(\G[\cc_m]) \cup \mathcal F(\G[\spt_1]) \cup \cdots \cup \mathcal F(\G[\spt_{m-1}])$. Then the complement of the induced subgraph $\G[\st]$ is a star. Consequently,
	\begin{equation}
		\label{eq:starQuadraticForm}
		\det(I_p + iT)[\st] = 1 + \sum\limits_{\substack{v_r, v_s \in \st \\ \{v_r, v_s\} \in \E(\G^*) \setminus \E(\G)}} t_{r,s}^2.
	\end{equation}
	
	Let $\tau = |\E(\G^*)| - |\E(\G)|$. By Theorem~\ref{thm:main}, (\ref{eq:detStar}), (\ref{eq:detStarSep}) and (\ref{eq:starQuadraticForm}), we have
	\begin{eqnarray}
		\mathcal I_{\G} (\beta, I_p) & = & \dfrac{1}{\pi^{\tau}} \bigintss_{{\mathbb S}(\G, \G^*)}  \dfrac{\prod\limits_{\mu=1}^m \det((I_p + iT)[\cc_\mu])^{-\beta - \frac{|\cc_\mu|+1}{2}} \Gamma_{|\cc_\mu|}\left(\beta + \frac{|\cc_\mu|+1}{2}\right)}{\prod\limits_{\nu=1}^{m-1} \det((I_p + iT)[\spt_\nu])^{-\beta - \frac{|\spt_\nu|+1}{2}}\Gamma_{|\spt_\nu|}\left(\beta + \frac{|\spt_\nu|+1}{2}\right)} \dd T \nonumber\\
		& = & \dfrac{\mathcal I_{\G^*}(\beta, I_p)}{\pi^{\tau}} \bigintss_{\mathbb R^{\tau}} \dfrac{\prod\limits_{\mu=1}^m \prod\limits_{\st \in \mathcal F(\cc_\mu)} (\det(I_p + iT)[\st])^{-\beta - \frac{|\cc_\mu|+1}{2}}  }{\prod\limits_{\nu=1}^{m-1} \prod\limits_{\st \in \mathcal F(\spt_\nu)} (\det(I_p + iT)[\st])^{-\beta - \frac{|\spt_\nu|+1}{2}}} \dd T \nonumber\\
		& = & \dfrac{\mathcal I_{\G^*}(\beta, I_p)}{\pi^{\tau}} \bigintss_{\mathbb R^{\tau}} \dfrac{\prod\limits_{\mu=1}^m \prod\limits_{\st \in \mathcal F(\cc_\mu)} (1 + \sum\limits_{\substack{v_r, v_s \in \st \\ \{v_r, v_s\} \in \E(\G^*) \setminus \E(\G)}} t_{r,s}^2)^{-\beta - \frac{|\cc_\mu|+1}{2}}  }{\prod\limits_{\nu=1}^{m-1} \prod\limits_{\st \in \mathcal F(\spt_\nu)} (1 + \sum\limits_{\substack{v_r, v_s \in \st \\ \{v_r, v_s\} \in \E(\G^*) \setminus \E(\G)}} t_{r,s}^2)^{-\beta - \frac{|\spt_\nu|+1}{2}}} \prod_{\substack{1 \leq r < s \leq p\\ \{v_r, v_s\} \in \E(\G^*) \setminus \E(\G)}} \dd t_{r, s}. \nonumber
	\end{eqnarray}
	
	\subsection{Proof of Corollary~\ref{coro:ell>2}} \label{sec:appell>2}
	
	Let $\G^* = \G^*(m; k_1, \ldots, k_\ell)$ be a chordal completion of $\G$ and it is clear that $\G$ has starry fill-ins. By Corollary~\ref{thm:star},
	\begin{eqnarray}
		& & \mathcal I_\G(\beta, I_{m+k_1+\cdots +k_\ell}) \nonumber \\
		& = & \dfrac{\mathcal I_{\G^*}(\beta, I_{m+k_1+\cdots +k_\ell})}{\pi^\ell}  \int_{\mathbb R^\ell} \dfrac{(1+t_1^2+\cdots + t_\ell^2)^{-\beta-\frac{m+1}{2}} (1+t_1^2)^{-\beta-\frac{k_1+3}{2}} \cdots (1+t_\ell^2)^{-\beta-\frac{k_\ell+3}{2}}}{(1+t_1^2)^{-\beta-\frac32} \cdots (1+t_\ell^2)^{-\beta-\frac32}} \dd t_1 \cdots \dd t_\ell \nonumber\\
		& = & \dfrac{\mathcal I_{\G^*}(\beta, I_{m+k_1+\cdots +k_\ell})}{\pi^\ell} 2^\ell  \int_0^\infty \cdots \int_0^\infty (1+t_1^2+\cdots+t_\ell^2)^{-\beta-\frac{m+1}{2}} (1+t_1^2)^{-\frac{k_1}{2}} \cdots (1+t_\ell^2)^{-\frac{k_\ell}{2}} \dd t_1 \cdots \dd t_\ell. \nonumber
	\end{eqnarray}
	As $\G^*$ is chordal, we have
	\begin{eqnarray}
		 \mathcal I_{\G^*}(\beta, I_{m+k_1+\cdots + k_\ell}) 
		& = & \pi^{k_1+\cdots + k_\ell} \Gamma_{k_1}\left(\beta+\frac{k_1+3}{2}\right) \cdots \Gamma_{k_\ell}\left(\beta+\frac{k_\ell+3}{2}\right) \Gamma_{m}\left(\beta+\frac{m+1}{2}\right). \nonumber
	\end{eqnarray}
	Finally, we simplify the $\ell$-dimensional integral. For positive real numbers $t_1, \ldots, t_\ell$, we substitute
	\begin{equation}
		y = \frac{(1+t_1^2 + \cdots + t_\ell^2)x}{2} \nonumber
	\end{equation}
	into the gamma function
	\begin{equation}
		\Gamma\left(\beta + \frac{m+1}{2}\right) = \int_0^\infty y^{\beta+\frac{m+1}{2}-1} e^{-y} \dd y \nonumber
	\end{equation}
	to obtain
	\begin{equation}
		(1+t_1^2+\cdots+t_\ell^2)^{-\beta - \frac{m+1}{2}} =  \int_0^\infty \dfrac{x^{\beta+\frac{m+1}{2} - 1} e^{-\frac{(1+t_1^2 +\cdots+t_\ell^2)x}{2}}}{\Gamma\left(\beta + \frac{m+1}{2} \right) 2^{\beta+\frac{m+1}{2}} } \dd x. \nonumber
	\end{equation}
	Hence,
	\begin{eqnarray}
		& & 2^\ell \int_0^\infty \cdots \int_0^\infty (1+t_1^2+\cdots+t_\ell^2)^{-\beta-\frac{m+1}{2}} (1+t_1^2)^{-\frac{k_1}{2}} \cdots (1+t_\ell^2)^{-\frac{k_\ell}{2}} \dd t_1 \cdots \dd t_\ell \nonumber\\
		& = & 2^\ell \int_0^\infty \cdots \int_0^\infty \dfrac{x^{\beta+\frac{m+1}{2} - 1} e^{-\frac{(1+t_1^2 +\cdots+t_\ell^2)x}{2}}}{\Gamma\left(\beta + \frac{m+1}{2} \right) 2^{\beta+\frac{m+1}{2}} }  (1+t_1^2)^{-\frac{k_1}{2}} \cdots (1+t_\ell^2)^{-\frac{k_\ell}{2}} \dd x \dd t_1 \cdots \dd t_\ell \nonumber\\
		& = & \int_0^\infty \dfrac{x^{\beta+\frac{m+1}{2} - 1} e^{-\frac{x}{2}}}{\Gamma\left(\beta + \frac{m+1}{2} \right) 2^{\beta+\frac{m+1}{2}} } \prod_{\mu=1}^\ell \left(2\int_0^\infty e^{-\frac{t_\mu^2x}{2}} (1+t_\mu^2)^{-\frac{k_\mu}{2}} \dd t_\mu\right) \dd x. \nonumber
	\end{eqnarray}
	For $\mu = 1, \ldots, \ell$, a change of variable $t_\mu \to \sqrt{t_\mu}$ gives
	\begin{equation}
		2\int_0^\infty e^{-\frac{t_\mu^2x}{2}} (1+t_\mu^2)^{-\frac{k_\mu}{2}} \dd t_\mu = \int_0^\infty e^{-\frac{t_\mu x}{2}} (1+t_\mu)^{-\frac{k_\mu}{2}} t_\mu^{-\frac12} \dd t_\mu = \pi^{\frac12} U\left(\frac12, \frac32 - \frac{k_\mu}{2}, \frac{x}{2}\right), \nonumber
	\end{equation}
	which implies
	\begin{eqnarray}
		\mathcal I_\G(\beta, I_{m+k_1+\cdots +k_\ell}) & = & \pi^{k_1+\cdots + k_\ell - \frac{\ell}{2}}\Gamma_{m}\left(\beta+\frac{m+1}{2}\right) \prod\limits_{\mu=1}^\ell \Gamma_{k_\mu}\left(\beta+\frac{k_\mu+3}{2}\right) \nonumber \\
		& & \times \int_0^\infty \dfrac{x^{\beta+\frac{m+1}{2} - 1} e^{-\frac{x}{2}}}{\Gamma\left(\beta + \frac{m+1}{2} \right) 2^{\beta+\frac{m+1}{2}} } \prod_{\mu=1}^\ell U\left(\frac12, \frac32 - \frac{k_\mu}{2}, \frac{x}{2}\right) \dd x.\nonumber
	\end{eqnarray}
	
	For computational purposes, we transform the domain of the integral into a bounded one using the fact that
	\begin{equation}
		\dfrac{x^{\beta+\frac{m+1}{2} - 1} e^{-\frac{x}{2}}}{\Gamma\left(\beta + \frac{m+1}{2} \right) 2^{\beta+\frac{m+1}{2}} } = f_{\chi^2(2\beta+m+1)}(x) = F'_{\chi^2(2\beta+m+1)}(x) \nonumber
	\end{equation}
	is the probability density function of chi-squared distribution with $(2\beta+m+1)$ degrees of freedom. It follows that
	\begin{eqnarray}
		\mathcal I_\G(\beta, I_{m+k_1+\cdots +k_\ell})
		& = & \pi^{k_1+\cdots + k_\ell - \frac{\ell}{2}}\Gamma_{m}\left(\beta+\frac{m+1}{2}\right) \prod\limits_{\mu=1}^\ell \Gamma_{k_\mu}\left(\beta+\frac{k_\mu+3}{2}\right) \nonumber \\
		& & \times \int_0^1 \prod_{\mu=1}^\ell U\left(\frac12, \frac32 - \frac{k_\mu}{2}, \frac{F^{-1}_{\chi^2(2\beta+m+1)}(x)}{2}\right) \dd x. \nonumber
	\end{eqnarray}
	
	\subsection{Proof of Corollary~\ref{coro:mfi2}} \label{sec:appmfi2}
	
	By assumption, $\{v_1, v_2\} \in \E(\G)$. Let $\G_0$ be the induced subgraph \begin{equation*}
		\G[(\NB_\G(v_1) \cap \NB_\G(v_2)) \cup \{v_1, v_2, v_3\}],
	\end{equation*}
	obtained from the common neighbours of $v_1$ and $v_2$, together with vertices $v_1, v_2$ and $v_3$. Let $\G_0^*$ be the corresponding chordal completion by adding the edges $e_1$ and $e_2$.  Let $\cc^{(0)}_1, \ldots, \cc^{(0)}_m$ be a perfect sequence of cliques of $\G_0^*$, and let $\spt^{(0)}_1, \ldots, \spt^{(0)}_{m-1}$ be the minimal separators.
	
	For $\mu = 1, 2$, let $\G_\mu$ be the induced subgraph \begin{equation*}
		\G[(\NB_\G(v_\mu) \cap \NB_\G(v_3)) \cup (\NB_\G(v_1) \cap \NB_\G(v_2)) \cup \{v_1, v_2, v_3\}],
	\end{equation*}
	obtained from the common neighbours of $v_\mu$ and $v_3$, and the common neighbours of $v_1$ and $v_2$, together with vertices $v_1, v_2$ and $v_3$. Let $\G_\mu^*$ be the corresponding chordal completion by adding the edges $e_1$ and $e_2$. Let $\cc^{(0)}_1, \ldots, \cc^{(0)}_m, \cc^{(\mu)}_1, \ldots, \cc^{(\mu)}_{k_\mu}$ be a perfect sequence of cliques of $\G_\mu^*$, and let $\spt^{(0)}_1, \ldots, \spt^{(0)}_{m-1}, \spt^{(\mu)}_1, \ldots, \spt^{(\mu)}_{k_\mu}$ be the minimal separators. (For any perfect sequence of cliques of $\G^*_\mu$, the running intersection property is not violated if $\cc^{(0)}_1, \ldots, \cc^{(0)}_m$ are moved to the front.)
	
	Note that the assumption that $\G$ is prime implies that every maximal clique of $\G^*$ contains at least one of the two missing edges. Therefore, $\cc^{(0)}_1, \ldots, \cc^{(0)}_m, \cc^{(1)}_1, \ldots, \cc^{(1)}_{k_1}, \cc^{(2)}_1, \ldots, \cc^{(2)}_{k_2}$ is a perfect sequence of cliques of $\G^*$, and the minimal separators are $\spt^{(0)}_1, \ldots, \spt^{(0)}_{m-1}$, $\spt^{(1)}_1, \ldots, \spt^{(1)}_{k_1}$, $\spt^{(2)}_1, \ldots, \spt^{(2)}_{k_2}$.
	Then, by Corollary~\ref{thm:star},
	\begin{eqnarray}
		\mathcal I_{\G} (\beta, I_p) &= &\dfrac{\mathcal I_{\G^*}(\beta, I_p)}{\pi^2} \bigintss_{\mathbb R^{2}} \dfrac{\prod\limits_{\nu=1}^m(1+t_1^2+t_2^2)^{-\beta-\frac{|\cc^{(0)}_\nu|+1}{2}} \prod\limits_{\nu=1}^{k_1}(1+t_1^2)^{-\beta-\frac{|\cc^{(1)}_\nu|+1}{2}} \prod\limits_{\nu=1}^{k_2}(1+t_2^2)^{-\beta-\frac{|\cc^{(2)}_\nu|+1}{2}}}{\prod\limits_{\nu=1}^{m-1}(1+t_1^2+t_2^2)^{-\beta-\frac{|\spt^{(0)}_\nu|+1}{2}} \prod\limits_{\nu=1}^{k_1}(1+t_1^2)^{-\beta-\frac{|\spt^{(1)}_\nu|+1}{2}} \prod\limits_{\nu=1}^{k_2}(1+t_2^2)^{-\beta-\frac{|\spt^{(2)}_\nu|+1}{2}}} \dd t_1 \dd t_2 \nonumber \\
		& = & \dfrac{\mathcal I_{\G^*}(\beta, I_p)}{\pi^2} \int_{\mathbb R^{2}} (1+t_1^2+t_2^2)^{-\gamma} (1+t_1^2)^{-r} (1+t_2^2)^{-s} \dd t_1 \dd t_2, \nonumber
	\end{eqnarray}
	where $r = \frac12 \sum\limits_{\nu=1}^{k_1} (|\cc^{(1)}_\nu| - |\spt^{(1)}_\nu|)$, $s = \frac12 \sum\limits_{\nu=1}^{k_2} (|\cc^{(2)}_\nu| - |\spt^{(2)}_\nu|)$, and
	\begin{equation}
		\gamma = \beta + \frac12 (|\cc_1| + \cdots + |\cc_m| - |\spt_1| - \cdots - |\spt_{m-1}| + 1). \nonumber
	\end{equation}
	
	We use Lemma~\ref{lem:inclusionExclusion} to write $r, s, \gamma$ in terms of $w, w_1, w_2$. With the graph $\G^*_0$, the lemma implies
	\begin{equation}
		|\cc^{(0)}_1| + \cdots + |\cc^{(0)}_m| - |\spt^{(0)}_1| - \cdots - |\spt^{(0)}_{m-1}| = w+3, \nonumber
	\end{equation}
	and so $\gamma = \beta + \frac{w+4}2$. For $\mu = 1, 2$, we again apply Lemma~\ref{lem:inclusionExclusion} (with the graph $\G^*_\mu$) to obtain
	\begin{equation}
		\sum_{\nu=1}^m |\cc^{(0)}_\nu| + \sum_{\nu=1}^{k_\mu} |\cc^{(\mu)}_\nu| - \sum_{\nu=1}^{m-1} |\spt^{(0)}_\nu| - \sum_{\nu=1}^{k_\mu} |\spt^{(\mu)}_\nu|= w_\mu+3, \nonumber
	\end{equation}
	which implies
	\begin{equation}
		\sum_{\nu=1}^{k_\mu} |\cc^{(\mu)}_\nu| - \sum_{\nu=1}^{k_\mu} |\spt^{(\mu)}_\nu|= w_\mu - w. \nonumber
	\end{equation}
	Thus, $r = \frac{w_1-w}2$ and $s = \frac{w_2-w}{2}$.
	
	It remains to solve the integral
	\begin{equation}
		\int_{\mathbb R^{2}} (1+t_1^2+t_2^2)^{-\gamma} (1+t_1^2)^{-r} (1+t_2^2)^{-s} \dd t_1 \dd t_2. \nonumber
	\end{equation}
	We use Euler's integral representation of the hypergeometric function ${}_2F_1$ to obtain
	\begin{eqnarray}
		& & \int_{\mathbb R^{2}} (1+t_1^2+t_2^2)^{-\gamma} (1+t_1^2)^{-r} (1+t_2^2)^{-s} \dd t_1 \dd t_2 \nonumber\\
		& = & \dfrac{\Gamma\left(\gamma+r-\frac12\right)\Gamma\left(\frac12\right)}{\Gamma\left(\gamma+r\right)} \int_0^1 t_2^{\gamma+s-\frac32} (1-t_2)^{-\frac12} {}_2F_1\left(\gamma, \frac12; \gamma+r; 1-t_2\right)\dd t_2.\nonumber
	\end{eqnarray}
	Finally, we substitute $1-t_2 \to t_2$ and the integral representation of ${}_3F_2$ implies
	\begin{eqnarray}
		& & \int_{\mathbb R^{2}} (1+t_1^2+t_2^2)^{-\gamma} (1+t_1^2)^{-r} (1+t_2^2)^{-s} \dd t_1 \dd t_2 \nonumber\\
		& = & \dfrac{\Gamma\left(\gamma+r-\frac12\right)\Gamma\left(\frac12\right)}{\Gamma\left(\gamma+r\right)} \int_0^1 t_2^{-\frac12} (1-t_2)^{\gamma+s-\frac32} {}_2F_1\left(\gamma, \frac12; \gamma+r; t_2\right)\dd t_2 \nonumber\\
		& = & \pi \dfrac{\Gamma\left(\gamma+r-\frac12\right)}{\Gamma\left(\gamma+r\right)} \dfrac{\Gamma\left(\gamma+s-\frac12\right)}{\Gamma\left(\gamma+s\right)} {}_3F_2\left(\gamma, \frac12, \frac12; \gamma+r, \gamma+s; 1 \right). \nonumber
	\end{eqnarray}
	We complete the proof by plugging in $\gamma = \beta+\frac{w+4}{2}$, $r = \frac{w_1-w}{2}$ and $s = \frac{w_2-w}{2}$.

\newpage

\section*{Supplementary Material for\\[1ex] \emph{A new way to evaluate $\G$-Wishart normalising constants via Fourier analysis} \\[1ex]
\normalfont{by Ching Wong, Giusi Moffa and Jack Kuipers}}

\renewcommand\thesection{S\arabic{section}}
\renewcommand\thefigure{S\arabic{figure}} 
\renewcommand\thetable{S\arabic{table}}
\renewcommand\thepage{S\arabic{page}}
\renewcommand{\theequation}{S\arabic{equation}}
\setcounter{equation}{0}
\setcounter{page}{1}
\setcounter{section}{0}
    
	\section{Prime graphs with few vertices} \label{sec:smallPrime}
	
	We give a summary on the integrals $\mathcal I_\G(\beta, I_p)$, where $\beta > -1$ is a real number and $p \leq 6$ is the number of vertices in connected prime graphs $\G$. Recall from Section~\ref{sec:contribution} that we obtain an explicit formula for the integral $\mathcal I_\G(\beta, I_p)$ for some graphs:
	\begin{itemize}
		\item [(B1)] $\G$ has a chordal completion $\G^*$ in which every triangle contains at most one edge from $\E(\G^*) \setminus \E(\G)$,
		\item [(B2)] $\G$ has minimum fill-in 2,
		\item [(B3)] $\G$ is complete $k$-partite.
	\end{itemize}
	
	All connected prime graphs on at most 5 vertices have minimum fill-in at most 2. Among the 24 connected prime graphs on 6 vertices, there are only 2 graphs which do no belong to the above classes, namely the cycle of length 6 and its complement. These two graphs both have minimum fill-in 3, and we can write the corresponding integral as a one-dimensional integral. We use solid edges to represent the graphs. Dashed edges are the missing edges.
	
	\subsection{Two vertices}
	
	There is one graph and it is in (B1):
	\begin{center}
		\GtwoA
	\end{center}
	
	\subsection{Three vertices}
	
	There is one graph and it is in (B1):
	\begin{center}
		\GthreeA
	\end{center}
	
	\subsection{Four vertices}
	
	There are two graphs, both in (B1):
	\begin{center}
		\GfourA \s \GfourB
	\end{center}
	
	\subsection{Five vertices}
	
	We have four graphs in (B1):
	\begin{center}
		\GfiveA \s \GfiveB \s \GfiveC \s \GfiveD
	\end{center}
	
	We have one graph in (B2):
	\begin{center}
		\GfiveE
	\end{center}
	
	\subsection{Six vertices}
	We have 10 graphs with minimum fill-in at most 1, all belong to (B1):
	\begin{center}
		\GsixA \s \GsixB \s \GsixC \s \GsixD \s \GsixE
		
		\vspace{0.5cm}
		
		  \GsixF \s \GsixG \s \GsixH \s \GsixI \s \GsixJ
	\end{center}
	
	We have 4 graphs with minimum fill-in 2 and they also belong to (B1):
	\begin{center}
		\GsixK \s \GsixL \s \GsixM \s \GsixN
	\end{center}
	
	The following 7 graphs have minimum fill-in 2 and belong to (B2):
	\begin{center}
		\GsixP \s \GsixQ \s \GsixR \s \GsixS
		
		\vspace{0.5cm}
		
		\GsixT \s \GsixU \s \GsixV
	\end{center}
	
	The graph $\com_{3,3}$ has minimum fill-in 3, and it is in (B3):
	\begin{center}
		\GsixO
	\end{center}
	
	Finally, there are two other graphs with minimum fill-in 3:
	\begin{center}
		\GsixW
	\end{center}
	For the cycle of length 6, we have (by Example~\ref{ex:C6})
	\begin{equation}
		\mathcal I_\G(\beta, I_6) = \dfrac{ \pi \Gamma_3(\beta+2)  \Gamma(\beta+2)^5}{\Gamma\left(\beta+\frac52\right)^2} \int_0^1 t^{-\frac12}(1-t)^{\beta+1} {}_2F_1\left(\beta+2, \frac12; \beta+\frac52; t \right)^2 \dd t .\nonumber
	\end{equation}
	
	\begin{center}
		\GsixX
	\end{center}
	For the complement of the cycle of length 6, we have (by Example~\ref{ex:C6complement})
	
	\begin{equation}
		\mathcal I_\G(\beta, I_6)
		= \dfrac{\pi \Gamma_4\left(\beta+\frac52\right)^5 \Gamma\left(\beta+\frac52\right)^4}{\Gamma(\beta+3)^2} \int_0^1 t^{\beta+2} (1-t)^{-\frac12} {}_2F_1\left(\frac12, \frac12; \beta+3; t \right)^2 \dd t. \nonumber
	\end{equation}

\section{Relating the normalising constants of different graphs, $D=I$} \label{sec:map}

We assume that $D=I$ in this Section and use Theorem~\ref{thm:main} to express the $\G$-Wishart normalising constant of a graph in terms of that of another graph (sometimes with a different $\beta$). This is useful when one of the constants has a simpler form than the other, and particularly useful for Bayesian samplers in the space of undirected graphs, since the acceptance probability of a move between graphs will just depend on their relative posterior probability.

In this Section, we give 2 examples. First, we show that the normalising constant of any graph with a chordal completion such that the missing edges form a triangle can be written as a one-dimensional integral, instead of three-dimensional. Second, we show that the normalising constant of any gear graph has the form of Equation (\ref{eq:specialForm}), instead of an integral with complex integrand.

\subsection{Example with triangular missing edges} \label{sec:missingTriangle}

Let $\G$ and $\G^*$ be the graphs as shown in Figure~\ref{fig:gear}. Let $\beta > -1$ be a real number. By Theorem~\ref{thm:main}, we have
\begin{eqnarray}
	& & \mathcal I_{\G}(\beta, I_{7}) 
	=  \Gamma_4\left(\beta+\frac 5 2\right) \Gamma(\beta+2)^3\nonumber\\
	& & \quad  \times \int_{\mathbb R^3} (1+t_1^2+t_2^2+t_3^2-2it_1t_2t_3)^{-\beta-\frac{5}{2}} (1+t_1^2)^{-\frac12} (1+t_2^2)^{-\frac12} (1+t_3^2)^{-\frac12} \dd t_1 \dd t_2 \dd t_3. \label{eq:H}
\end{eqnarray}
Let $\HH$ be the cycle of length 6. Note that there are three different (up to isomorphism) chordal completions of $\HH$ using 3 added edges; see Figure~\ref{fig:C6star}.
\begin{figure} 
\centering
	\CsixB\s 
	\CsixA \s
	\CsixC	
	\caption{Solid edges represent the graph $\HH$, i.e., the cycle of length 6. Together with the dashed edges, they are chordal completions of $\HH$. Left: $\HH^*_1$, middle: $\HH^*_2$, right: $\HH^*_3$.}
	\label{fig:C6star}
\end{figure}
When we apply Theorem~\ref{thm:main} with $\HH^*_2$ or $\HH^*_3$, we get the same formula; see Example~\ref{ex:C6}. However, if we apply Theorem~\ref{thm:main} with $\HH^*_1$, we get an integral very similar to the one appearing in Equation (\ref{eq:H}). Indeed, we have
\begin{equation}
	\mathcal I_\HH(\beta, I_6) =  \Gamma_3(\beta+2) \Gamma(\beta+2)^3 \int_{\mathbb R^3} (1+t_1^2+t_2^2+t_3^2-2it_1t_2t_3)^{-\beta-2} (1+t_1^2)^{-\frac12} (1+t_2^2)^{-\frac12} (1+t_3^2)^{-\frac12} \dd t_1 \dd t_2 \dd t_3. \label{eq:C62}
\end{equation}
Hence, Equations (\ref{eq:H}) and (\ref{eq:C62}) together imply that
\begin{equation}
	\mathcal I_\G(\beta, I_7) 
	 = \Gamma_4\left(\beta+\frac 5 2\right) \Gamma(\beta+2)^3  \dfrac{\mathcal I_{\HH}\left(\beta + \frac12, I_6\right)}{ \Gamma_3\left(\beta+\frac52\right) \Gamma\left(\beta+\frac52\right)^3} \dfrac{\pi^{\frac32}\Gamma(\beta+2)^3\Gamma(\beta+1)}{\Gamma\left(\beta+\frac52\right)^3} \mathcal I_{\HH}\left(\beta+\frac12, I_6\right). \nonumber
\end{equation}
Now, if we use the result obtained from Example~\ref{ex:C6} in Equation (\ref{eq:C6}), we have
\begin{equation}
	\mathcal I_\G(\beta, I_7) =  \dfrac{\Gamma_4\left(\beta+\frac 5 2\right)\Gamma\left(\beta+\frac 5 2\right)^2 \Gamma(\beta+2)^3 }{\pi\Gamma(\beta+3)^2 } \int_0^1 t^{-\frac12}(1-t)^{\beta+\frac32} {}_2F_1\left(\beta+\frac52, \frac12; \beta+3; t \right)^2 \dd t. \label{eq:Gear3}
\end{equation}

The above argument works for any graph with a chordal completion such that the missing edges form a triangle, for which we now provide details.

\subsection{Missing triangular edges}	\label{sec:apptriangle}
	
Let $\G$ be a prime graph with $p$ vertices and let $\G^*$ be a chordal completion. Suppose that the missing edges form a triangle, say $e_1 = \{v_1, v_2\}$, $e_2 = \{v_2, v_3\}$ and $e_3 = \{v_3, v_1\}$. In this section, we aim to write the integral $\mathcal I_\G(\beta, I_p)$ as a one-dimensional integral, for real number $\beta > -1$.
	
Since $\G$ is prime, every maximal clique of $\G^*$ contains either one missing edge or all three of them. Let $\cc_1, \ldots, \cc_k$ be the maximal cliques that contain all three missing edges. For $1 \leq \mu \leq 3$, let $\cc_{1}^{(\mu)}, \ldots, \cc_{k_\mu}^{(\mu)}$ be the maximal cliques that contains the edge $e_\mu$ but not the other two missing edges. Note that $k \geq 1$ .
	
It is easy to see that among the minimal separators of $\G^*$, exactly $k-1$ of them (say $\spt_1, \ldots, \spt_{k-1}$) contain all three missing edges, and $k_\mu$ of them (say $\spt^{(\mu)}_1, \ldots, \spt^{(\mu)}_{k_\mu}$) contain the edge $e_\mu$ but not the other two missing edges, for $\mu = 1, 2, 3$.
By Theorem~\ref{thm:main}, we have
\begin{eqnarray}
	\mathcal I_{\G} (\beta, I_p) 
	& = & \dfrac{\mathcal I_{\G^*}(\beta, I_p)}{\pi^{3}} \bigintss_{\mathbb S(\G, \G^*)}  \dfrac{\prod\limits_{\nu=1}^k \det((I_p+iT)[\cc_\nu])^{-\beta - \frac{|\cc_\nu|+1}{2}} \prod\limits_{\mu=1}^3 \prod\limits_{\nu=1}^{k_\mu} \det((I_p+iT)[\cc_\nu^{(\mu)}])^{-\beta - \frac{|\cc^{(\mu)}_\nu|+1}{2}}}{\prod\limits_{\nu=1}^{k-1} \det((I_p+iT)[\spt_\nu])^{-\beta - \frac{|\spt_{\nu}|+1}{2}} \prod\limits_{\mu=1}^3 \prod\limits_{\nu=1}^{k_\mu} \det((I_p+iT)[\spt^{(\mu)}_\nu])^{-\beta - \frac{|\spt^{(\mu)}_{\nu}|+1}{2}}} \dd T \nonumber\\
	& = & \dfrac{\mathcal I_{\G^*}(\beta, I_p)}{\pi^{3}} \bigintss_{\mathbb R^3}  \dfrac{\prod\limits_{\nu=1}^k (1+t_1^2+t_2^2+t_3^2-2it_1t_2t_3)^{-\beta - \frac{|\cc_\nu|+1}{2}} \prod\limits_{\mu=1}^3 \prod\limits_{\nu=1}^{k_\mu} (1+t_\mu^2)^{-\beta - \frac{|\cc^{(\mu)}_\nu|+1}{2}}}{\prod\limits_{\nu=1}^{k-1} (1+t_1^2+t_2^2+t_3^2-2it_1t_2t_3)^{-\beta - \frac{|\spt_{\nu}|+1}{2}} \prod\limits_{\mu=1}^3 \prod\limits_{\nu=1}^{k_\mu} (1+t_\mu^2)^{-\beta - \frac{|\spt^{(\mu)}_{\nu}|+1}{2}}} \dd t_1 \dd t_2 \dd t_3 \nonumber\\
	& = & \dfrac{\mathcal I_{\G^*}(\beta, I_p)}{\pi^{3}} \int_{\mathbb R^3}  (1+t_1^2+t_2^2+t_3^2-2it_1t_2t_3)^{-\beta - \frac{\gamma}{2}} (1+t_1^2)^{-\frac{\gamma_1}{2}} (1+t_2^2)^{-\frac{\gamma_2}{2}} (1+t_3^2)^{-\frac{\gamma_3}{2}}\dd t_1 \dd t_2 \dd t_3, \nonumber
\end{eqnarray}
where $\gamma = \sum\limits_{\nu=1}^k (|\cc_\nu| + 1) - \sum\limits_{\nu=1}^{k-1} (|\spt_\nu| + 1)$ and $\gamma_\mu = \sum\limits_{\nu=1}^{k_\mu} (|\cc_\nu^{(\mu)}| + 1) - \sum\limits_{\nu=1}^{k_\mu} (|\spt^{(\mu)}_\nu| + 1) > 0$, for $\mu = 1, 2, 3$. We note that $\gamma \geq 4$ because $|\cc_1| \geq 3$ and $|\spt_\nu| \leq |\cc_{\nu+1}|$ for all $1 \leq \nu \leq k-1$.
	
Now, we define a graph $\HH_1^* = \HH_1^*(\gamma_1, \gamma_2, \gamma_3)$ with vertex set
\begin{equation}
	\V(\HH_1^*) = \{v_1, v_2, v_3\} \sqcup \{u_\mu : 1 \leq \mu \leq k_1\} \sqcup \{w_\mu : 1 \leq \mu \leq k_2\} \sqcup \{x_\mu : 1 \leq \mu \leq k_3\} \nonumber
\end{equation}
and the edge set is defined by connecting pairs of vertices inside one of the following three sets: $\{u_\mu : 1 \leq \mu \leq \gamma_1\} \sqcup \{v_1, v_2\}$, $\{w_\mu : 1 \leq \mu \leq \gamma_2\} \sqcup \{v_2, v_3\}$, $\{x_\mu : 1 \leq \mu \leq \gamma_3\} \sqcup \{v_3, v_1\}$. Furthermore, we define $\HH = \HH(\gamma_1, \gamma_2, \gamma_3)$ to be the graph obtained from $\HH_1^*$ by removing the three edges $\{v_1, v_2\}, \{v_2, v_3\}, \{v_3, v_1\}$. We define one last graph $\HH^*_2 = \HH_2^*(\gamma_1, \gamma_2, \gamma_3)$ on the same vertex set, by adding the edges $\{v_1, v_2\}, \{v_1, v_3\}$ and $\{v_1, w_1\},\ldots, \{v_1, w_{\gamma_2}\}$ to the graph $\HH$. It is clear that both $\HH_1^*$ and $\HH_2^*$ are chordal completions of $\HH$. Figure~\ref{fig:H} gives an example of these graphs.
	
\begin{figure}
  \centering
	\HA \quad \HB
	\caption{For both graphs, the solid edges represent those in $\HH(1,2,3)$. Together with the dashed edges, the graph on the left is $\HH_1^*(1,2,3)$, the graph on the right is $\HH_2^*(1,2,3)$.} \label{fig:H}
\end{figure}
	
As above, we find the integral $\mathcal I_\HH(\tilde \beta, I_{m})$, where $m := 3+\gamma_1+\gamma_2+\gamma_3$, using these two chordal completions, for real number $\tilde\beta > -1$. First, we use Theorem~\ref{thm:main} with $\HH_1^*$:
	\begin{eqnarray}
		\mathcal I_{\HH} (\tilde\beta, I_m) 		& = & \dfrac{\mathcal I_{\HH_1^*} (\tilde\beta, I_m)}{\pi^3} \bigintss_{\mathbb R^3}  \dfrac{(1+t_1^2+t_2^2+t_3^2-2it_1t_2t_3)^{-\beta -2} \prod\limits_{\mu=1}^3(1+t_\mu^2)^{-\tilde\beta-\frac{\gamma_\mu+3}{2}}}{\prod\limits_{\mu=1}^3(1+t_\mu^2)^{-\tilde\beta-\frac{3}{2}}}\dd t_1 \dd t_2 \dd t_3 \nonumber\\
		& = & \dfrac{\mathcal I_{\HH_1^*} (\tilde\beta, I_m)}{\pi^3} \int_{\mathbb R^3}  (1+t_1^2+t_2^2+t_3^2-2it_1t_2t_3)^{-\tilde\beta -2} (1+t_1^2)^{-\frac{\gamma_1}{2}} (1+t_2^2)^{-\frac{\gamma_2}{2}} (1+t_3^2)^{-\frac{\gamma_3}{2}} \dd t_1 \dd t_2 \dd t_3.\nonumber
	\end{eqnarray}
	Comparing the above two integrals, we take $\tilde\beta = \beta+\frac{\gamma}{2}-2 > -1$ and obtain
	\begin{equation}
		\mathcal I_{\G} (\beta, I_p) = \dfrac{\mathcal I_{\G^*}(\beta, I_p) \mathcal I_{\HH} (\beta+\frac{\gamma}{2}-2, I_m)}{\mathcal I_{\HH_1^*} (\beta+\frac{\gamma}{2}-2, I_m)}. 	\label{eq:HandH1}
	\end{equation}
	
	Next, we find the integral $\mathcal I_\HH(\beta, I_{m})$ using Theorem~\ref{thm:main} with $\HH_2^*$:
	\begin{eqnarray}
		\mathcal I_{\HH} (\tilde\beta, I_m) 
		& = & \dfrac{\mathcal I_{\HH_2^*} (\tilde\beta, I_m)}{\pi^{\gamma_2+2}}  \bigintss_{\mathbb R^{\gamma_2+2}}  \dfrac{\left(1+\sum\limits_{\mu=1}^{\gamma_2+1}t_\mu^2\right)^{-\tilde\beta -\frac{\gamma_2+3}{2}} \left(1+\sum\limits_{\mu=2}^{\gamma_2+2}t_\mu^2\right)^{-\tilde\beta -\frac{\gamma_2+3}{2}} (1+t_{\gamma_2+2}^2)^{-\tilde\beta-\frac{\gamma_3+3}{2}}}{(1+t_1^2)^{-\tilde\beta-\frac{3}{2}} (1+t_{\gamma_2+2}^2)^{-\tilde\beta-\frac{3}{2}}\left(1+\sum\limits_{\mu=2}^{\gamma_2+1}t_\mu^2\right)^{-\tilde\beta -\frac{\gamma_2+2}{2}}}\dd \boldsymbol{t} \nonumber\\
		& = & \dfrac{\mathcal I_{\HH_2^*} (\tilde\beta, I_m)}{\pi^{\gamma_2+2}} \bigintsss_{\mathbb R^{\gamma_2}}  \left(1+\sum\limits_{\mu=2}^{\gamma_2+1}t_\mu^2\right)^{\tilde\beta +\frac{\gamma_2+2}{2}} \left(\bigintsss_{\mathbb R} \left(1+\sum\limits_{\mu=1}^{\gamma_2+1}t_\mu^2\right)^{-\tilde\beta -\frac{\gamma_2+3}{2}} (1+t_1^2)^{-\frac{\gamma_1}{2}} \dd t_1\right) \nonumber\\
		& &\s \s \times \left(\bigintsss_{\mathbb R} \left(1+\sum\limits_{\mu=2}^{\gamma_2+2}t_\mu^2\right)^{-\tilde\beta -\frac{\gamma_2+3}{2}} (1+t_{\gamma_2+2}^2)^{-\frac{\gamma_3}{2}} \dd t_{\gamma_2+2}\right)\dd t_2 \cdots \dd t_{\gamma_2+1}. \nonumber
	\end{eqnarray}
	By Euler's integral representation of the hypergeometric function ${}_2F_1$, we have
	\begin{eqnarray}
		& & \bigintsss_{\mathbb R} \left(1+\sum\limits_{\mu=1}^{\gamma_2+1}t_\mu^2\right)^{-\tilde\beta -\frac{\gamma_2+3}{2}} (1+t_1^2)^{-\frac{\gamma_1}{2}} \dd t_1 \nonumber\\
		& = & \dfrac{\pi^{\frac12}\Gamma\left(\tilde\beta +\frac{\gamma_1+\gamma_2+2}{2}\right)}{\Gamma\left(\tilde\beta +\frac{\gamma_1+\gamma_2+3}{2}\right)} {}_2F_1\left(\tilde\beta +\frac{\gamma_2+3}{2}, \tilde\beta +\frac{\gamma_1+\gamma_2+2}{2}; \tilde\beta +\frac{\gamma_1+\gamma_2+3}{2}; -\sum\limits_{\mu=2}^{\gamma_2+1}t_\mu^2\right) \nonumber
	\end{eqnarray}
	and
	\begin{eqnarray}
		& & \int_{\mathbb R} \left(1+\sum\limits_{\mu=2}^{\gamma_2+2}t_\mu^2\right)^{-\tilde\beta -\frac{\gamma_2+3}{2}} (1+t_{\gamma_2+2}^2)^{-\frac{\gamma_3}{2}} \dd t_{\gamma_2+2} \nonumber \\
		& = & \dfrac{\pi^{\frac12}\Gamma\left(\tilde\beta +\frac{\gamma_3+\gamma_2+2}{2}\right)}{\Gamma\left(\tilde\beta +\frac{\gamma_3+\gamma_2+3}{2}\right)} {}_2F_1\left(\tilde\beta +\frac{\gamma_2+3}{2}, \tilde\beta +\frac{\gamma_3+\gamma_2+2}{2}; \tilde\beta +\frac{\gamma_3+\gamma_2+3}{2}; -\sum\limits_{\mu=2}^{\gamma_2+1}t_\mu^2\right). \nonumber
	\end{eqnarray}
	It follows that
	\begin{eqnarray}
		\mathcal I_{\HH} (\tilde\beta, I_m) 
		& = & \dfrac{\mathcal I_{\HH_2^*} (\tilde\beta, I_m)}{\pi^{\gamma_2+1}} \dfrac{\Gamma\left(\tilde\beta +\frac{\gamma_1+\gamma_2+2}{2}\right)}{\Gamma\left(\tilde\beta +\frac{\gamma_1+\gamma_2+3}{2}\right)} \dfrac{\Gamma\left(\tilde\beta +\frac{\gamma_3+\gamma_2+2}{2}\right)}{\Gamma\left(\tilde\beta +\frac{\gamma_3+\gamma_2+3}{2}\right)} \nonumber\\
		& &  \times \bigintsss_{\mathbb R^{\gamma_2}}  \left(1+\sum\limits_{\mu=2}^{\gamma_2+1}t_\mu^2\right)^{\tilde\beta +\frac{\gamma_2+2}{2}} {}_2F_1\left(\tilde\beta +\frac{\gamma_2+3}{2}, \tilde\beta +\frac{\gamma_1+\gamma_2+2}{2}; \tilde\beta +\frac{\gamma_1+\gamma_2+3}{2}; -\sum\limits_{\mu=2}^{\gamma_2+1}t_\mu^2\right) \nonumber\\
		& & \s \s {}_2F_1\left(\tilde\beta +\frac{\gamma_2+3}{2}, \tilde\beta +\frac{\gamma_3+\gamma_2+2}{2}; \tilde\beta +\frac{\gamma_3+\gamma_2+3}{2}; -\sum\limits_{\mu=2}^{\gamma_2+1}t_\mu^2\right) \dd t_2 \cdots \dd t_{\gamma_2+1} \nonumber\\
		& = & \dfrac{\mathcal I_{\HH_2^*} (\tilde\beta, I_m)}{\pi^{\gamma_2+1}} \dfrac{\Gamma\left(\tilde\beta +\frac{\gamma_1+\gamma_2+2}{2}\right)}{\Gamma\left(\tilde\beta +\frac{\gamma_1+\gamma_2+3}{2}\right)} \dfrac{\Gamma\left(\tilde\beta +\frac{\gamma_3+\gamma_2+2}{2}\right)}{\Gamma\left(\tilde\beta +\frac{\gamma_3+\gamma_2+3}{2}\right)} \dfrac{2\pi^{\frac{\gamma_2}{2}}}{\Gamma\left(\frac{\gamma_2}{2}\right)} \nonumber\\
		& & \times \int_0^\infty r^{\gamma_2-1} (1+r^2)^{\tilde\beta+\frac{\gamma_2+2}{2}} {}_2F_1\left(\tilde\beta +\frac{\gamma_2+3}{2}, \tilde\beta +\frac{\gamma_1+\gamma_2+2}{2}; \tilde\beta +\frac{\gamma_1+\gamma_2+3}{2}; -r^2\right) \nonumber\\
		& & \s \s {}_2F_1\left(\tilde\beta +\frac{\gamma_2+3}{2}, \tilde\beta +\frac{\gamma_3+\gamma_2+2}{2}; \tilde\beta +\frac{\gamma_3+\gamma_2+3}{2}; -r^2\right) \dd r. \nonumber
	\end{eqnarray}
	
	By (\ref{eq:HandH1}), we have
	\begin{eqnarray}
		\mathcal I_{\G} (\beta, I_p) & = &\dfrac{2\mathcal I_{\G^*}(\beta, I_p) \mathcal I_{\HH_2^*} (\beta+\frac{\gamma}{2}-2, I_m)\Gamma\left(\tilde\beta +\frac{\gamma_1+\gamma_2+2}{2}\right)\Gamma\left(\beta +\frac{\gamma+\gamma_3+\gamma_2-2}{2}\right)}{\pi^{\frac{\gamma_2}{2}+1}\mathcal I_{\HH_1^*} (\beta+\frac{\gamma}{2}-2, I_m)\Gamma\left(\beta +\frac{\gamma+\gamma_1+\gamma_2-1}{2}\right)\Gamma\left(\beta+\frac{\gamma+\gamma_3+\gamma_2-1}{2}\right)\Gamma\left(\frac{\gamma_2}{2}\right)} \nonumber\\
		& & \times \int_0^\infty r^{\gamma_2-1} (1+r^2)^{\tilde\beta+\frac{\gamma_2+2}{2}} {}_2F_1\left(\tilde\beta +\frac{\gamma_2+3}{2}, \tilde\beta +\frac{\gamma_1+\gamma_2+2}{2}; \tilde\beta +\frac{\gamma_1+\gamma_2+3}{2}; -r^2\right) \nonumber\\
		&& \s \s {}_2F_1\left(\tilde\beta +\frac{\gamma_2+3}{2}, \tilde\beta +\frac{\gamma_3+\gamma_2+2}{2}; \tilde\beta +\frac{\gamma_3+\gamma_2+3}{2}; -r^2\right) \dd r.\nonumber
	\end{eqnarray}

\subsection{Gear graphs} \label{sec:gear}
For integer $m \geq 3$, the \textit{gear graph} with $2m+1$ vertices, denoted by $\G_m$ in this section, has vertex set $\V(\G_m) = \{v_1, \ldots, v_{2m+1}\}$ and edge set $\E(\G_m)$ is
\begin{equation}
	\{\{v_{2m+1}, v_\mu\} : 1 \leq \mu \leq 2m-1, \textrm{ $\mu$ is odd}\} \sqcup \{\{v_\mu, v_{\mu+1}\}:1 \leq \mu \leq 2m\} \sqcup \{\{v_{2m}, v_1\}\}. \nonumber
\end{equation}
In words, the first $2m$ vertices form an induced cycle and the last vertex $v_{2m+1}$ is adjacent to alternate vertices in the cycle. The gear graph $\G_3$ is shown in Figure~\ref{fig:gear}, and the integral $\mathcal I_{\G_3}(\beta, I_7)$ can be expressed as a one-dimensional integral, see Equation (\ref{eq:Gear3}).

For $\G_m$, let $\G_m^*$ be a supergraph with added edges
\begin{equation}
	\{\{v_{2\mu+1}, v_{2\mu+3}\}: 0 \leq \mu \leq m-2\} \sqcup \{\{v_1, v_\mu\}: 5 \leq \mu \leq 2m-1, \textrm{ $\mu$ is odd}\}. \nonumber
\end{equation}
In words, the neighbours of $v_{2m+1}$ in $\G$ form a cycle (in the same ordering) and the vertex $v_1$ is connected to every vertex in this cycle. An example of $\G^*_4$ is shown in Figure~\ref{fig:gear4}. It is straightforward to verify that $\G^*_m$ is chordal.

\begin{corollary} \label{coro:gear}
	Let $m \geq 4$ be an integer. Let $\G_m$ be the gear graph on $2m+1$ vertices and let $\G_m^*$ be a chordal completion defined above. 
	Let $\beta > -1$ be a real number. Then,
	\begin{equation}
		\mathcal I_{\G_m}(\beta, I_{2m+1}) = \dfrac{\mathcal I_{\G_m^*}(\beta, I_{2m+1})}{\mathcal I_{\HH^*_m}\left(\beta+\frac12, I_{2m}\right)} \mathcal I_{C_{2m}}\left(\beta+\frac12, I_{2m}\right). \nonumber
	\end{equation}
	where $C_{2m}$ is the cycle graph on $2m$ vertices, and $\HH_{m}^*$ is the induced subgraph obtained from $\G^*_m$ by removing the vertex with degree $m$ ($v_{2m+1}$ in the above definition).
\end{corollary}

\begin{proof}
	Since $\G_m^*$ is chordal, the graph $\HH_m^*$ is a chordal completion of the cycle $C_{2m}$. We compare the integrals obtained from Theorem~\ref{thm:main} using the pair $\G_m, \G_m^*$ and the pair $C_{2m}$, $\HH^*_m$.
	Note that a maximal clique $\cc$ in $\G^*$ contains the vertex $v_{2m+1}$ if and only if $\cc$ contains more than 3 vertices. Similarly, a minimal separator $\spt$ in $\G^*$ contains the vertex $v_{2m+1}$ if and only if $\spt$ contains more than 2 vertices. 
	
	Following the same argument as in Supplementary Section~\ref{sec:missingTriangle}, it is easy to see that
	\begin{equation}
		\mathcal I_{\G_m}(\beta, I_{2m+1}) = \dfrac{\mathcal I_{\G_m^*}(\beta, I_{2m+1})}{\pi^{2m-3}} \dfrac{\pi^{2m-3} \mathcal I_{C_{2m}}\left(\beta+\frac12, I_{2m}\right)}{\mathcal I_{\HH^*_m}\left(\beta+\frac12, I_{2m}\right)}. \nonumber
	\end{equation}
\end{proof}

In the above result, both $\mathcal I_{\G_m^*}(\beta, I_{2m+1})$ and $\mathcal I_{\HH^*_m}\left(\beta+\frac12, I_{2m}\right)$ have an explicit formula from Equation (\ref{eq:chordalA}). Since $C_{2m}$ has starry fill-ins, Corollary~\ref{thm:star} implies that the integral $\mathcal I_{C_{2m}}\left(\beta+\frac12, I_{2m}\right)$ has the form of Equation (\ref{eq:specialForm}).

\section{Bayesian analysis of Fisher's Iris Virginica dataset} \label{suppsec:iris}

For the Fisher's Iris Virginica dataset of 4 features [sepal length (SL), sepal width (SW), petal length (PL) and petal width (PW)] measured for 50 Iris Virginica, there are $2^{6} = 64$ possible Gaussian graphical models. To perform model selection, we need to compute
\begin{equation}
	p(Z \given \G) = \dfrac{2^{-6}}{(2\pi)^{100}} \dfrac{\const_\G(\delta + 50, U+D)}{\const_\G(\delta, D)}, \qquad 
	U = \begin{pmatrix}
		19.8128 && \hfill 4.5944 && 14.8612 && \hfill 2.4056 \\
		\hfill 4.5944 && \hfill 5.0962 && \hfill 3.4976 && \hfill 2.3338 \\
		14.8612 && \hfill 3.4976 && 14.9248 && \hfill 2.3924 \\
		\hfill 2.4056 && \hfill 2.3338 && \hfill 2.3924 && \hfill 3.6962
	\end{pmatrix} \nonumber
\end{equation}
where $U$ is the scatter matrix (sample covariance rescaled by 49) of the dataset $Z \in \mathbb R^{4 \times 50}$, and $\G$ is any of the 64 graphs on 4 vertices. In line with \cite{Atay-Kayis, Roverato}, we take $\delta = 3$ and $D = I_4$.

Results were previously known for the chordal graphs, while the approach in Section~\ref{sec:generalA} now allows us to complete the analysis and compute the marginal likelihood of the non-chordal graphs on 4 vertices which are the three 4-cycles, $\G_1$, $\G_2$, $\G_3$ (as in Figure~\ref{fig:C4}), with minimum fill-in 1. 

\begin{figure}
\centering
	\begin{tabular}{ccccc}
		& & \includegraphics[width=0.15\textwidth]{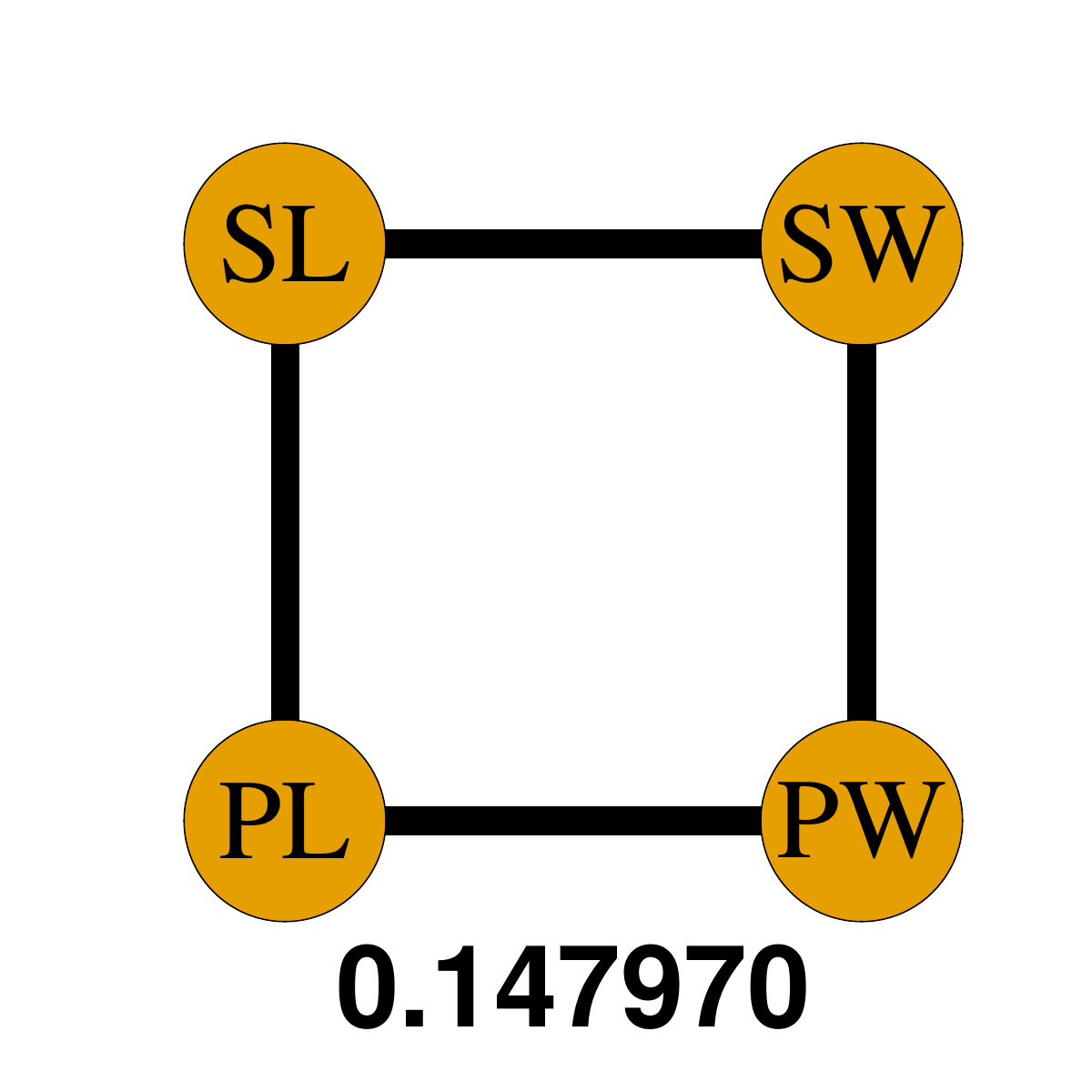} & & \\
		\includegraphics[width=0.15\textwidth]{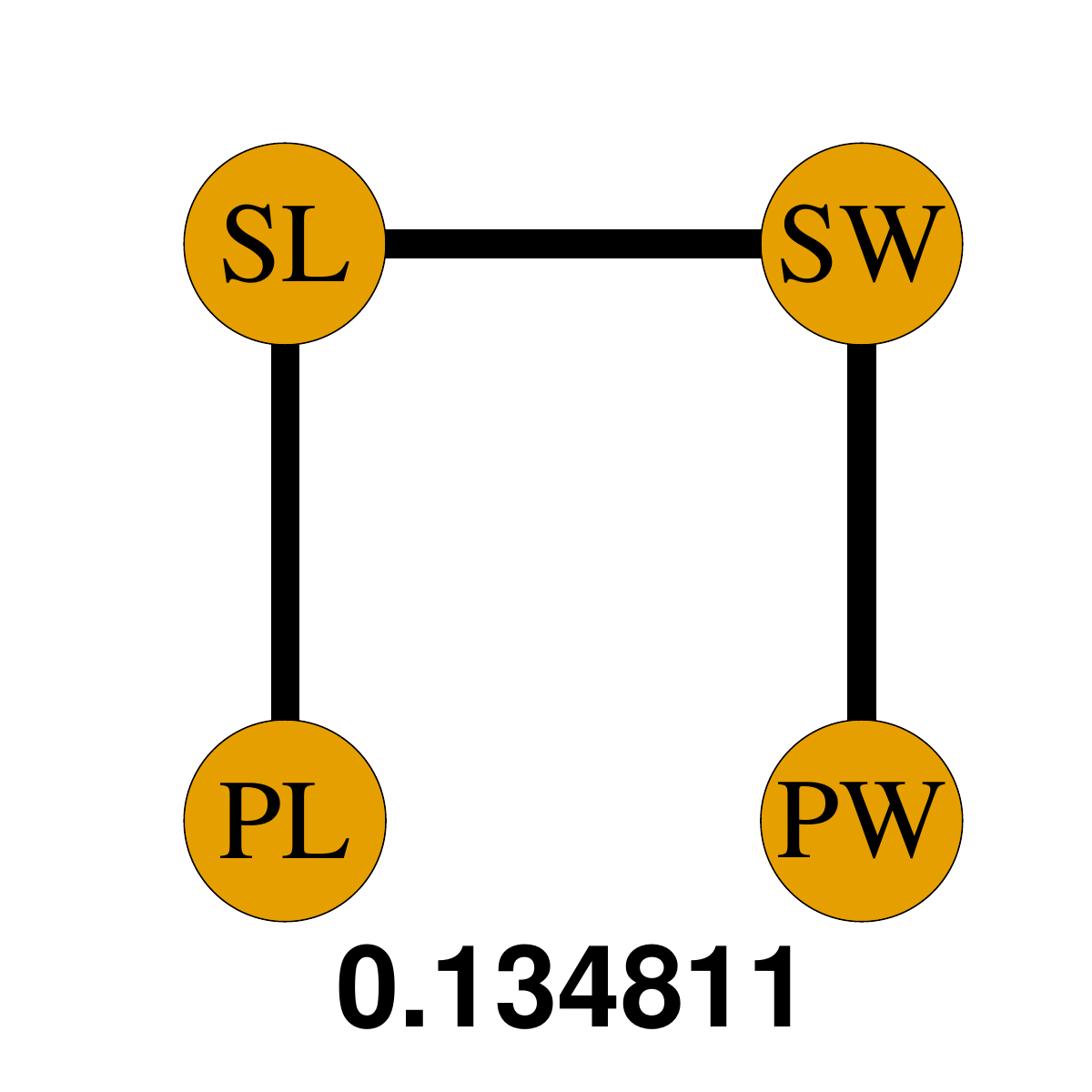} & \includegraphics[width=0.15\textwidth]{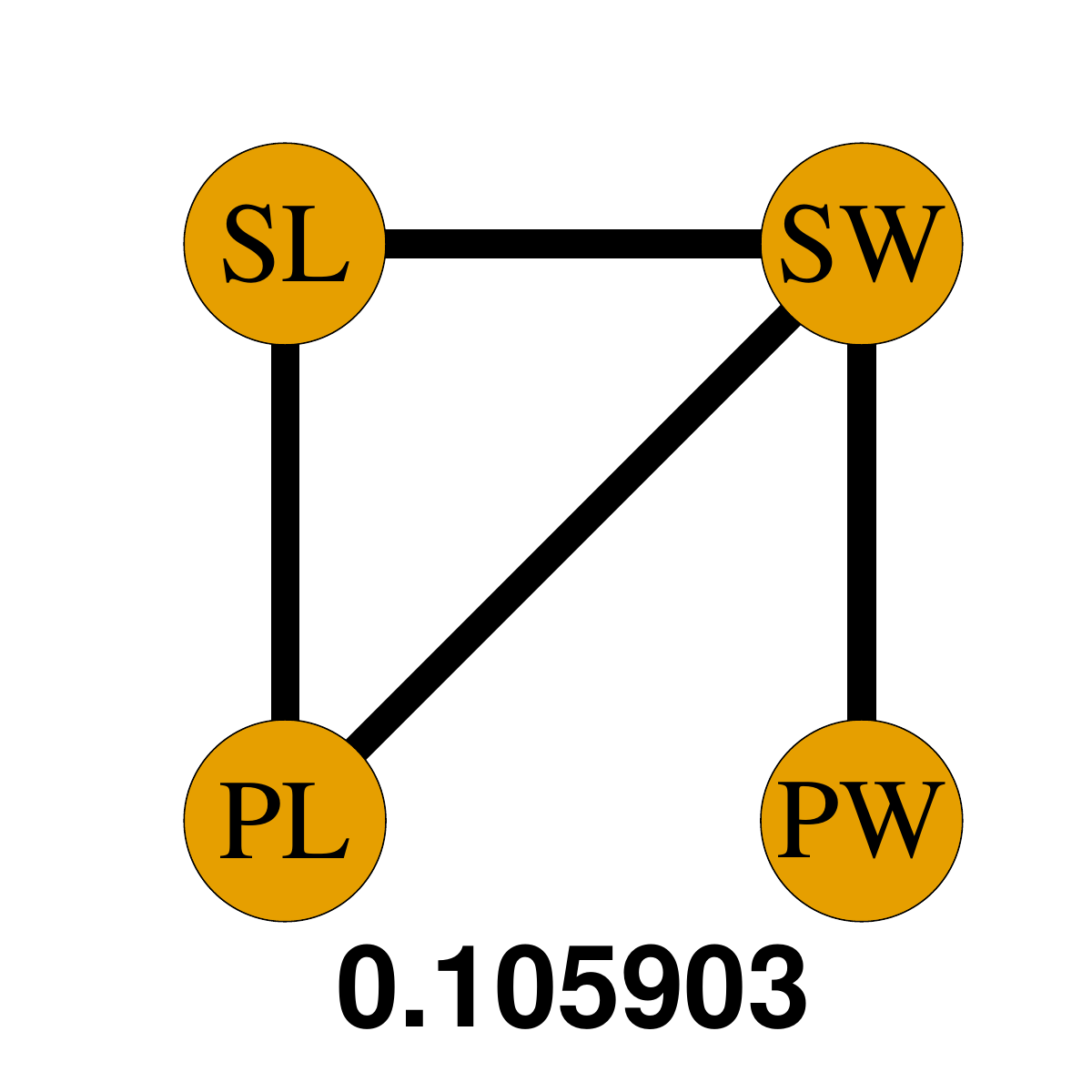} & \includegraphics[width=0.15\textwidth]{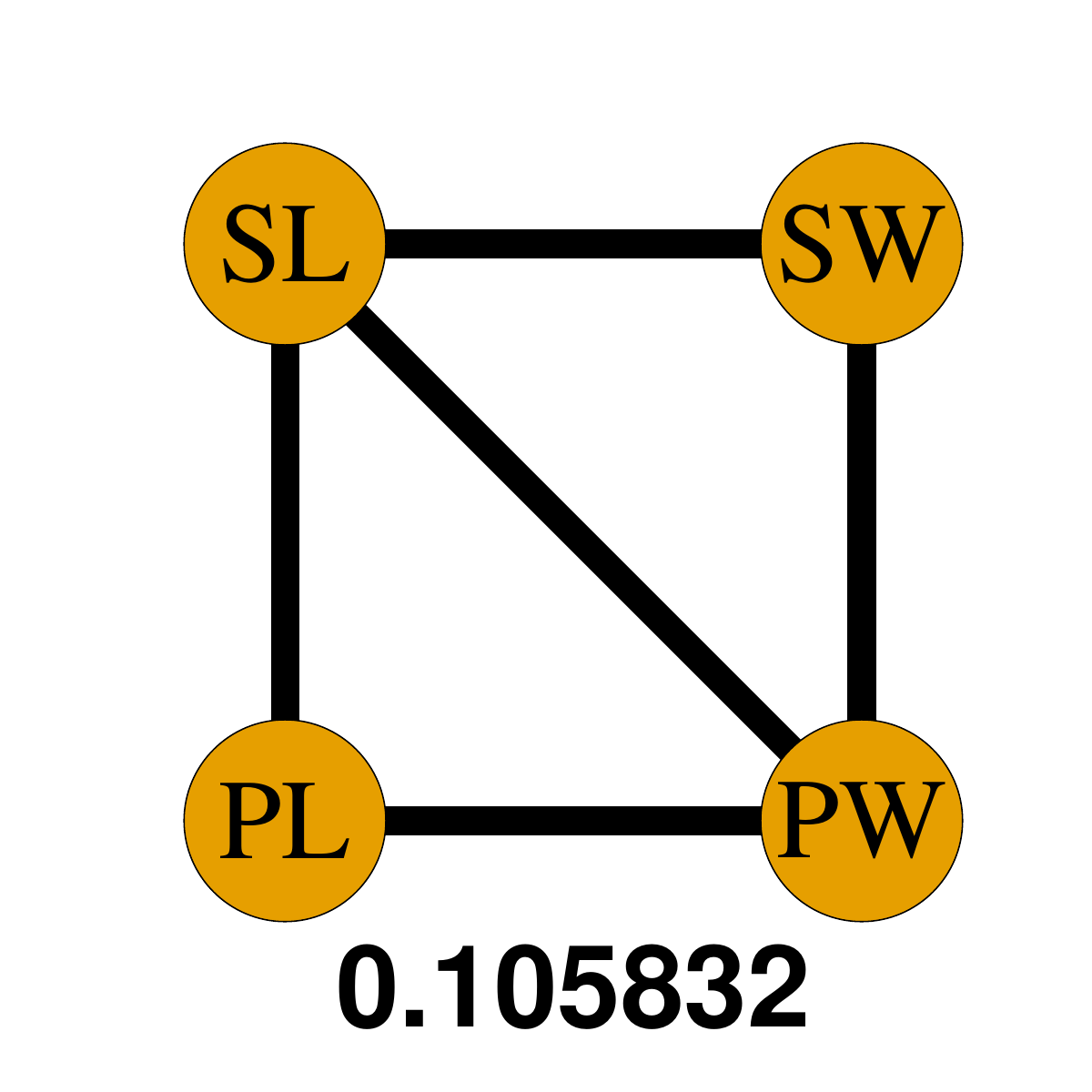} & \includegraphics[width=0.15\textwidth]{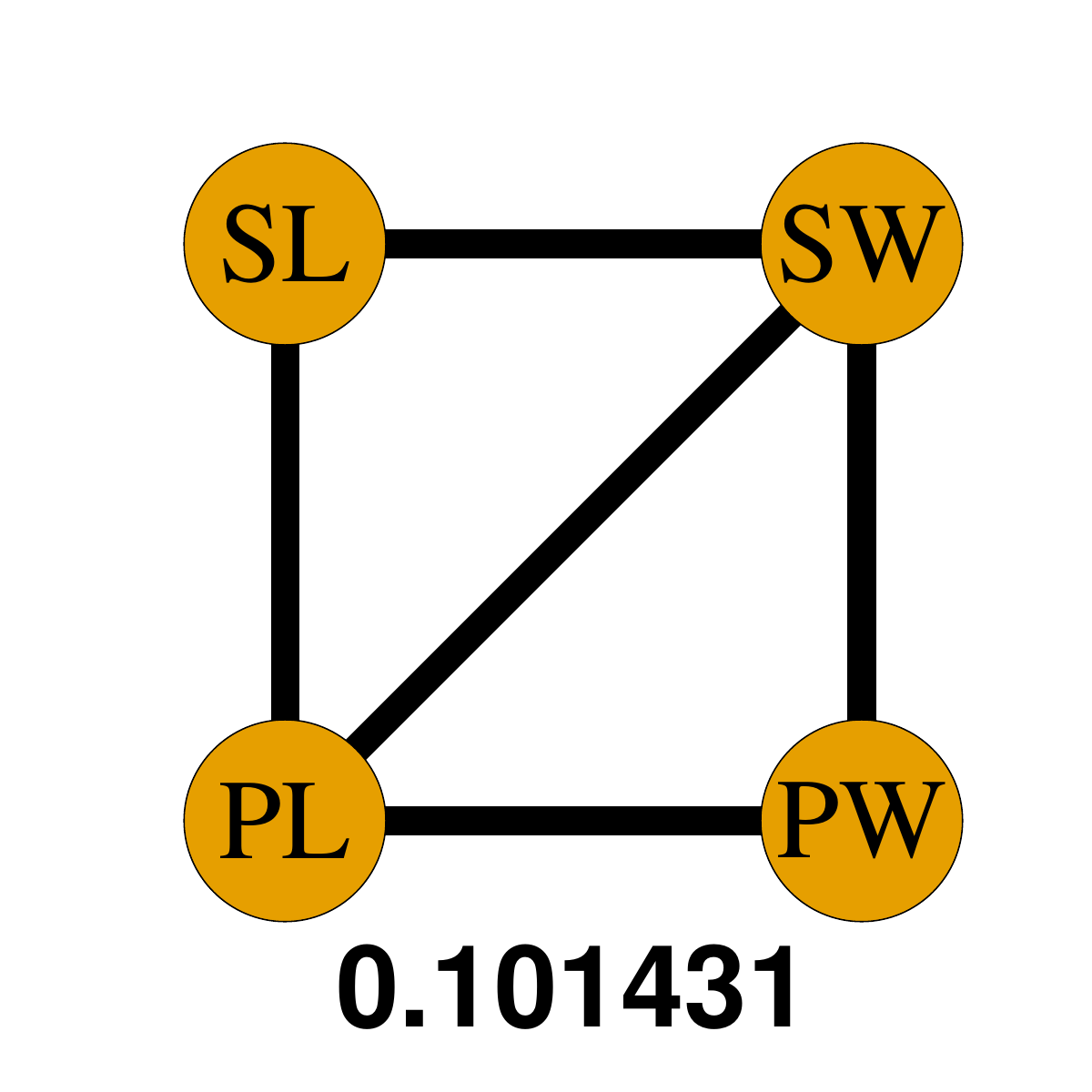} &
		\includegraphics[width=0.15\textwidth]{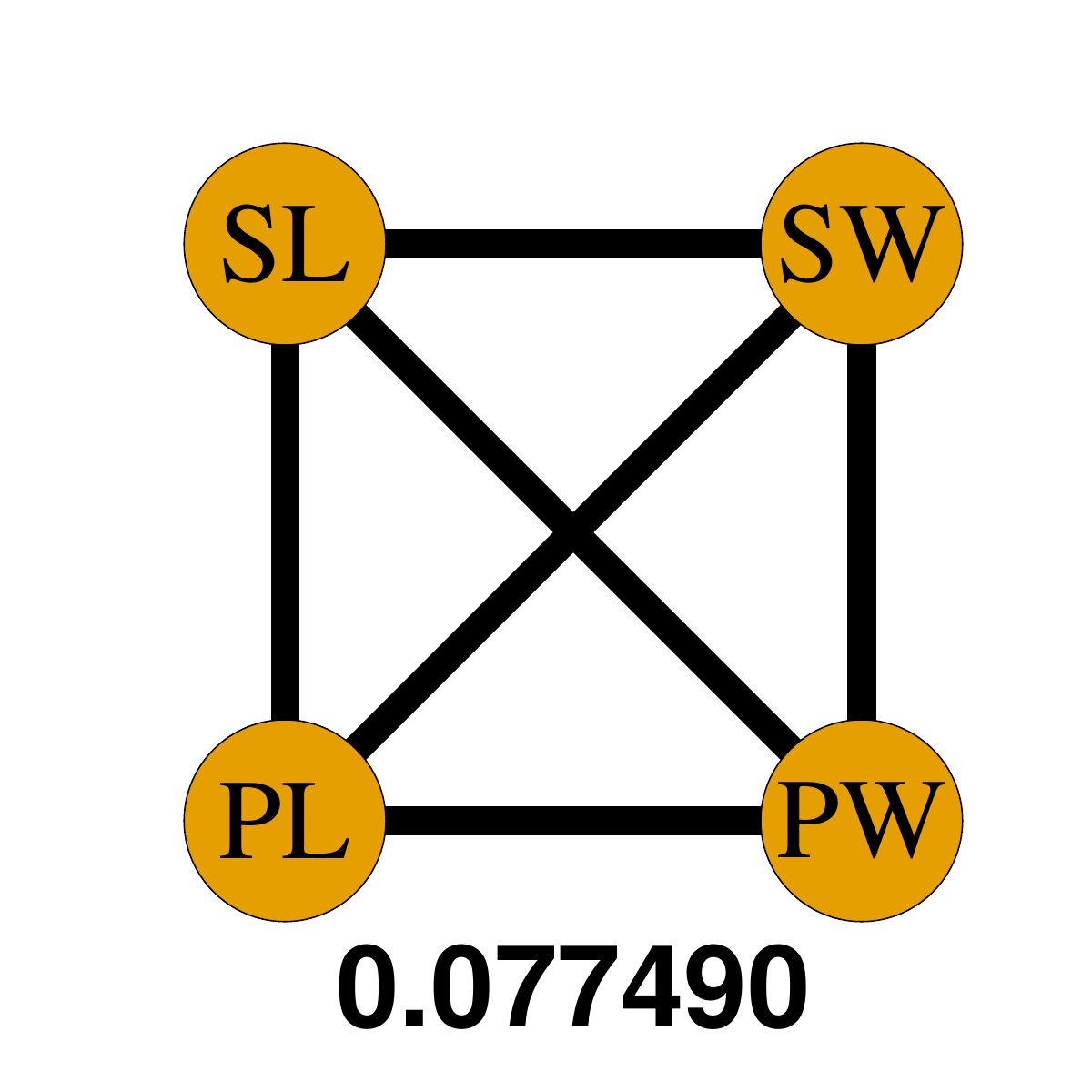} \\
		\includegraphics[width=0.15\textwidth]{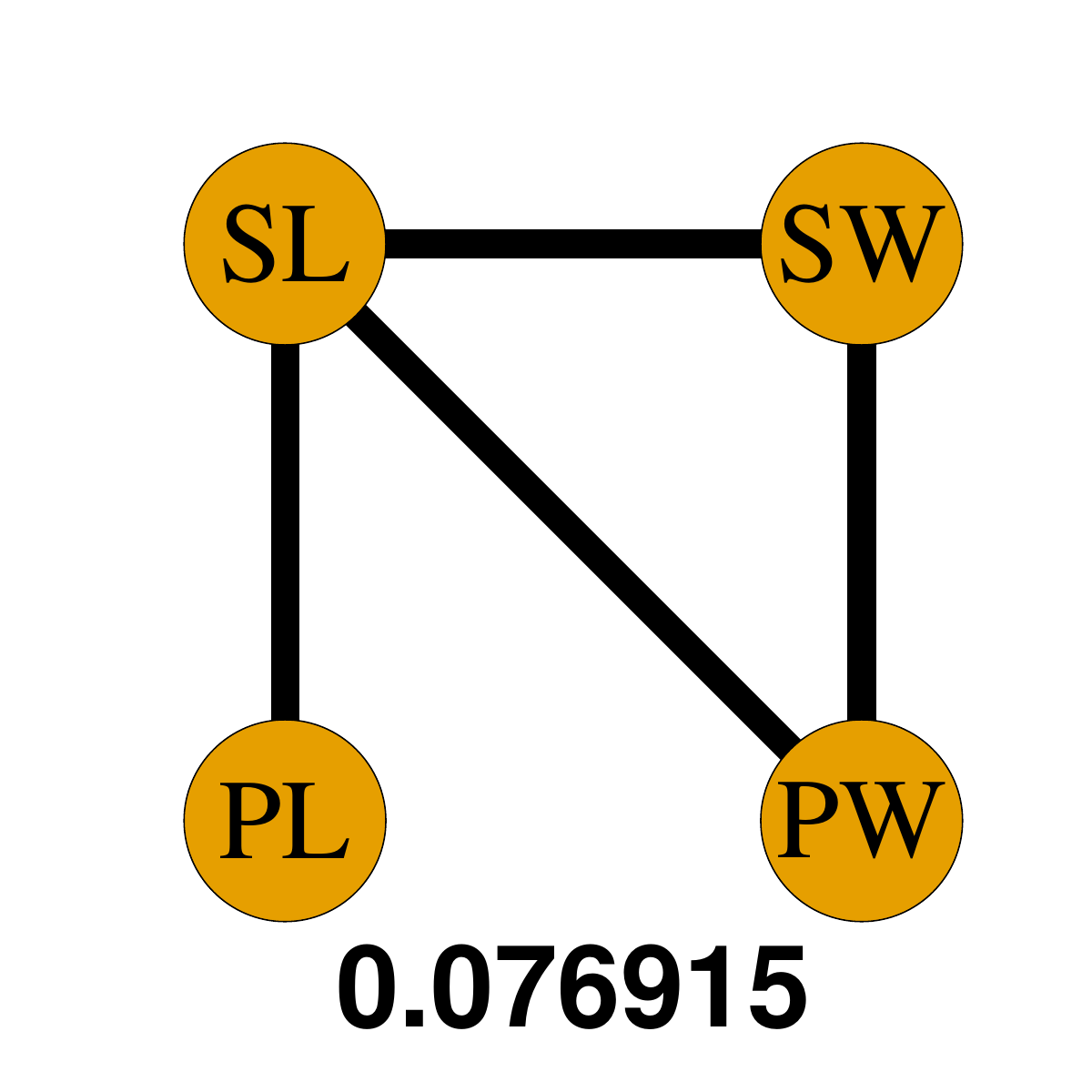} & \includegraphics[width=0.15\textwidth]{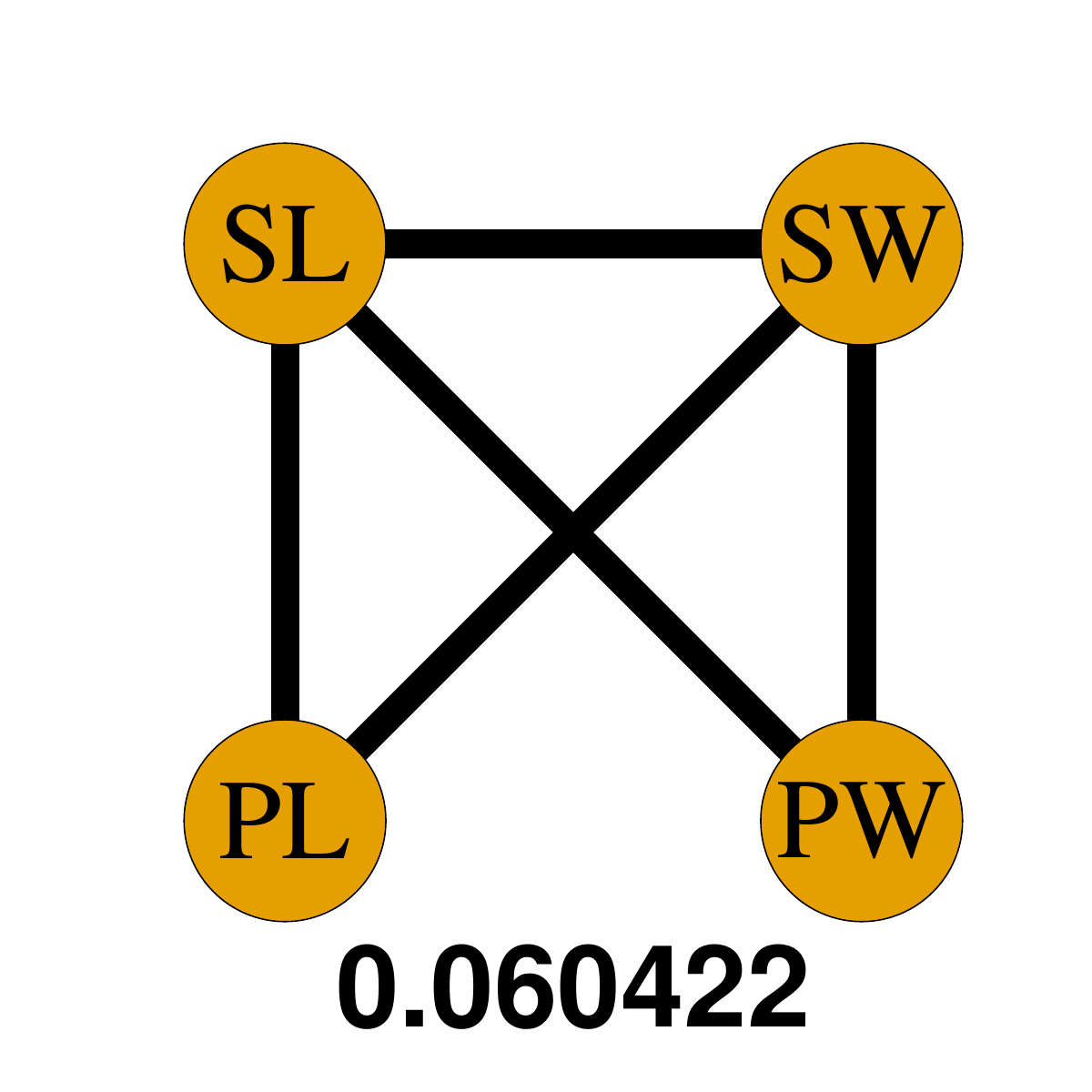} & \includegraphics[width=0.15\textwidth]{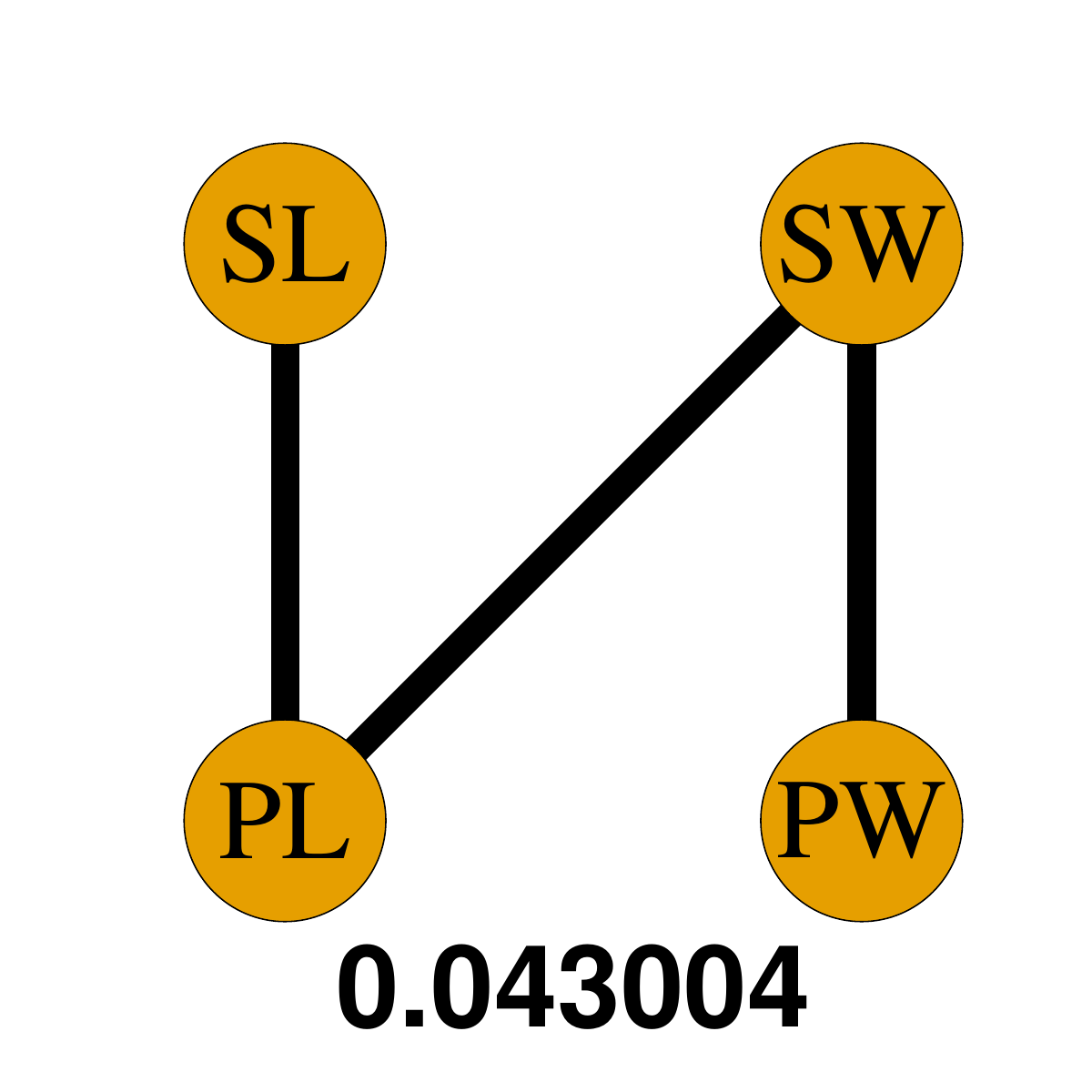} & \includegraphics[width=0.15\textwidth]{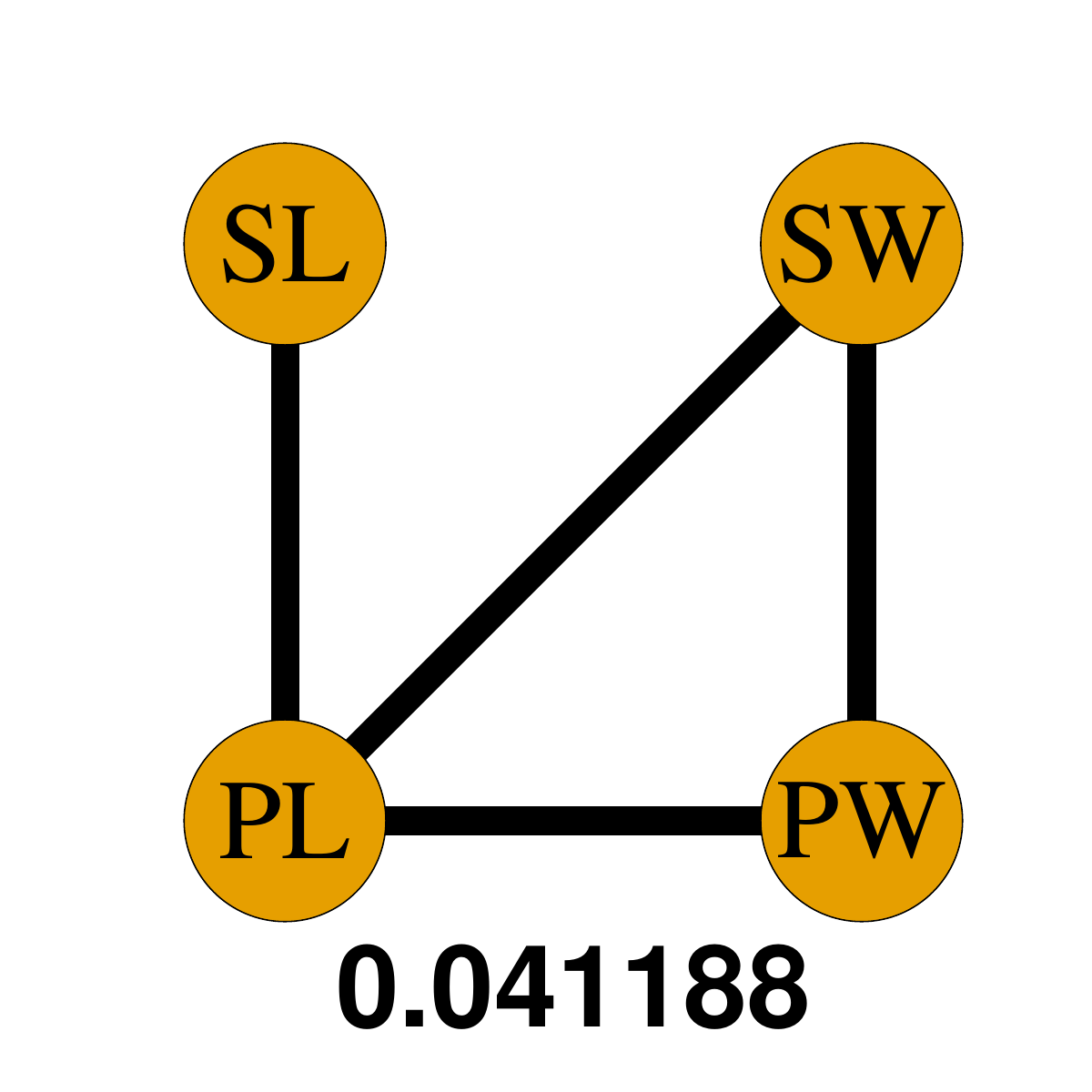} &
		\includegraphics[width=0.15\textwidth]{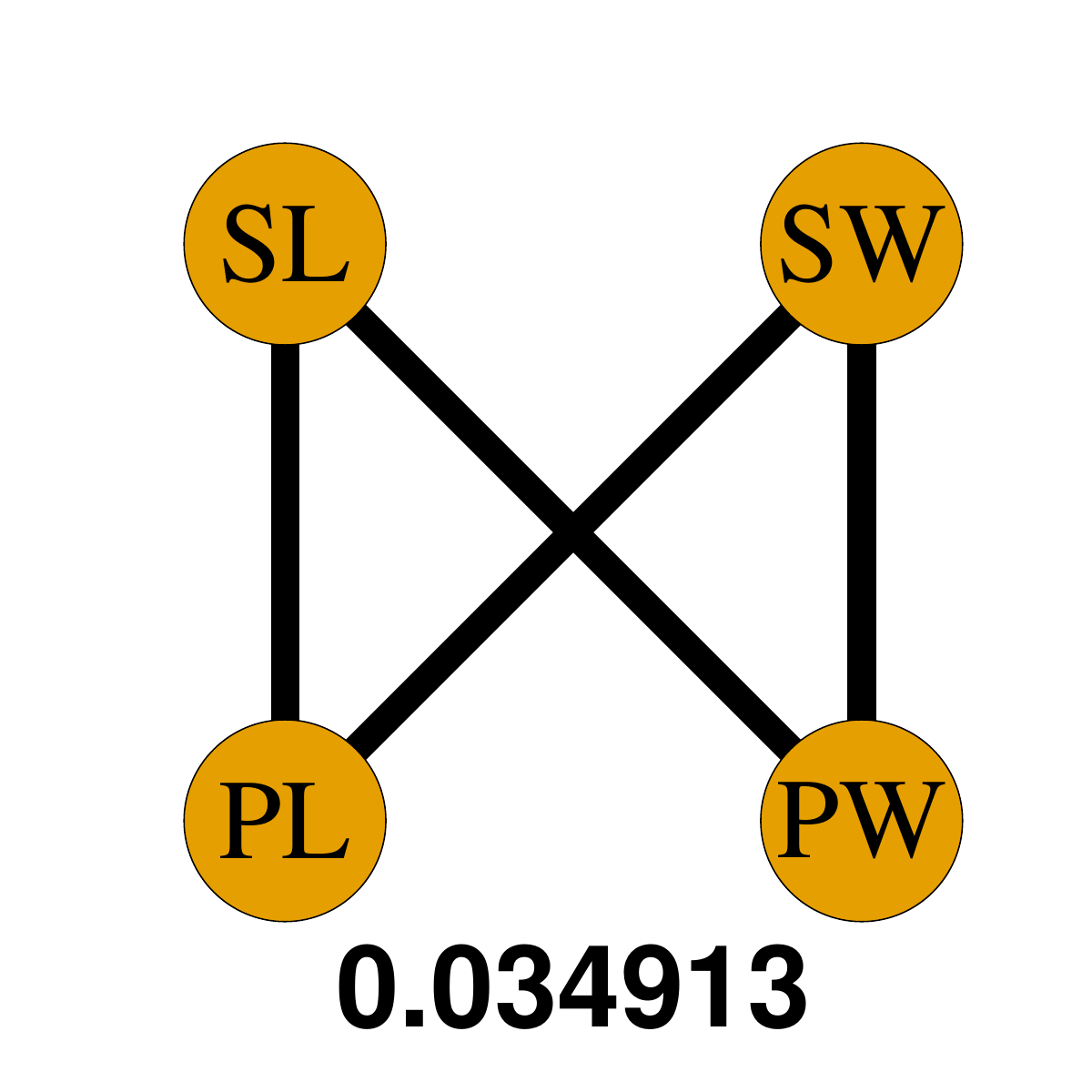} \\
		\includegraphics[width=0.15\textwidth]{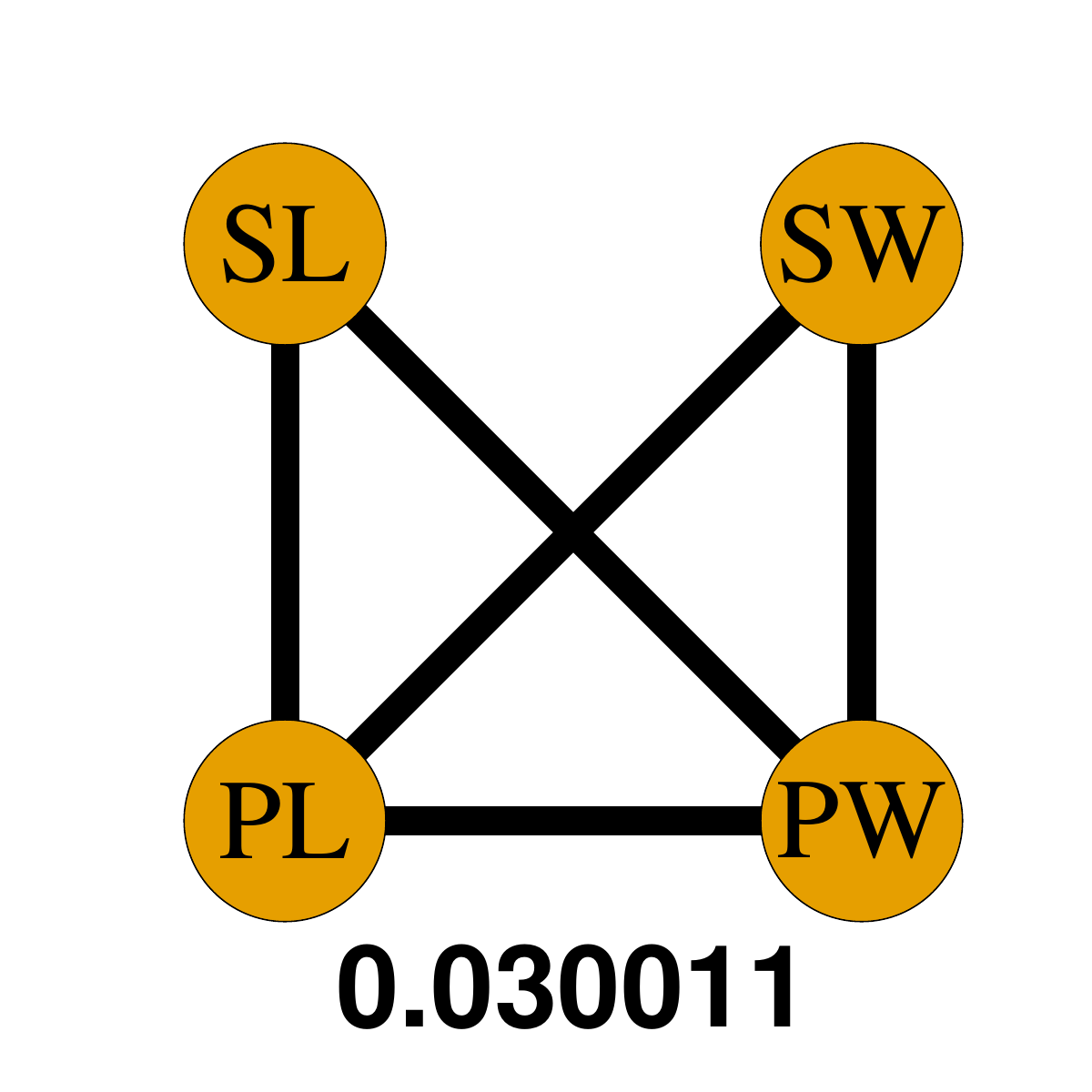} & \includegraphics[width=0.15\textwidth]{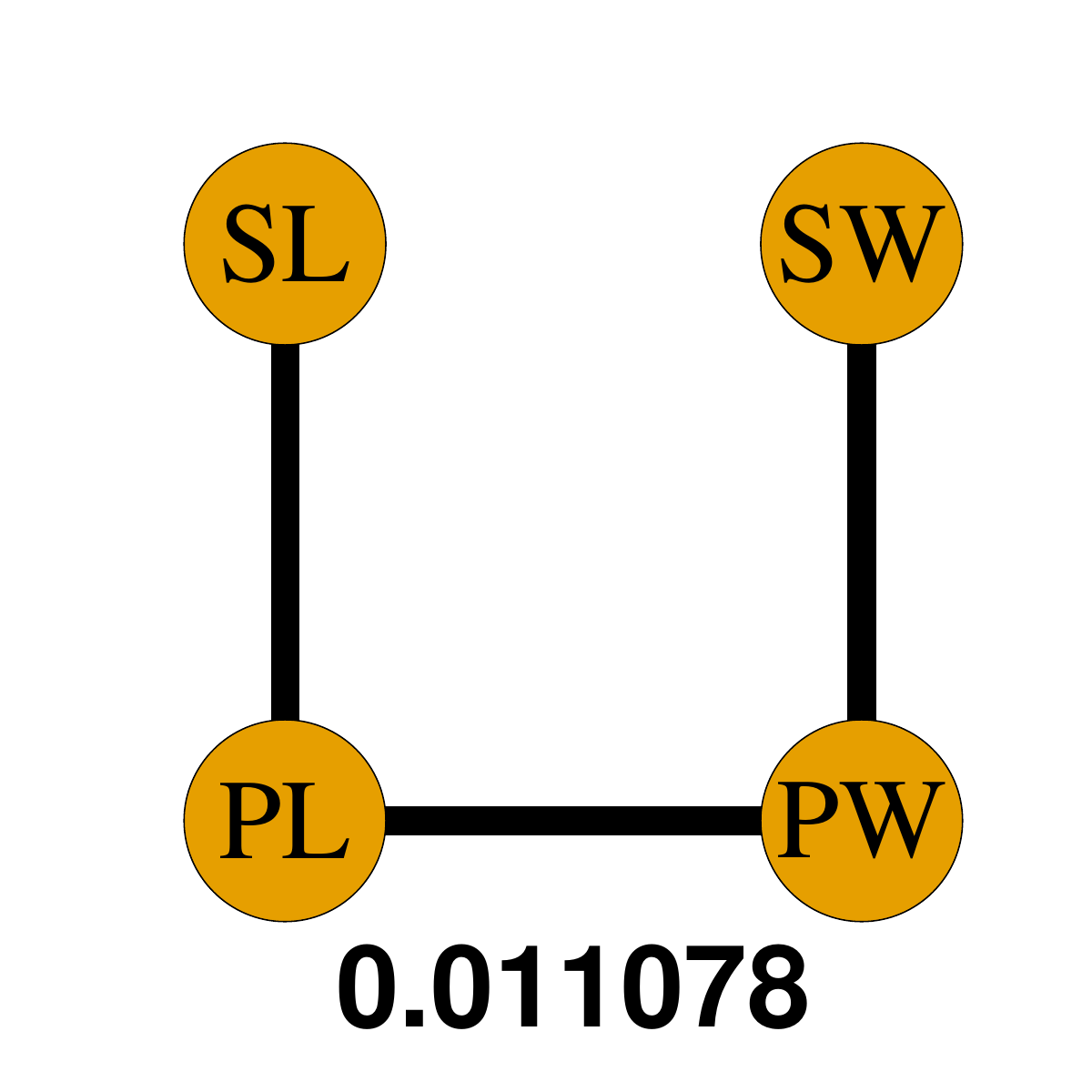} & \includegraphics[width=0.15\textwidth]{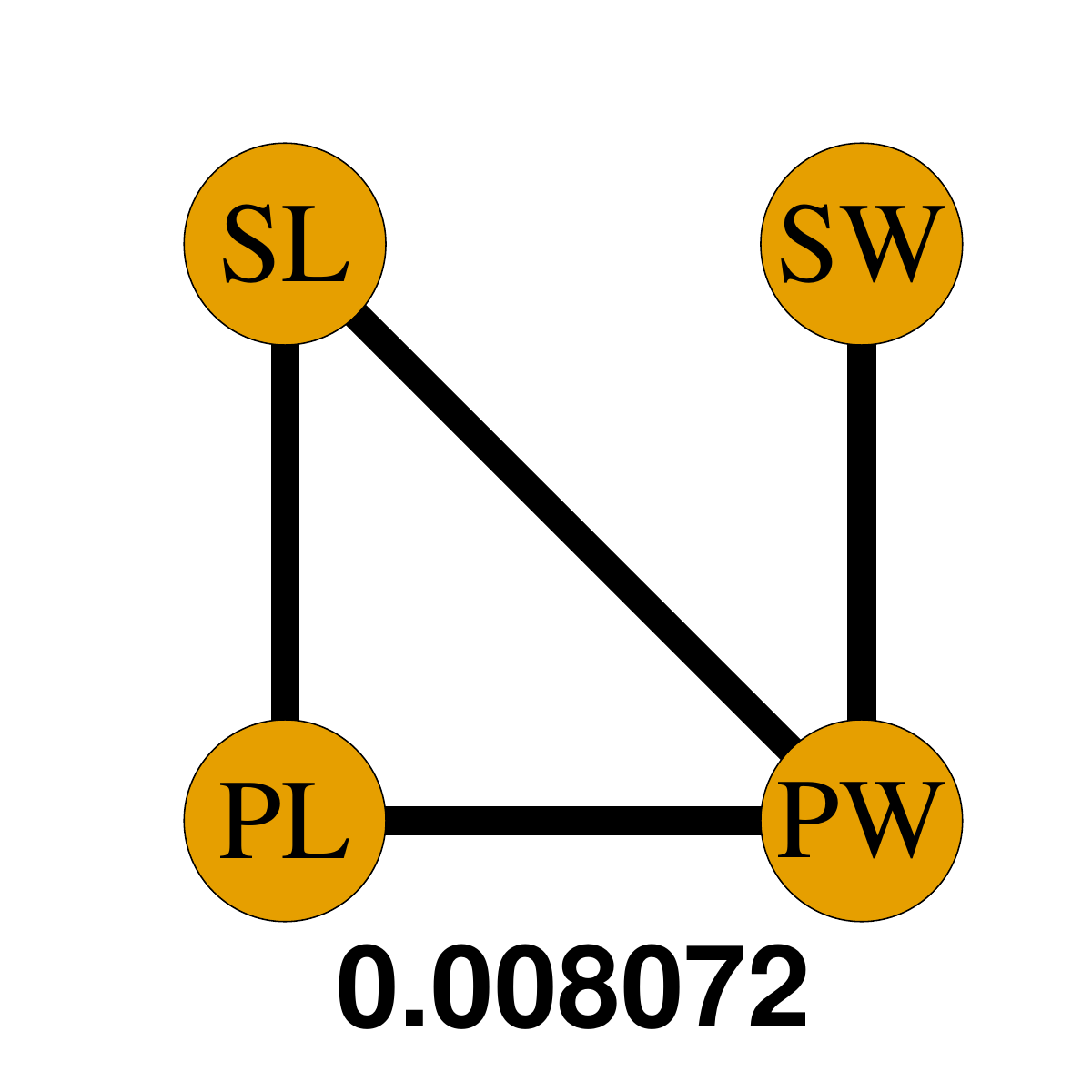} & \includegraphics[width=0.15\textwidth]{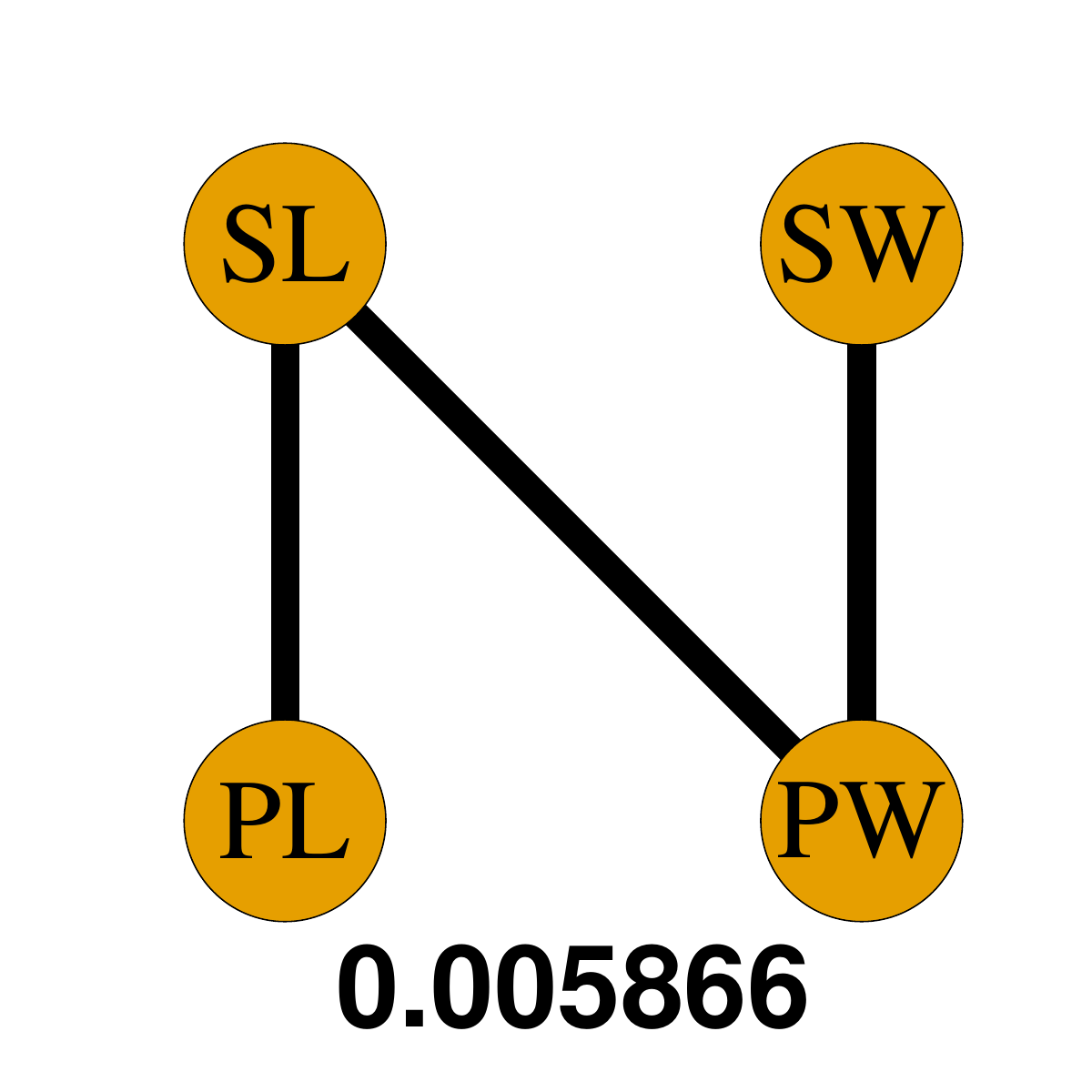} &
		\includegraphics[width=0.15\textwidth]{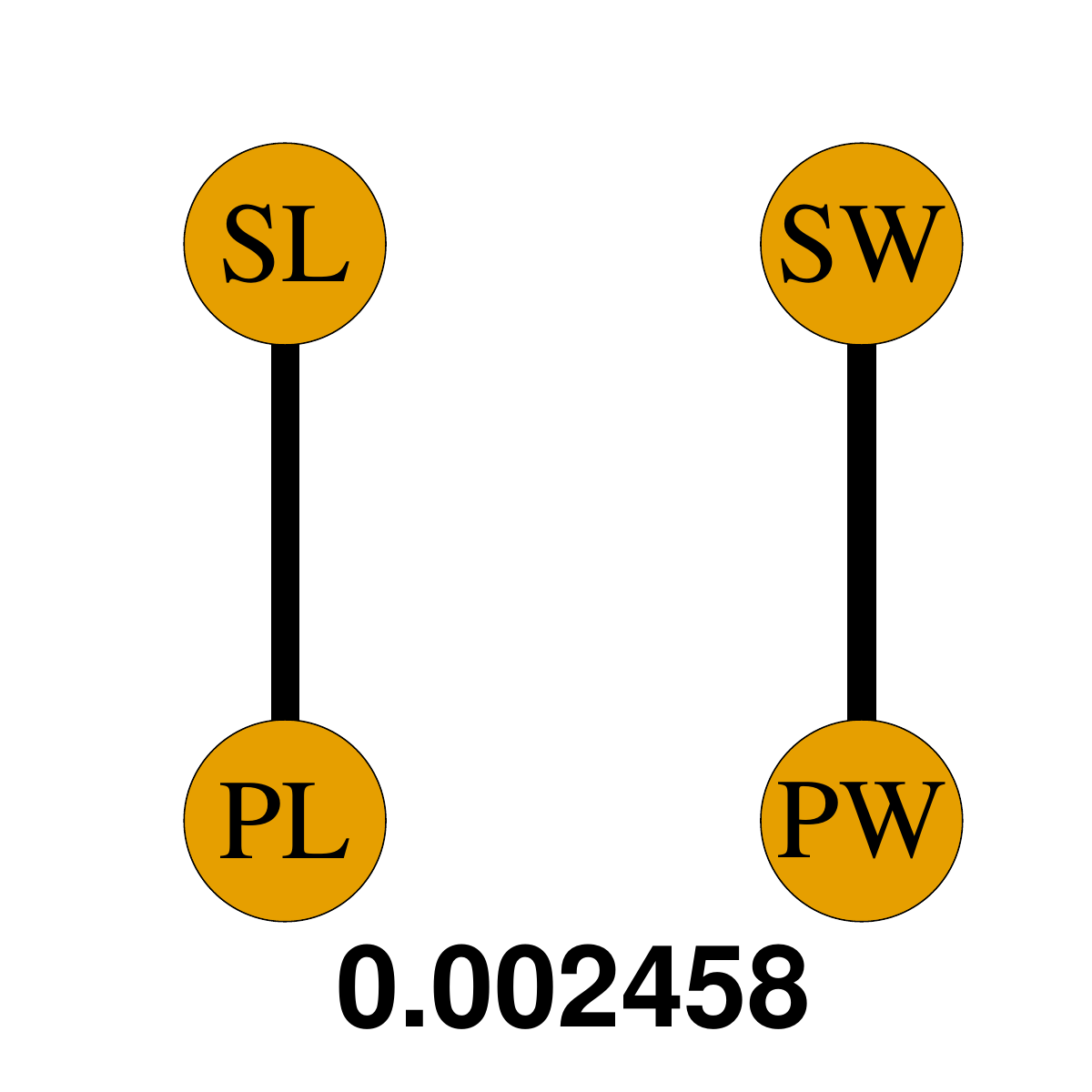} 
	\end{tabular}
	\caption{Top 16 models for Fisher's Iris Virginica data with their posterior probabilities.}
	\label{fig:irisProb}
\end{figure}

We compare the performance of finding these three marginal likelihoods $\const_{\G_j}(53, U+I_4)$, for $j = 1, 2, 3$, using our proposed method with the Monte Carlo method in \cite{Atay-Kayis} (as implemented in the  \textit{gnorm} function from the R package \textit{BDgraph} \cite{mw19} with $10^6$ samples, for 200 different seeds). The results are shown in Table~\ref{tab:irisData} and Figure~\ref{fig:C4}. It can be seen, particularly for the last case ($\G_3$), that the ratio of interest struggles to converge when using the Monte Carlo method, whereas our method gives an accurate estimate. Moreover, our approach is hundreds of times faster than the Monte Carlo method, as shown in Table~\ref{tab:irisTime}. In Figure~\ref{fig:irisProb}, we show the top 16 models for this dataset, each with its associated posterior probability for a uniform prior on graphs. \rev{These are the exact posterior for each graph, computed with our approach involving a one-dimensional numerical integral (Section~\ref{sec:generalA}) for the three non-chordal graphs.}

\begin{table}[ht] 
	\caption{Summary of the estimated values of $\log(p(Z \given \G_j))$, for $j = 1,2,3$, using Monte Carlo integration \cite{Atay-Kayis,mw19} with $10^6$ samples and for 200 different seeds, and using our approach.}
	\begin{tabular}{l|S[table-format=3.4]|S[table-format=3.4]|S[table-format=4.4]}
		& {$\log(p(Z \given \G_1))$} & {$\log(p(Z \given \G_2))$} & {$\log(p(Z \given \G_3))$}\\
		\hline
		{min(MC)} & -84.4427 & -85.8881 & -114.6686 \\
		{max(MC)} & -84.4394 & -85.8834 & -111.0994 \\
		{sd(MC)} & 0.0006 & 0.0008 & 0.5142 \\
		{mean(MC)} & -84.4413 & -85.8854 & -113.7590 \\
		\hline
		{our method} & -84.4412 & -85.8854 & -113.5226
	\end{tabular} \label{tab:irisData}
\end{table}

\clearpage

\begin{table}[hb!]
	\caption{The average time (in milliseconds) needed to compute the normalising constant $\const_{\G_j}(53, U+I_4)$, for $j = 1, 2, 3$.}
	\begin{tabular}{l|S[table-format=3.4]|S[table-format=3.4]|S[table-format=3.4]}
		& {$\const_{\G_1}(53, U+I_4)$} & {$\const_{\G_2}(53, U+I_4)$} & {$\const_{\G_3}(53, U+I_4)$}\\
		\hline
		{MC \cite{Atay-Kayis,mw19}} & 433.75 & 440.53 & 432.15 \\
		\hline
		{our method} & 0.51 & 0.66 & 1.45
	\end{tabular} \label{tab:irisTime}
\end{table}

\begin{figure}
\centering
\includegraphics[width=0.66\textwidth]{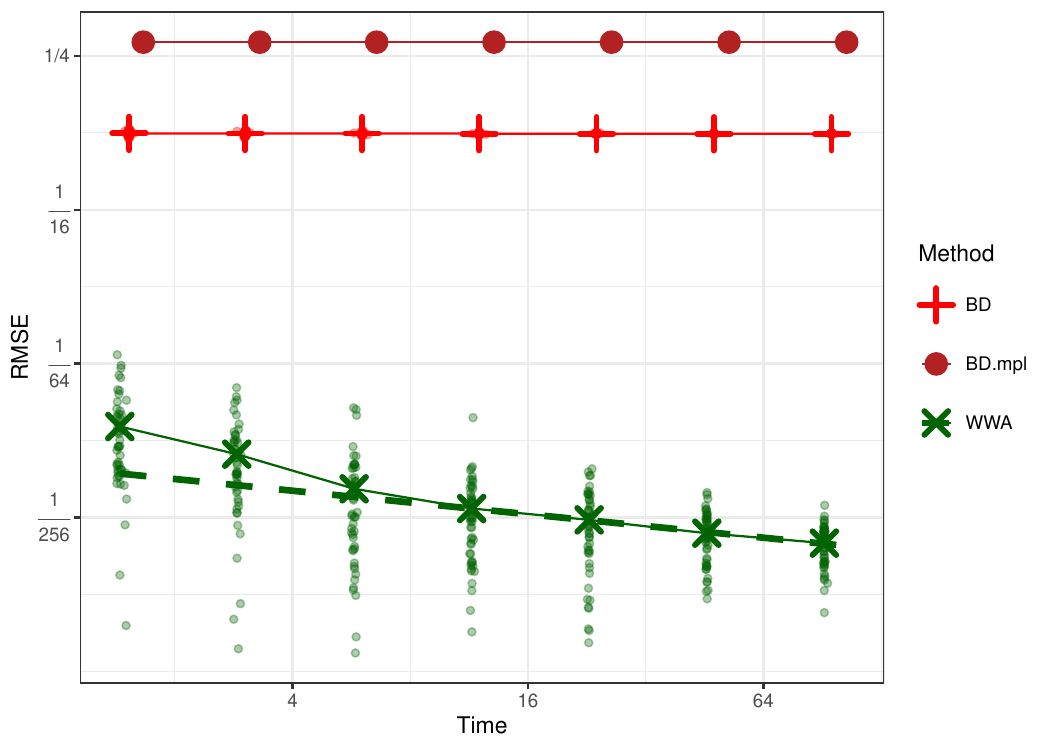}
\caption{\rev{Convergence diagnostics of Bayesian samplers from \cite[][BD]{mml23}, \cite[][BD.mpl]{mohammadi2025scalable} and \cite[][WWA]{Willem}, in terms of the root mean square error (RMSE) from the true edge posterior (Table~\ref{tab:IrisPostProbs}) as a function of the runtime. The approximations employed in BD and BD.mpl lead to clear biases, while the convergence fit of WWA (dashed line, fit on the later time points) has a slope (around -0.15) well below the theoretical behaviour (-0.5).} }
\label{fig:irisConverge}
\end{figure}

\rev{As we now have the exact posterior of each graph, we can obtain the posterior edge probabilities which we display in Table~\ref{tab:IrisPostProbs} to high precision for reference \cite[c.f.~Table S1 of][]{Willem}. Moreover, we can use these exact results to test the quality of Bayesian samplers, and we compare the convergence to the posterior for two approaches employing approximations from the \textit{BDgraph} \cite{mw19} package, one using an approximation for the ratio of normalising constants \cite[][which we label BD]{mml23} and a faster approach which uses marginal pseudo-likelihoods \cite[][labelled BD.mpl]{mohammadi2025scalable}, along with the WWA sampler \cite[][labelled WWA]{Willem}.

Plotting the error as a function of the time (indicative of the chain length) in Figure~\ref{fig:irisConverge}, we observe that even the seemingly innocuous normalising constant ratio approximation of BD leads to a notable error in the posterior, while the pseudo-likelihood approximation of BD.mpl leads to a large bias. While the WWA sampler seems to converge to the true posterior, when we fit a regression line through the average error for longer times (dashed line in Figure~\ref{fig:irisConverge}), rather than the slope of $-\frac{1}{2}$ we would expect from Monte Carlo estimates, we observed a slope of around $-0.15$. This very slow convergence may hint at heavy-tailed events dominating the sampling (for example through the exchange algorithm), bias or both. Code to reproduce these results is hosted at \url{https://github.com/jackkuipers/GWnorm}.}

\begin{table}[t]
\caption{\rev{The posterior probabilities of each edge for the Iris data example.}}
\begin{tabular}{l|cccccc}
Edge & SL--SW & SL--PL & SL--PW & SW--PL & SW--PW & PL--PW \\
\hline
Posterior probability & 0.8214037 & 1.0000000 & 0.4059322 & 0.5009369 & 0.9873611 & 0.5318798 \\
\end{tabular} \label{tab:IrisPostProbs}
\end{table}

\section{Comparison of Monte Carlo approaches} \label{suppsec:GW_BDcomp}

\begin{figure}
\centering
\includegraphics[width=0.66\textwidth]{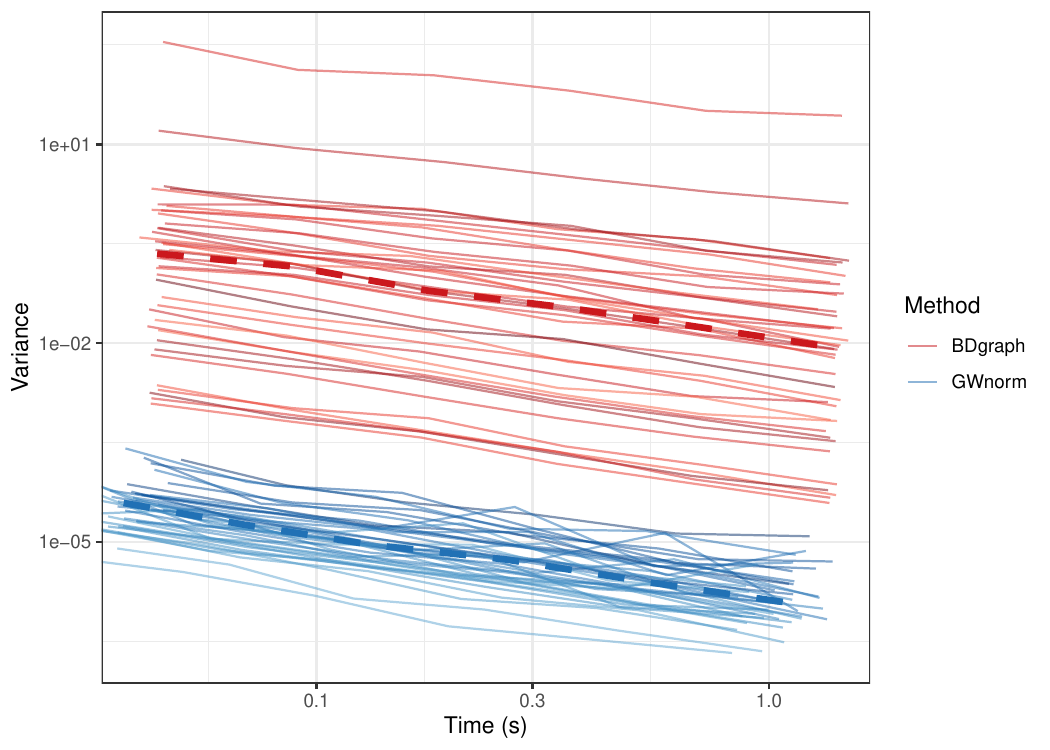}
\caption{\rev{Convergence diagnostics of estimating the normalising constant with our approach implemented in the \textit{GWnorm} package against the sampler from \cite{Atay-Kayis} in the \textit{BDgraph} package \cite{mw19}. The dashed lines represent the medians.}}
\label{fig:simConverge}
\end{figure}

\begin{figure}
\centering
\includegraphics[width=0.66\textwidth]{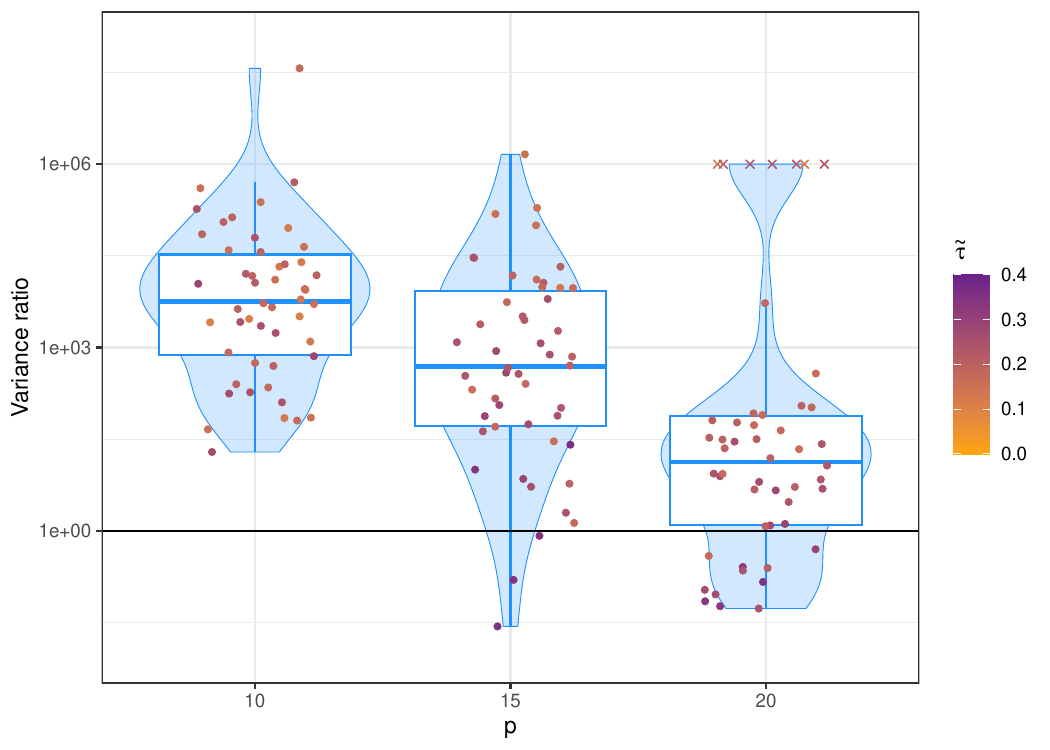}
\caption{\rev{Relative efficiency of \textit{GWnorm} against \textit{BDgraph} \cite{mw19} for prime graphs of different sizes. Where \textit{BDgraph} failed to compute the normalising constant is indicated with diagonal crosses placed at $10^{6}$. The points are coloured by the fraction of missing edges $\tilde{\tau}$.}}
\label{fig:simPrime}
\end{figure}

\rev{To benchmark the Monte Carlo estimation of the normalising constant developed in Section~\ref{sec:MC}, which uses the Fourier analysis to integrate over the missing edges in the graph, we compare to the direct sampler approach \cite{Atay-Kayis} implemented in the \textit{BDgraph} \cite{mw19} package. All the code to reproduce the following results is available at \url{https://github.com/jackkuipers/GWnorm}.

First we look at networks of size $p=10$ and sample random prime graphs with a density (expected neighbourhood size) of $\zeta=2$. We then sample data by sampling a precision matrix from the $\G$-Wishart distribution (with \textit{BDgraph::rgwish}), sample $n=20$ observations from a multivariate normal with that precision matrix, and compute $D$ as the sample correlation matrix.

We estimate the $\G$-Wishart normalising constant with our \textit{GWnorm} package for Monte Carlo sample sizes of $1000\cdot2^{i}$ for $i = 0,\ldots,5$ and with the \textit{BDgraph} package with twice the number of samples to have similar computational time for each approach. With the parameter $\delta = 3$ (corresponding to $\beta = 1$), we computed the variance in the estimates of $\log \const_\G(\delta, D)$ over 100 repetitions. We repeated the process over 40 different random networks each with a random $D$ matrix.

The results (Supplementary Figure~\ref{fig:simConverge}) show a clear decrease in variance with increasing time, with slopes in the log-log plot of $-0.96$ for \textit{BDgraph} and $-1.00$ for \textit{GWnorm}, aligning well with the theoretical value of -1 for an inverse relationship between variance and sample size. We also observe a clear separation between the methods, with our approach implemented in \textit{GWnorm} being orders of magnitude more efficient.

With the linear behaviour, the product of the variance and the time $(V\cdot t)$ represents the quality of each sample for a given time budget. We therefore extract the relative efficiency of the two methods as the following ratio
\begin{equation} \nonumber
\mathrm{Variance~ratio} = \frac{(V\cdot t)_{\mathrm{BiDAG}}}{(V\cdot t)_{\mathrm{GWnorm}}}
\end{equation}
which accounts for any time differences in the various runs, and pairs the results for each simulated example.

To explore the efficiency in more detail, we rerun the simulations with $5\cdot10^{5}$ samples for \textit{GWnorm} and $10^{6}$ for \textit{BDgraph}, with 40 repetitions per network to estimate the variances and over 50 different networks. The results (Supplementary Figure~\ref{fig:simPrime}, left column) show that our approach is typically several orders of magnitude more efficient. 

We additionally colour the points in the plot by the dimension of the integral in the Fourier space represented as the fraction of missing edges: $\tilde{\tau} = \tau / \binom{p}{2}$. There is a slight pattern then when there are fewer missing edges (Supplementary Figure~\ref{fig:simPrime}, lighter orange dots), and our approach has a lower dimensional integral, we have a higher gain in efficiency that when there are more missing edges (Supplementary Figure~\ref{fig:simPrime}, darker purple dots), as would be expected.}

\rev{We extended the simulation to networks of size $p=15$, reducing the number of Monte Carlo samples for \textit{BDgraph} to $6\cdot10^{5}$ to balance the time, and sampling $n=45$ observations (setting $n = \frac{p^2}{2}$), and to network of size $p=20$ with $3\cdot10^{5}$ samples for \textit{BDgraph} and $n=80$ observations.

The results (Supplementary Figure~\ref{fig:simPrime}) show that while our approach is typically still much more efficient, there are some cases (6\% at $p=15$ and 22\% at $p=20$) where the direct sampler of \textit{BDgraph} is better. However, at the larger size, \textit{BDgraph} had numerical issues and failed to produce an estimate at all in 14\% of the cases.  There were some other cases where \textit{BDgraph} could not produce an estimate, with an overall rate of 18.3\%. At the same time, there were 8.5\% of cases where the difference between the importance sample and integral was such that also \textit{GWnorm} did not produce an estimate. This highlights that estimating and computing $\G$-Wishart normalising constants can be challenging, and aligns with the comment in \cite{mml23} about limitations of the approach of \cite{Atay-Kayis} for larger networks.}

\begin{figure}
\centering
\includegraphics[width=\textwidth]{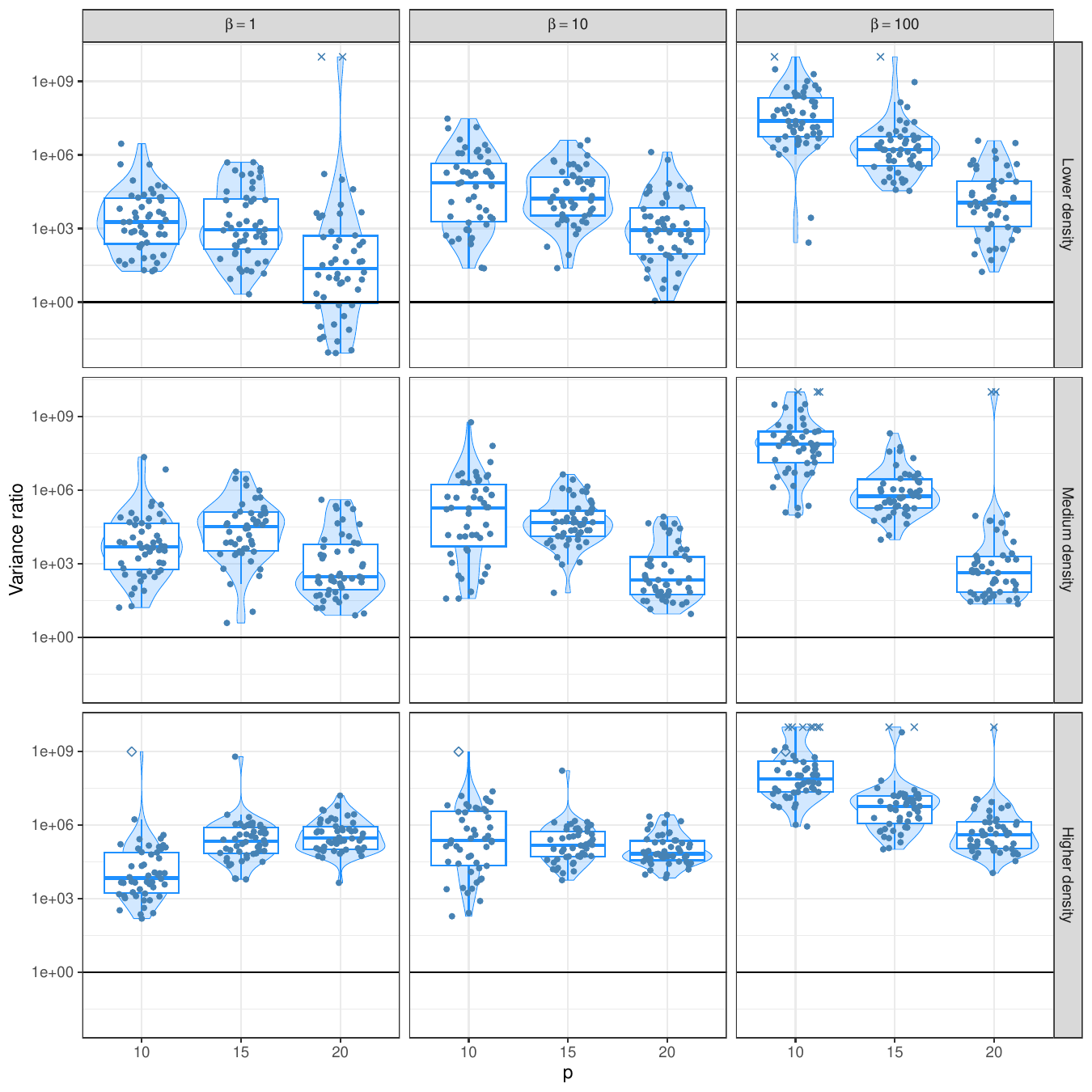}
\caption{\rev{Relative efficiency of \textit{GWnorm} against \textit{BDgraph} \cite{mw19} for random graphs of different sizes and densities for a range of parameters $\beta =\frac{\delta-2}{2}$. Where \textit{BDgraph} failed to compute the normalising constant is indicated with diagonal crosses placed at $10^{10}$. Where \textit{GWnorm} returned an exact result is indicated with diamonds placed at $10^{9}$.}}
\label{fig:simDensity}
\end{figure}

\rev{Finally, we extend our simulation study to the more realistic setting where we remove the restriction to prime graphs, and use prime decomposition to reduce each normalising constant to a set of lower-dimensional ones (for both methods). We observe (Supplementary Figure~\ref{fig:simDensity}, top left) similar behaviour to the prime case of Supplementary Figure~\ref{fig:simPrime}, but with far fewer cases where \textit{BDgraph} fails.

We additionally explored the effect of the $\delta$ parameter, which in the posterior is increased from the prior value by the number of observations $n$. For larger $\delta$, which therefore correspond to more realistic values, we see a clear relative improvement our approach (Supplementary Figure~\ref{fig:simDensity}, top row).

The other parameter we varied was the network density. Along with the lower density case above, we simulated the medium density where the probability of any edge is $\frac{1}{2}$ and where we took the complement of the lower density networks to arrive at higher density ones. In every one of these cases (Supplementary Figure~\ref{fig:simDensity}, bottom rows), our approach was better than using \textit{BDgraph}, and typically by several orders of magnitude.

At the higher values of $\delta$, \textit{BDgraph} often failed to return a value, even for the smaller networks. At the highest value for the larger networks, even accounting for the higher variance of \textit{BDgraph}, the values it produced seemed to start to become unaligned with our results. Our results, however, remained consistent with large $\delta$ asymptotics \cite{wong2025conjecture}, and given the failures even for smaller networks, may simply suggest the accumulation of numerical imprecision in the \textit{BDgraph} approach.}


\clearpage


\renewcommand\thepage{\arabic{page}}
\setcounter{page}{25}

\end{document}